\documentstyle[12pt,epsf]{article}
\begin{document}
     
%
     
\newcommand\ie {{\it i.e.}}
\newcommand\eg {{\it e.g.}}
\newcommand\etc{{\it etc.}}
\newcommand\cf {{\it cf.}}
\newcommand\etal {{\it et al.}}
\newcommand{\be}{\begin{eqnarray}}
\newcommand{\ee}{\end{eqnarray}}
\newcommand{\jp}{$ J/ \psi $}
\newcommand{\pp}{$ \psi^{ \prime} $}
\newcommand{\ppp}{$ \psi^{ \prime \prime } $}
\newcommand{\dd}[2]{$ #1 \overline #2 $}
\newcommand\noi {\noindent}

\begin{flushright}
LBNL-42294
\end{flushright}
\vspace{1cm}
     
\begin{center}
     
{\Large { The $x_F$ Dependence of $\psi$ and Drell-Yan Production}}
\vspace{2mm}

R. Vogt\footnote{This work was supported in part by
the Director, Office of Energy Research, Division of Nuclear Physics
of the Office of High Energy and Nuclear Physics of the U. S.
Department of Energy under Contract Number DE-AC03-76SF00098.} \\
\vspace{2mm}
{\it Nuclear Science Division,} \\
{\it Lawrence Berkeley National Laboratory,} \\
{\it  Berkeley, CA 94720} \\
{\it and}\\
{\it Physics Department,}\\
{\it University of California at Davis} \\
{\it Davis, CA 95616}\\
\vspace{2mm}

     
\vspace{1cm}
{\bf Abstract}\\
\begin{quote} \begin{small}
We discuss the nuclear 
dependence of $\psi$ and $\psi'$ production in hadron-nucleus interactions
as a function of longitudinal momentum fraction $x_F$.
Nuclear effects such as final-state absorption, interactions with comovers, 
shadowing of parton distributions, energy loss, and intrinsic heavy-quark 
components are described separately and incorporated into the model which is 
then compared to the preliminary E866 data.  
The resulting nuclear dependence of Drell-Yan production at 800 GeV
and proposed measurements of $\psi$, $\psi'$ and Drell-Yan production
at 120 GeV are also calculated.\\[2ex] 
\end{small} \end{quote} 
\end{center}
\newpage

\section{Introduction}

The factorization theorem of perturbative QCD \cite{CSS}, separates the
perturbatively-calculable short-distance quark and gluon interactions from
the nonperturbative dynamics underlying the parton distribution functions
in the hadron.  The effectiveness of the factorization theorem in nuclear
targets with mass number $A$ can be obtained by a comparison of the
perturbative production cross sections of hard processes in nuclei to those
in a free proton, or, since nuclear isospin effects are generally small, a
nucleon.  The dependence of particle production on atomic mass
number $A$ 
is conventionally parameterized by a power law as 
\cite{E772DY,NA3,E537,E772,E789,na50,e866,MJL}
\be
\sigma_{pA} = \sigma_{pN} A^\alpha \, \, , \label{alfint}
\ee where $\sigma_{pA}$ and $\sigma_{pN}$ are the integrated particle
production cross sections in proton-nucleus and
proton-nucleon interactions respectively.  
If factorization is satisfied, then particle production should be 
independent of the
presence of nuclear matter and $\sigma_{pA}$ 
would grow linearly with $A$, implying $\alpha = 1$.  
Drell-Yan production, integrated over all kinematic variables,
shows this linear growth to rather high precision
\cite{E772DY}.  A number of experiments have measured 
a less than linear $A$ dependence for $\psi$ and $\psi'$ production
\cite{NA3,E537,E772,E789,na50,e866,MJL}.  Typical values of the exponent
$\alpha$ in Eq.~(\ref{alfint}) are between 0.9 and 1.  
This nonlinear growth of the $\psi$ cross section with $A$ has been used
to determine an effective nuclear absorption cross section \cite{na50,RV,klns}
under the assumption that the deviation of $\alpha$ from unity is solely due
to $\psi$ dissociation by nucleons.  However,
attributing the entirety of the integrated nuclear dependence to final-state
absorption neglects other possible contributions, perhaps
resulting in an overestimate of the nuclear absorption cross section.
This paper identifies a number of possible nuclear effects on $\psi$ and,
consequently, Drell-Yan production and examines their contributions to
Eq.~(\ref{alfint}). 

Any dependence on the kinematic variables such as projectile energy or
longitudinal momentum fraction, $x_F$, would reveal the importance of going
beyond a simple $A$ scaling for production
and a constant absorption cross section for $\psi$ production.  
Indeed, it has long been known 
that in quarkonium production $\alpha$ decreases as a function of $x_F$
\cite{NA3,E537,E772}.  There are a number of effects which could contribute to
the $x_F$ dependence.  The nuclear parton densities are systematically
different from those in a deuteron or a free proton as a function of parton
momentum fraction $x$ \cite{Arn}.  Such alterations of the parton densities 
are universal because they are independent of the final state, affecting both
$\psi$ and Drell-Yan production.  Another possible universal
component is energy loss by the incoming parton as it traverses the
nucleus \cite{GM,BH}. The remaining mechanisms primarily
affect $\psi$ and $\psi'$
production.  Absorption of the produced $\psi$ or $c \overline c$ state
by interactions with nucleons and/or produced particles has been claimed to 
be responsible for all $\psi$ suppression in nuclear collisions, 
at least up to S+U interactions
\cite{RV,klns,GV,Armesto}.  The importance 
of these absorption effects depends on the
production mechanism and the magnitude of the interaction cross sections.
Energy loss by the final-state color octet $c \overline c$ has also been 
suggested \cite{KS}.  Finally, the presence of 
intrinsic heavy-quark component of the projectile wavefunction are also
considered \cite{intc,BandH}. 
 
We study the $x_F$ dependence using a two-component model employing concepts
developed in Ref.\ \cite{VBH1}.  The first component, based on perturbative
QCD \cite{HPC}, is a hard-scattering approach that 
would yield an approximately linear $A$ dependence, as in dilepton
production by the Drell-Yan mechanism.  The $A$ dependent effects associated
with hard scattering are
final-state interactions, nuclear shadowing in the target and energy loss in
the projectile.  
The second component of the $x_F$ dependence arises from
intrinsic $c \overline c$ pairs in the projectile wavefunction
\cite{intc,BandH}. 
Since the charm quark mass is large, these intrinsic heavy quark pairs 
carry a significant fraction
of the longitudinal momentum and contribute at large $x_F$  
whereas perturbative production decreases strongly with $x_F$.  
The light spectator quarks in the intrinsic \dd{c}{c} state 
interact on the nuclear surface, leading to an approximate $A^{2/3}$ dependence
\cite{BandH}.  

Such a separation of production mechanisms was first proposed by
the NA3 collaboration \cite{NA3} when they divided their data into hard and
diffractive components so that
\be  \frac{d\sigma_{pA}}{dx_F}  & = &
A^{\alpha^\prime} \frac{d\sigma_h}{dx_F} + A^\beta \frac{d\sigma_d}{dx_F} \, \,
. \label{twocom} \ee 
The `hard' component, $\sigma_h$, includes nuclear shadowing and parton
energy loss which can alter its effective $A$ dependence.  
Final state dissociation of the $\psi$ or $c \overline c$
state by nucleons and secondaries, which does not affect the parton densities,
also contributes to the effective exponent
$\alpha'$.  The `diffractive' component, $\sigma_d$, is 
assumed to be due to intrinsic charm and only contributes
significantly at $x_F > 0.25$.  The NA3 Collaboration found $\alpha^\prime =
0.97$ and $\beta = 0.71$
for proton projectiles \cite{NA3}. Taken together, these components give the 
effective $\alpha$ of Eq.~(\ref{alfint}). 
We will discuss each mechanism in detail in the following
sections.  Since Drell-Yan production would also be affected by nuclear
shadowing in the target and energy loss by the projectile partons, 
its effective $A$ dependence should also depend on 
$x_F$.  Thus we will include the Drell-Yan $A$ dependence in our study.

\begin{figure}[htbp]
\setlength{\epsfxsize=\textwidth}
\setlength{\epsfysize=0.3\textheight}
\centerline{\epsffile{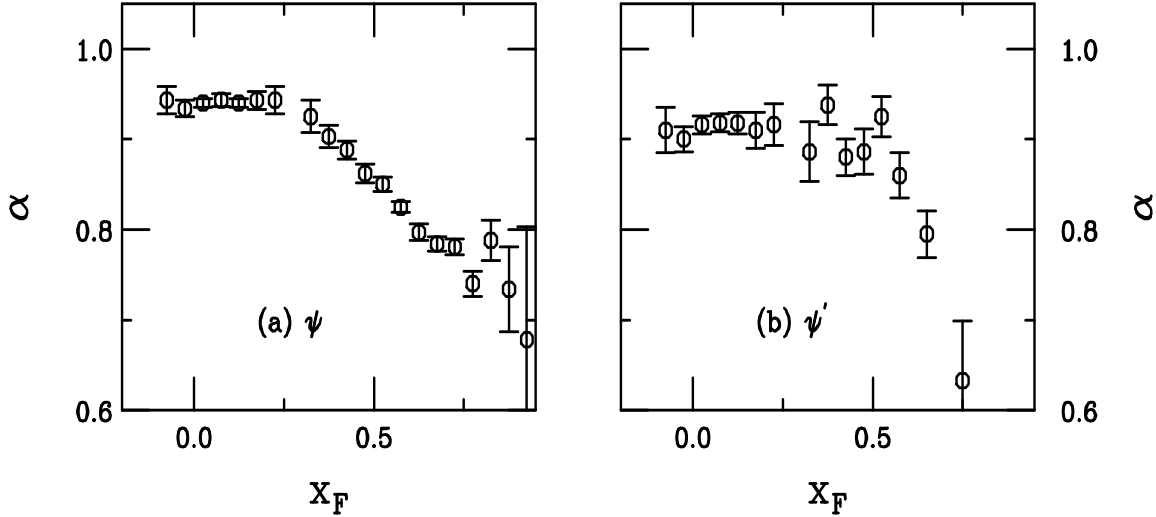}}
\caption[]{The preliminary E866 data \protect\cite{MJL}
for the $\psi$ and $\psi'$ $A$ dependence
as a function of $x_F$. } 
\label{data}
\end{figure}

We will calculate $\alpha(x_F)$ for
each effect individually and then compare the shapes of $\alpha(x_F)$ 
with the preliminary E866
$\psi$ and $\psi'$ 800 GeV $pA$
data \cite{MJL}, shown in Fig.~\ref{data}, when all
the mechanisms are combined.  
For the $\psi$ $\alpha \approx 0.94$ until $x_F \approx 0.25$, decreasing to
$\alpha \approx 0.7$ at large $x_F$ and for the $\psi'$  $\alpha \approx 0.91$,
effectively constant, until $x_F \approx 0.5$.  
The two values of $\alpha$ are essentially compatible with each other
within the experimental uncertainties.  No decrease at negative $x_F$ is
observed, contrary to previous results \cite{E789,e866}.
The drop to $\alpha \approx 0.7$
at $x_F \approx 0.9$ is similar to previous measurements \cite{NA3,E772}.
Since the E866 targets are tungsten,
W $(A=184)$, and beryllium, Be $(A=9)$, we calculate the cross section per
nucleon for each target according to Eq.~(\ref{twocom}), obtaining from
Eq.~(\ref{alfint}), 
\be \alpha(x_F) = 1 + \frac{\ln[(d\sigma_{p{\rm W}}/A_{\rm W} 
dx_F)/(d\sigma_{p{\rm 
Be}}/A_{\rm Be} dx_F)]}{\ln(A_{\rm W}/A_{\rm Be})} \, \, . \label{alfdef} \ee

We first discuss $\psi$ and Drell-Yan production in QCD and then each nuclear
effect in turn.  Quarkonium production by color evaporation and in
non-relativistic QCD is discussed in Section 2 and dilepton
production by the Drell-Yan mechanism is briefly touched upon in Section 3.
Three different models of nuclear absorption are discussed in
Section 4.  The first two, absorption of pure
color octet and color singlet states
respectively, are used in conjunction with the color evaporation model of
quarkonium production.  The last, 
a combination of octet and singlet absorption,
is coupled to quarkonium production in non-relativistic QCD.
A discussion of quarkonium dissociation by comoving secondaries is presented in
Section 5.  Three different parameterizations of nuclear shadowing are
described in Section 6.  Several models of energy loss are discussed in 
Section 7.  The intrinsic charm model is introduced in Section 8.
The combined results are given in Section 9, along with
predictions for projected lower energy measurements at 120 GeV.  A nontrivial
combination of effects is required to understand the $\psi$ and $\psi'$
data.
 
\section{Charmonium Production}

There are two basic models of quarkonium hadroproduction that have enjoyed
considerable phenomenological success.  
The first, the color evaporation model, treats all charmonium production
identically to $c \overline c$ production below the $D \overline D$ threshold.
The more recent non-relativistic QCD approach involves an expansion of
quarkonium production in powers of $v$, the relative $Q$-$\overline
Q$ velocity within the bound state.  Each approach
will be described in turn and the
$x_F$ distributions in $pp$ interactions will be presented to provide a basis
for understanding $d\sigma_h/dx_F$, Eq.~(\ref{twocom}), before nuclear effects
are included.  We will also show
the relative contributions from $gg$ fusion, $q \overline q$ annihilation,
and $gq$ scattering.

\subsection{The Color Evaporation Model}

In the color evaporation model, CEM,
quarkonium production is treated identically to
open heavy quark production except that
the invariant mass of the heavy quark pair is restricted to be less than twice
the mass of the lightest meson that can be formed with one heavy
constituent quark.  For
charmonium the upper limit on the $c \overline c$ pair mass is then $2m_D$.
The hadroproduction of heavy quarks at leading order (LO) 
in perturbative QCD is the sum of contributions from \dd{q}{q} annihilation 
and $gg$ fusion.  
The hadroproduction cross section
is a convolution of the $q \overline q$ and $gg$ partonic cross
sections with the parton densities in projectile $A$ and target $B$.
If $x_F$ is the \dd{c}{c} longitudinal momentum fraction 
in the $AB$ center-of-mass frame and 
$\sqrt{s}$ is the center-of-mass energy of a nucleon-nucleon collision, 
the cross section for production of
free \dd{c}{c} pairs with mass $m$ is \cite{Bkp2}
\be      \frac{d \sigma^{c \overline c}}{dx_F dm^2} & = & 
\int_0^1 dx_1 dx_2 \, \delta(x_1x_2 s - m^2) \,
\delta ( x_F - x_1 + x_2 )
 H_{AB}(x_1,x_2;m^2) \label{cemdef} \\
& = & \frac{H_{AB}(x_{01},x_{02};m^2)}{\sqrt{x_F^2 s^2 + 4m^2s}} \, \, , \ee 
where $x_1$ and $x_2$ are the fractions of the hadron momentum carried by the
projectile and target partons respectively.  After integration over the delta
functions in Eq.~(\ref{cemdef}), $x_{01,02}
= \frac{1}{2}(\pm x_F + \sqrt{x_F^2 + 4m^2/s})$.  The convolution of the
partonic cross sections and the parton densities is
\be
\lefteqn{H_{AB}(x_1,x_2; m^2) = f_g^A(x_1,m^2)f_g^B(x_2,m^2)\sigma_{gg}(m^2)}
\\    & & \mbox{} 
+ \sum_{q=u,d,s} [f_q^A(x_1,m^2) f_{\overline q}^B(x_2,m^2) + f_{\overline 
q}^A(x_1,m^2) f_q^B(x_2,m^2)] \sigma_{q \overline q}(m^2) 
\nonumber \ee
where the parton densities $f_i(x,m^2)$ are evaluated
at momentum fraction $x$ and scale $m^2 = x_1x_2 s$ and $m$ is 
the invariant mass of the $c
\overline c$ pair. The sum over $q$ includes only light quark flavors. 
The LO partonic cross sections are \cite{Bkp2} 
\be
\sigma_{gg}(m^2) & = & \frac{\pi \alpha_s^2(m^2)}{3m^2} \left\{ \left(1 + 
\frac{4m_c^2}{m^2} + \frac{m_c^4}{m^4} \right)
\ln \left( \frac{1 + \lambda}{1 - \lambda} \right) 
- \frac{1}{4} \left( 7 + \frac{31m_c^2}{m^2} \right)
\lambda \right\} \, \, , \\
\sigma_{q \overline q}(m^2) & = & \frac{8 \pi
\alpha_s^2(m^2)}{27 m^2} \left( 1 + \frac{2m_c^2}{m^2} \right) \lambda \, \, , 
\ee where $\lambda = \sqrt{1 - 4m_c^2/m^2}$. 

The LO charmonium cross section for charmonium state $i$, 
$\tilde{\sigma}_i$, is then obtained by integrating
the free \dd{c}{c} cross section over the pair mass from the \dd{c}{c}
production threshold, $2m_c$, to the open charm threshold,
$2m_D = 3.74$ GeV.  Then \be \frac{d \tilde{\sigma}_i}{d x_F} = 2F_i
\int_{2m_c}^{2m_D} m \, dm \,  \frac{d \sigma^{c \overline c}}{dx_F dm^2}
 \, \, , \label{cevap} \ee  
where $F_i$ is the fraction of $\sigma^{c \overline c}$
that produces the final-state $c \overline c$ resonance.

The CEM assumes that the 
quarkonium dynamics are identical to those of low invariant 
mass $c \overline c$ pairs.  The
hadronization of the charmonium states from the $c \overline c$ pairs is
nonperturbative, involving the emission of one or more soft gluons.
A different nonperturbative matrix element is needed
for the direct production of each charmonium state.
Each nonperturbative matrix element is represented by a 
single universal factor $F_i$ which depends on the charm quark mass, $m_c$, 
the scale of $\alpha_s$, $\mu$, and the 
parton densities\footnote{The mass and scale parameters are $m_c = 1.2$ GeV and
$\mu=2m_c$ for the MRST LO \cite{mrsg}, CTEQ 4L \cite{cteq}, and CTEQ 3L
\cite{cteq3} parton distributions and $m_c = 1.3$ GeV and $\mu = m_c$ with
the GRV 94 LO \cite{GRV94} parton distributions.  At next-to-leading order,
the fraction $F_\psi$ is
2.54\% for all parton distributions \cite{HPC}.}. In our calculations with the
CEM, we use the leading order
MRST LO parton distributions \cite{mrsg}.  This set,
more recent than the GRV 94 LO densities, has a low initial $Q^2$, $Q_0 = 1$
GeV. Once $F_i$ 
has been determined for each state, {\it e.g.}\ $\psi$, $\psi'$ or $\chi_{cJ}$,
the model successfully predicts the energy and momentum dependencies.  
We note that $F_{\psi}$ includes both direct
$\psi$ production and indirect production through radiative decays of the
$\chi_{cJ}$ states and hadronic $\psi'$ decays. 

Since $F_i$ must be a constant for
the model to have any predictive power,
the relative differential and
integrated quarkonium production rates should be independent of
projectile, target, and energy.  This appears to be true for the charmonium
production ratios $\sum_J \chi_{cJ}/ \psi \approx 0.4$ 
and $\psi^\prime/\psi \approx
0.14$ \cite{Teva,CarlosEPS,Anton1,Anton2,Ronceux}.  See Ref.~\cite{HPC} for
more details.

The next-to-leading order (NLO) quarkonium production cross section in the
CEM \cite{HPC}
was calculated using the $Q \overline Q$ production code of 
Ref.\ \cite{MNR} with the mass cut in Eq.~(\ref{cevap}). 
When the NLO contribution is included, the $p_T$ dependence of 
$\psi$ production at the Tevatron has been shown to agree with the
CEM calculations \cite{SchulV}. 
The LO and NLO calculations agree equally well with the energy and $x_F$
dependent data if $F_i^{\rm LO}$ is defined as $F_i^{\rm NLO}$ multiplied 
by a theoretical $K$ factor,
the ratio of the NLO to LO cross sections \cite{HPC}.

\begin{figure}[htbp]
\setlength{\epsfxsize=\textwidth}
\setlength{\epsfysize=0.3\textheight}
\centerline{\epsffile{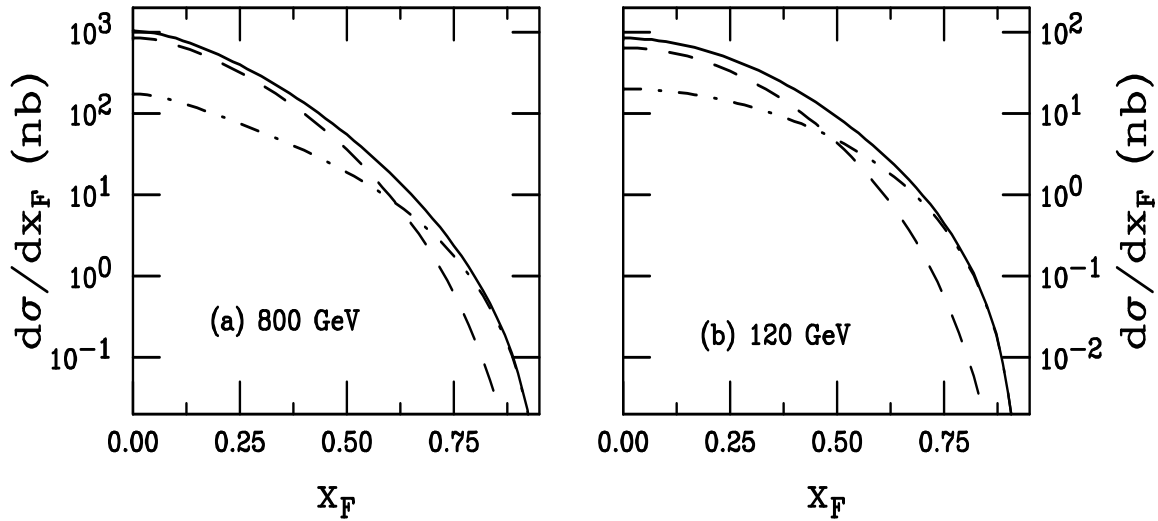}}
\caption[]{The $\psi$ $x_F$ distributions at (a) 800 GeV and (b) 120 GeV in
the CEM.  The contributions from $gg$ fusion (dashed) and
$q \overline q$ annihilation (dot-dashed) are given along with the total
(solid).} 
\label{cem}
\end{figure}

Figure~\ref{cem} shows the forward $x_F$ distributions for $\psi$ production
in $pp$ collisions\footnote{The $x_F$ distributions are symmetric around $x_F =
0$ in $pp$ production.} 
in the CEM at 800 GeV and 120 GeV using the MRST LO \cite{mrsg} parton
distributions.   
The $\psi'$
distributions are identical except for the relative fraction of $\psi'$
production below the $D \overline D$ threshold and are thus not shown.  
Note that at large $x_F$, 
$x_F \geq 0.6$ at 800 GeV and $\geq 0.5$ at 120 GeV, $q
\overline q$ annihilation is the most important contribution to 
the cross section.  

\subsection{Quarkonium Production in Non-Relativistic QCD}

An alternative model of quarkonium production, the color singlet model
\cite{baru}, predicted that high $p_T$ $\psi$ production would be dominated
by $\chi_{cJ}$ decays.  It also predicted 
that direct $\psi$ and $\psi'$ production would be rare
because a hard gluon emission was required to make a color singlet $^3S_1$
state on a perturbative timescale.  On the other hand, the CEM is an average
over the color and spin of the produced $c \overline
c$ pair and cannot make such predictions.  
Soon after the high $p_T$ Tevatron data \cite{ppbar}
made clear that the hard gluon emission constraint in the color singlet
model severely underpredicted
direct $\psi$ and $\psi'$ production, the non-relativistic QCD, NRQCD, 
approach to quarkonium
production was formulated \cite{bbl}.  This approach does not restrict
the angular momentum or color of the quarkonium state to only the leading
singlet state.  For example, the final-state $\psi$ may produced as a $^3P_0$
color octet state which becomes a $\psi$ through nonperturbative soft gluon
emissions.  Thus the NRQCD model is similar in spirit to the CEM 
albeit with more nonperturbative parameters, as we
will see.

The $x_F$ distribution of a charmonium state, $C$, in NRQCD is
\be      \frac{d \sigma^C}{dx_F} & = & \sum_{i,j} 
\int_0^1 dx_1 dx_2 \delta ( x_F - x_1 + x_2 ) f_i^A(x_1,\mu^2)f_j^B(x_2,\mu^2)
\widehat{\sigma}(ij \rightarrow C) \label{signrqcd} \\
\widehat{\sigma}(ij \rightarrow C) & = & \sum_n C^{ij}_{Q \overline Q \, [n]} 
\langle {\cal O}_n^C \rangle \, \, , \ee 
where the $C$ production cross section, $\widehat{\sigma}(ij \rightarrow C)$,
is the product of expansion coefficients, $C^{ij}_{Q \overline Q \, [n]}$, 
calculated perturbatively in powers of $\alpha_s(\mu^2)$ 
and nonperturbative parameters,
$\langle {\cal O}_n^C \rangle$, describing the hadronization of the 
charmonium state.  We
use the parameters determined by Beneke and Rothstein for fixed-target
hadroproduction of charmonium with $m_c = 1.5$ GeV and $\mu = 2m_c$ and
the CTEQ 3L parton densities \cite{benrot}. 
The total $\psi$ $x_F$ distribution includes radiative decays of the
$\chi_{cJ}$ states and hadronic decays of the $\psi'$,
\be \frac{d \sigma_{\psi}}{dx_F} =  \frac{d \sigma_{\psi}^{\rm dir}}{dx_F}
+ \sum_{J=0}^2 B(\chi_{cJ} \rightarrow \psi X) \frac{d 
\sigma_{\chi_{cJ}}}{dx_F}
+ B(\psi' \rightarrow \psi X) \frac{d\sigma_{\psi'}}{dx_F} \label{dirpsi}
\, \, . \ee In contrast, in the CEM, the $x_F$ distributions of all states are
assumed to be the same.  Thus $F_\psi$ in Eq.~(\ref{cevap}) implicitly
includes the $\chi_{cJ}$ and $\psi'$
decay contributions given explicitly in Eq.~(\ref{dirpsi}).

For completeness, we now present the individual contributions to $\psi'$ and 
$\psi$ production also given in Ref.~\cite{benrot}.  
To simplify the cross sections, we define the coefficients 
proportional to $\alpha_s^2$ and $\alpha_s^3$ as
\be B_2 = \frac{\pi \alpha_s^2}{(2m_c)^3 s} \,\,\,\,\,\,\,\,\,\,\,\,
 B_3 = \frac{\pi \alpha_s^3}{(2m_c)^5} \ee
and denote the delta and theta functions of argument $x_1x_2 - 4m_c^2/s$
by $\delta_x$ and $\theta_x$. 
Direct $\psi$ production has only contributions
from $gg$ fusion and $q \overline q$ annihilation \cite{benrot}, as in the CEM,
\be \widehat{\sigma}(gg \rightarrow \psi) & = & \frac{5}{12} B_2 \delta_x
\Delta_8(\psi) + \frac{20}{81} B_3 \theta_x \langle {\cal O}_1^\psi ( ^3S_1 )
\rangle z^2 \left[ \frac{1-z^2 + 2z\ln z}{(1-z)^2} \right. \label{psigg}
\nonumber \\
 & & \mbox{} + \left. \frac{1 - z^2 - 2z\ln z}{(1+z)^2} \right] \, \, , \\
 \widehat{\sigma}(q \overline q \rightarrow \psi) & = & \frac{16}{27} B_2 
\delta_x  \langle {\cal O}_8^\psi ( ^3S_1 ) \rangle \label{psiqq} \, \, , \ee
where $z = 4m_c^2/(x_1x_2 s)$.  Note that in the terms proportional to 
$\delta_x$, the momentum fractions
$x_1$ and $x_2$ are the same as $x_{10}$ and $x_{20}$ in the CEM
for $m=2m_c$.  The $\psi'$ cross sections are identical except
for the values of the matrix elements $\Delta_8$, $\langle {\cal O}_1 ( ^3S_1)
\rangle$, and  $\langle {\cal O}_8 ( ^3S_1)
\rangle$.  The nonperturbative matrix elements 
for $\psi$ and $\psi'$ production in Eqs.~(\ref{psigg}) and (\ref{psiqq}) are
\cite{benrot} 
\begin{equation}  \begin{array}{cc}
 \Delta_8(\psi) = 0.03 \, {\rm GeV}^3 \, \, , & \Delta_8(\psi') = 0.0052 \,
{\rm GeV}^3 \, \, , \\
 \langle {\cal O}_1^\psi ( ^3S_1 )
\rangle = 1.16 \, {\rm GeV}^3 \, \, , &  \langle {\cal O}_1^\psi ( ^3S_1 )
\rangle = 0.76 \, {\rm GeV}^3 \, \, , \\
 \langle {\cal O}_8^\psi ( ^3S_1 ) \rangle = 
0.0066 \, {\rm GeV}^3 \, \, , & \langle {\cal O}_8^\psi ( ^3S_1 ) \rangle = 
0.0046 \, {\rm GeV}^3 \, \, .
\end{array}  \label{psime} \end{equation}
At 800 GeV $\approx 63$\% of the $\psi'$ cross section is color
octet while 85\% of direct $\psi$ production comes from the octet
contribution.  This is because $\Delta_8(\psi) \approx 5.8 \Delta_8(\psi')$,
see Eq.~(\ref{psime}).   

Most of the octet contribution to $\psi$ production is from direct
production.  The singlet contribution, $\approx 40$\% of the total production
cross section at 800 GeV, is due to the $\chi_{c1}$.  The $\chi_{cJ}$
cross sections from Ref.~\cite{benrot} are given below according to their 
production mechanisms.
The following symmetry relations,
\be \langle {\cal O}_1^{\chi_{cJ}}( ^3P_J ) \rangle & = & (2J+1) \langle 
{\cal O}_1^{\chi_{c0}}( ^3P_0 ) \rangle \, \, , \label{chissp} \\
 \langle {\cal O}_8^{\chi_{cJ}}( ^3S_1 ) \rangle & = & (2J+1) \langle 
{\cal O}_8^{\chi_{c0}}( ^3S_1 ) \rangle \, \, , \label{chisss} \, \, \ee
are particularly useful for calculating the $\chi_{cJ}$ contributions to the 
total $\psi$
production cross section.  The $gg$ components of the $\chi_{cJ}$ cross
sections, all singlets, are 
\be \widehat{\sigma}(gg \rightarrow \chi_{c0}) & = & \frac{2}{3} B_2 \delta_x
\frac{\langle {\cal O}_1^{\chi_{c0}} ( ^3P_0 ) \rangle}{m_c^2} \, \, , 
\label{chigg0}\\
 \widehat{\sigma}(gg \rightarrow \chi_{c1}) & = & \frac{2}{9} B_3 \theta_x
\frac{\langle {\cal O}_1^{\chi_{c1}} ( ^3P_1 ) \rangle}{m_c^2}
\left[ \frac{4z^2 \ln z f_1(z)}{(1+z)^5 (1-z)^4} + \frac{f_2(z)}{3 (1+z)^4
(1-z)^3} \right] \, \, , \label{chigg1} \\
 \widehat{\sigma}(gg \rightarrow \chi_{c2}) & = & \frac{4}{15} 5 \, 
\widehat{\sigma}(gg \rightarrow \chi_{c0}) \label{chigg2} \, \, , \ee
where
\be f_1(z) & = & z^8 + 9z^7 + 26z^6 + 28z^5 + 17z^4 + 7z^3 - 40z^2 -4z -4 \, \,
, \nonumber \\
f_2(z) & = & z^9 + 39z^8 + 145z^7 + 251z^6 + 119z^5 - 153z^4 - 17z^3 - 147z^2 -
8z + 10 \, \, . \nonumber \ee
The $\chi_{c1}$ also has a singlet contribution from $gq$ scattering,
\be  \widehat{\sigma}(gq \rightarrow \chi_{c1}) & = & \frac{8}{81} B_3 \theta_x
\frac{\langle {\cal O}_1^{\chi_{c1}} ( ^3P_1 ) \rangle}{m_c^2}
\left[ -z^2 \ln z + \frac{4z^3 - 9z + 5}{3} \right] \label{chigq1} \, \, . \ee
The $q \overline q$ contributions to $\chi_{cJ}$ production are all color
octets, 
\be \widehat{\sigma}(q \overline q \rightarrow \chi_{c0}) & = & 
\frac{16}{27} B_2 \delta_x
\langle {\cal O}_8^{\chi_{c0}} ( ^3S_1 ) \rangle \, \, , \label{chiqq0} \\ 
 \widehat{\sigma}(q \overline q \rightarrow \chi_{cJ}) & = & (2J+1)
\widehat{\sigma}(q \overline q \rightarrow \chi_{c0}) \label{chiqq12} \, \,
. \ee 
The symmetry relations have been used to obtain the $gg$ contribution to
$\chi_{c2}$ production, Eq.~(\ref{chigg2}), and the $q \overline q$
contributions to $\chi_{c1}$ and $\chi_{c2}$ 
in Eq.~(\ref{chiqq12}).  The relevant nonperturbative
matrix elements for $\chi_{cJ}$ production are 
\be
\frac{\langle {\cal O}_1^{\chi_{c0}} ( ^3P_0 ) \rangle}{m_c^2} = 0.044 \, {\rm
GeV}^3 \,\, , 
\,\,\,\,\,\,\,\, \langle {\cal O}_8^{\chi_{c0}} ( ^3S_1 ) \rangle =
0.0032 \, {\rm GeV}^3 \, \, . \label{chime} \ee

In Ref.~\cite{benrot}, the singlet matrix elements in Eqs.~(\ref{psime}) and
(\ref{chime}) were calculated from the
quarkonium wavefunctions at the origin.  The octet matrix elements were fit to
Tevatron production data and $\Delta_8$ was obtained from 
a fit to total cross sections data at fixed-target energies.  
Note that in NRQCD, three parameters are needed to fix the
$\psi'$ production cross section while eight are needed for the total $\psi$
cross section.  Only one parameter for each state is needed in the CEM, a
considerable reduction.

\begin{figure}[htbp]
\setlength{\epsfxsize=\textwidth}
\setlength{\epsfysize=0.6\textheight}
\centerline{\epsffile{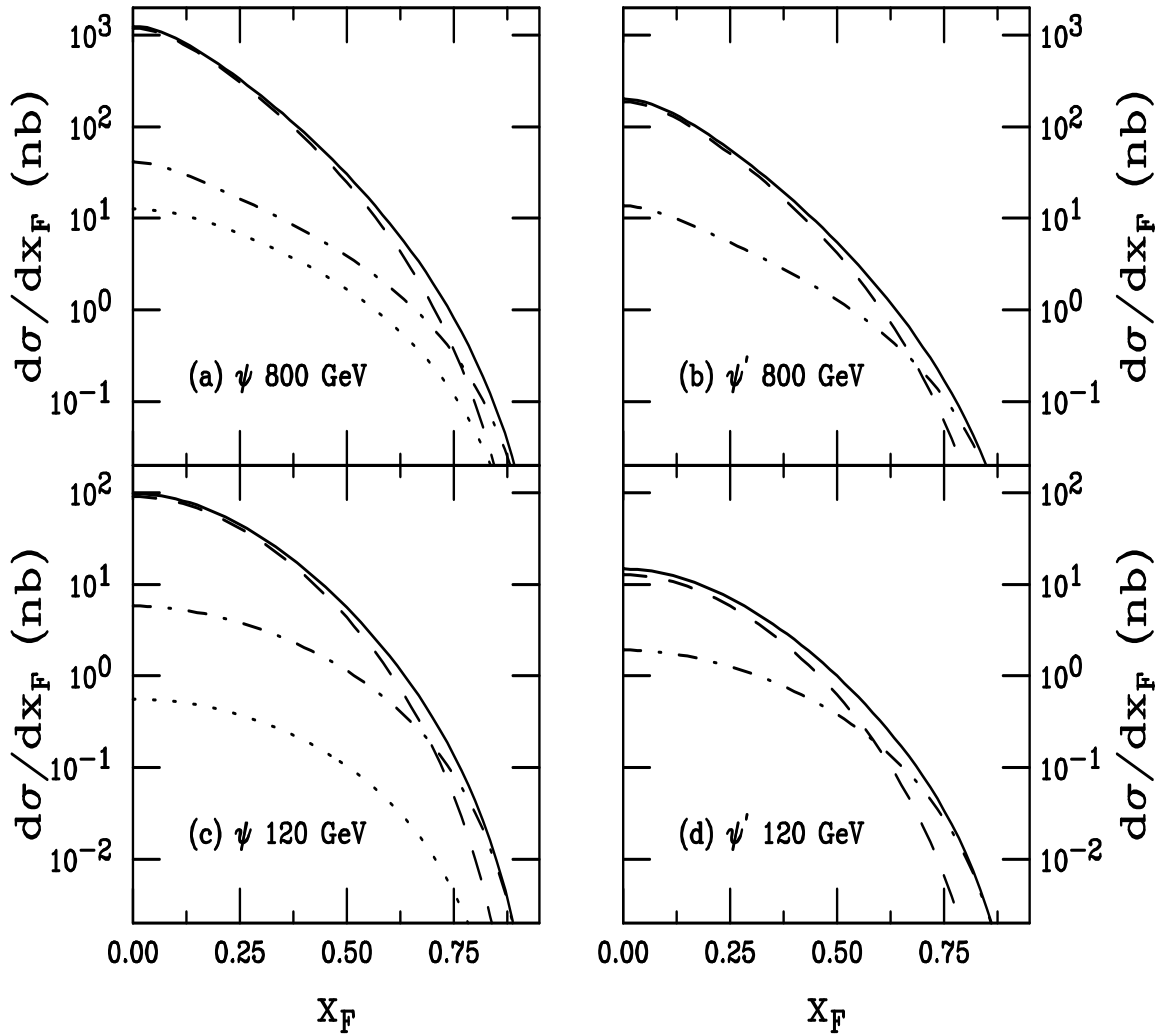}}
\caption[]{The $\psi$ $x_F$ distributions at (a) 800 GeV and (c) 120 GeV in
NRQCD.  The contributions from $gg$ fusion (dashed),
$q \overline q$ annihilation (dot-dashed), $gq$ scattering (dotted) and 
the total (solid) are given. 
The corresponding $\psi'$ distributions are given in (b) and (d).
The curves show $gg$ fusion (dashed), $q \overline q$ annihilation 
(dot-dashed), and the total (solid).} 
\label{nrqcd}
\end{figure}

The total $\psi$ forward
$x_F$ distributions\footnote{Note that, as in the CEM, the $pp$ $x_F$
distributions are symmetric around $x_F = 0$.} 
at 800 GeV and 120 GeV, Eq.~(\ref{dirpsi}), are shown in
Fig.~\ref{nrqcd}(a) and (c) respectively.  
Since the $\chi_{c0}$ branching
ratio to $\psi$ is less than 1\%, its contribution is virtually negligible.
However, $\approx 27$\% of the produced $\chi_{c1}$ states decay to $\psi$,
including the $gq$ scattering contribution, Eq.~(\ref{chigq1}), shown in the
dotted curves.  At 800 GeV this component is only a factor of 2-3 less than the
total $q \overline q$ contribution to the full $\psi$ cross section.
The $gg$ contribution from $\chi_{c1}$ decays,
Eq.~(\ref{chigg1}), and the smaller $\chi_{c2}$ decay contribution, $\approx
14$\%, provide most of the singlet component of total $\psi$ production.
Interestingly, when the $\chi_{cJ}$ decays are included, the octet
contribution to the total $\psi$ production cross section is 60\%, close to
the $\psi'$.  At 120 GeV, the percentage of octet production is larger for
both the total $\psi$ and $\psi'$, 67\% and 78\% respectively.
The $\psi'$ $x_F$ distributions at 800 GeV and 120 GeV
are given in Fig.~\ref{nrqcd}(b) and (d), including the individual 
contributions
from $gg$ fusion and $q \overline q$ annihilation.  Here the $q \overline q$
contribution is largest at $x_F \approx 0.7$ at 800 GeV and $\approx 0.6$ at
120 GeV, similar to the CEM.  However, for the total $\psi$ $x_F$ distribution,
the $q \overline
q$ contribution is significantly smaller than in the CEM, resulting in a
narrower $\psi$ $x_F$ distribution in NRQCD. The NRQCD cross section is a
factor of 3 smaller than the CEM cross section at $x_F \approx 0.9$ and a
factor of 1.5 smaller at $x_F \approx 0.5$ where the two are equal at $x_F =
0$.  The $q \overline q$ component does
not dominate the total $\psi$ distribution until $x_F \approx 0.8$
at 800 GeV and $\approx 0.7$ at 120 GeV.
These relative differences can influence the strength of
nuclear effects that depend on the nuclear quark and gluon distributions.

\section{Drell-Yan Production}

Lepton pairs are produced by the Drell-Yan process,
$q \overline q$ annihilation
into a virtual photon at leading order, $q \overline q \rightarrow
\gamma^\star \rightarrow l^+ l^-$ \cite{hpdy}.  
The partonic cross section for Drell-Yan production is
\be
\frac{d\widehat{\sigma}}{dM} = \frac{8 \pi \alpha^2}{9M} e_q^2 
\delta(\widehat{s} - M^2)
\ee
where $\widehat{s} = x_1x_2s$. 
To obtain the hadroproduction cross section as a function of
pair mass, $M$, and $x_F$, we must fold the partonic cross
section with the quark and antiquark densities evaluated at $M$,
here taken to be in the range $4 < M < 9$ GeV, between the $\psi$ and
$\Upsilon$ family regions.  Then
\be
\lefteqn{ \frac{d\sigma^{\rm DY}}{dx_F dM} =  
\frac{8 \pi \alpha^2}{9M} \int_0^1 dx_1 dx_2 \, \delta(x_1x_2 s - M^2)
\, \delta ( x_F - x_1 + x_2)} 
\nonumber \\ 
&   & \mbox{} \times \sum_q e_q^2 [f_q^p(x_1,M^2)f_{\overline q}^A
(x_2,M^2) + f_{\overline q}^p(x_1,M^2)f_q^A (x_2,M^2) ] \, \, .
\ee
After integrating the delta functions, the LO cross section, including the 
isospin of the target nucleus, is
\be \frac{d\sigma^{\rm DY}}{dx_F dM} & = & \frac{8 \pi \alpha^2}{9M} 
\frac{1}{\sqrt{x_F^2s^2 + 4M^2 s}}  
\sum_q e_q^2 [f_q^p(x_{01},M^2) (z_A f_{\overline q}^p 
(x_{02},M^2) + n_A f_{\overline q}^n (x_{02},M^2)) \nonumber \\
&  & \mbox{} + f_{\overline q}^p(x_{01},M^2) (z_A f_q^p 
(x_{02},M^2) + n_A f_q^n (x_{02},M^2)) ] \, \, ,
\label{dysc} \ee
where $z_A = Z/A$ and $n_A = N/A$ are, respectively, the fractions of protons 
and neutrons in the target nucleus.
 
When this leading order cross section is compared to data, it falls short
by an approximately constant factor, known as the $K$ factor.  Experimentally,
it is $\approx 1.7-2.5$, depending on the energy, mass range, and parton
distribution functions.  At NLO, the Compton and
annihilation processes, $q g \rightarrow q \gamma^\star$ and $q \overline q
\rightarrow g \gamma^\star$ respectively, contribute in addition to virtual 
corrections to the
LO cross section, resulting in a theoretical $K$ factor --- the ratio of the
NLO to the LO cross sections --- of approximately $1.4 - 2$, somewhat less
than that obtained by comparison to the data \cite{hpdy}.  
This theoretical $K$ factor serves the same purpose as
the adjustment of $F_\psi$ between the LO and NLO calculations in the CEM,
discussed earlier. 

\begin{figure}[htbp]
\setlength{\epsfxsize=\textwidth}
\setlength{\epsfysize=0.6\textheight}
\centerline{\epsffile{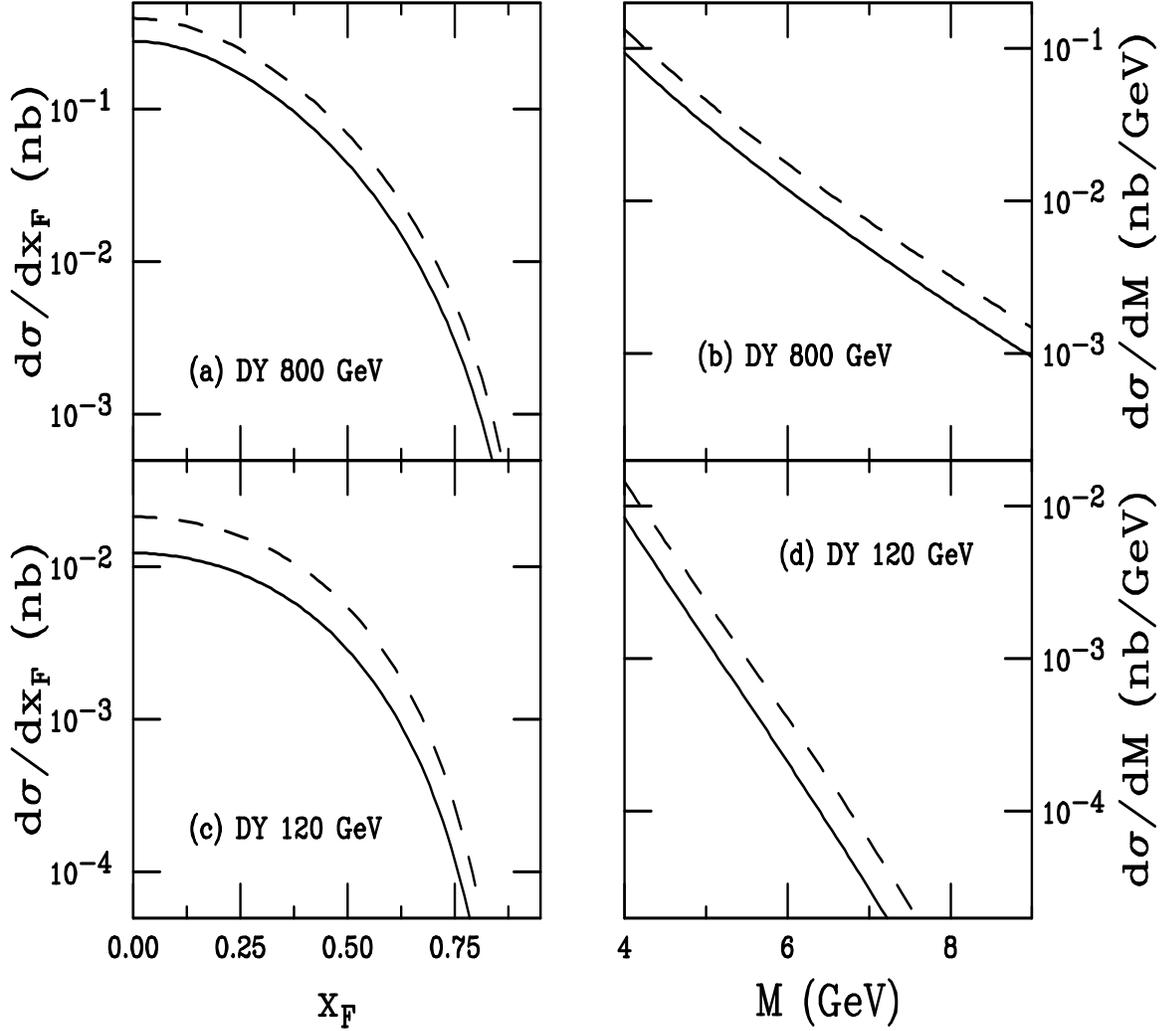}}
\caption[]{The Drell-Yan $x_F$ distributions for $4<M<9$ GeV
at (a) 800 GeV and (c) 120 GeV.  The Drell-Yan mass distributions, integrated
over $x_F$ are shown in (b) and (d) at 800 and 120 GeV respectively.  The
leading order results are given by the solid curves, the next-to-leading order
results are shown in the dashed curves.}
\label{dydists}
\end{figure}
In Fig.~\ref{dydists}(a) and (c) the Drell-Yan $x_F$ distribution is shown for
masses between 4 and 9 GeV at 800 GeV and 120 GeV.  The 120 GeV $x_F$
distribution does not extend over all $x_F$ since
the phase space for high $x_F$ and high mass pairs becomes limited.  
Because the quark
distributions have a harder $x$ dependence than the gluon, the Drell-Yan $x_F$
distribution is broader than the $\psi$ $x_F$ distributions shown in
Figs.~\ref{cem} and \ref{nrqcd}.  For comparison, both the LO and the NLO
distributions are shown.  There is some dependence of the $K$ factor on $x_F$.
At 800 GeV and $x_F \sim 0$, the theoretical $K$ factor is 1.4, increasing 
to 2.1 at $x_F = 0.9$.  The calculated $K$ factor is 20\% larger at 120 GeV.
The change with $x_F$ reflects the increasing importance of the 
Compton process with increasing $x_F$, corresponding to an increase in the
gluon density at low $x_2$.  The Drell-Yan mass distribution, integrated over
the corresponding $x_F$ ranges given in Fig.~\ref{dydists}(a) and (c), is 
shown in Fig.~\ref{dydists}(b) and (d) at 800 and 120 GeV respectively.  Since
$x_F$ is integrated over, the lower $x_F$ values are the most important for
the determination of the $K$ factor as a function of mass.  The theoretical
$K$ factor is $\approx 1.4$ at 800 GeV and 1.8 at 120 GeV
and does not change more than a few percent with
mass. 

Since the Drell-Yan mechanism produces lepton pairs which only interact
electroweakly, the $A$ dependence is expected to be weak because no
final-state interactions affect the lepton pair.  However, initial-state
interactions such as shadowing and energy loss
may influence the $A$ dependence, as we discuss in Sections 6 and 7.

\section{Nuclear Absorption in $pA$ Interactions}

The \dd{c}{c} pair may interact with nucleons and be dissociated or
absorbed before it can escape the target. 
The effect of nuclear absorption alone on the $\psi$ 
production cross section in
$pA$ collisions may be expressed as \be \sigma_{pA} =
\sigma_{pN} \int d^2b \, T_A^{\rm eff}(b) \, \, , 
\label{sigfull} \ee where $b$ is the impact parameter and
$T_A^{\rm eff}(b)$ is the effective nuclear profile function, \be T_A^{{\rm
eff}}(b) = \int_{-\infty}^{\infty}\, dz \, \rho_A (b,z) S^{\rm abs} \, \, .
\label{taeff} \ee The probability for the \dd{c}{c} pair to avoid
nuclear absorption and form a $\psi$, called the nuclear absorption
survival probability,
$S^{\rm abs}$, is \be S^{\rm abs} = \exp \left\{
-\int_z^{\infty} dz^{\prime} \rho_A (b,z^{\prime}) \sigma_{\rm abs}(z^\prime
-z)\right\} \label{nsurv} \ee where $\sigma_{\rm abs}$ 
is the charmonium (or $c \overline c g$ \cite{KSoct})
nucleon absorption cross section.  The nuclear density profile 
is $T_A(b) = \int_{-\infty}^\infty dz \rho_A(b,z)$ so that $T_A^{\rm
eff}(b)=T_A(b)$ when $S^{\rm abs} = 1$.  Nuclear charge density
distributions from data are used for $\rho_A$ \cite{JVV}.
Note that expanding $S^{\rm abs}$, integrating
Eq.~(\ref{sigfull}), and reexponentiating the results assuming $A$ is large
leads to Eq.~(\ref{alfint}) with $\alpha = 1 - 9\sigma_{\rm abs}/(16 \pi
r_0^2)$. 

We consider three different models of nuclear absorption:  either all 
quarkonium states are produced as
color octets or color singlets or as a combination of octet and singlet states.
When pure octet or pure singlet
absorption is considered, $\psi$ and
$\psi'$ production are calculated in the CEM.  When a combination of 
octet and singlet
absorption is assumed, NRQCD is used to obtain the correct balance between 
octet and singlet production.  Both the CEM and NRQCD model parameters
are tuned to fit $pp$ production.   In this section, 
we only give examples of a range of cross sections for each absorption model.
The actual values of $\sigma_{\rm abs}$ are set in Section 9
after initial state effects have also been included.

In our considerations
of $\psi$ absorption, we include the $\approx 30$\% contribution from 
$\chi_{cJ}$ decays \cite{Teva} and the $\approx 12$\% contribution from   
$\psi'$ decays \cite{HPC} decays.  Then the total $\psi$ survival probability,
including indirect production, is
\be S_\psi^{\rm  abs} = 0.58 S_{\psi, \, {\rm dir}}^{\rm abs} + 0.3
S_{\chi_{cJ}}^{\rm abs} + 0.12 S_{\psi'}^{\rm abs} \, \, . \label{psisurv} \ee
The $\psi'$ itself is only produced directly since
other, more massive, charmonium resonances decay to $D \overline D$
pairs.  

\begin{figure}[htbp]
\setlength{\epsfxsize=\textwidth}
\setlength{\epsfysize=0.6\textheight}
\centerline{\epsffile{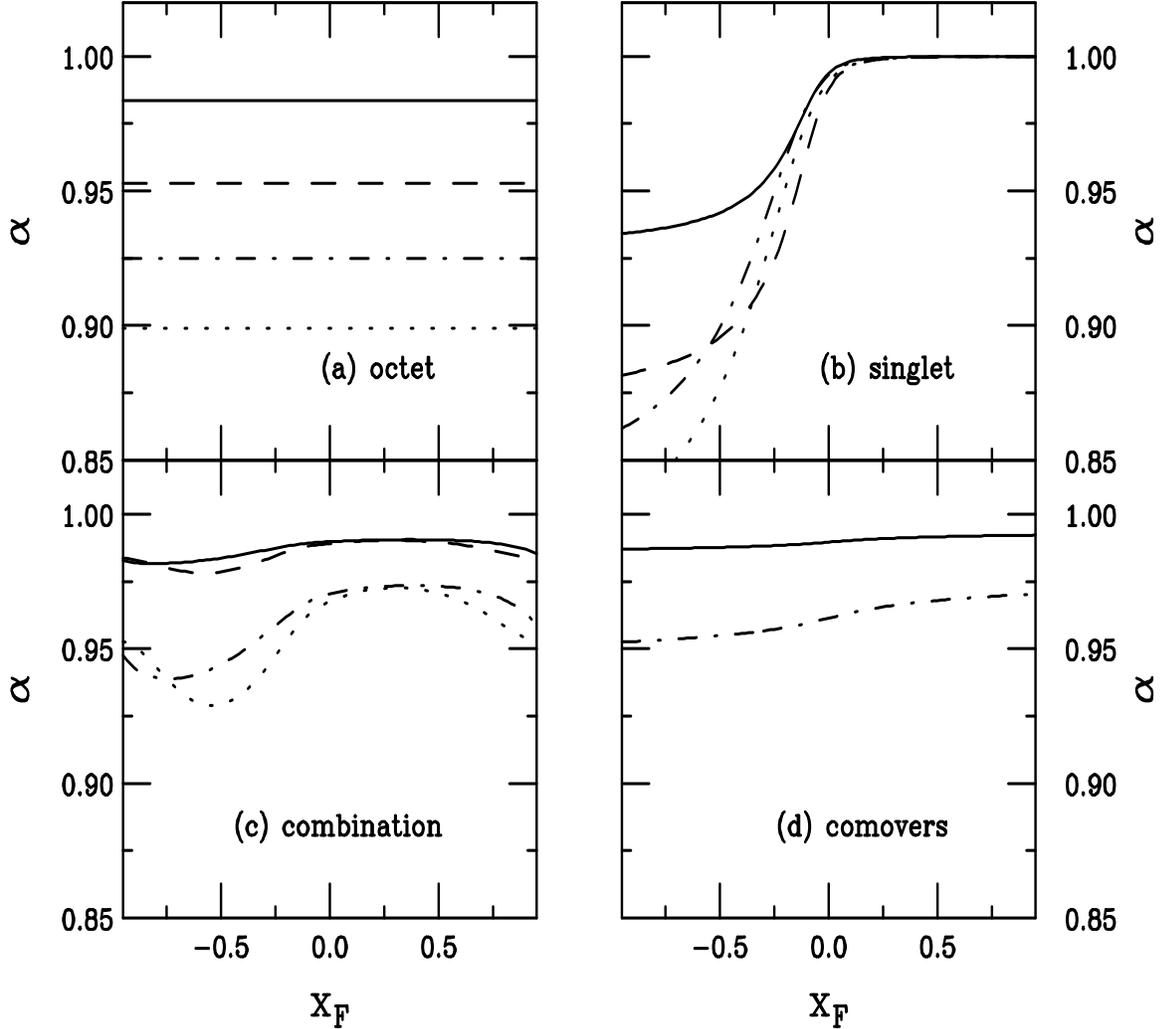}}
\caption[]{The $A$ dependence of nuclear absorption models is given in (a), (b)
and (c) and the comover $A$ dependence is shown in (d).  In (a), octet cross
sections of 1 mb (solid), 3 mb (dashed), 5 mb (dot-dashed) and 7 mb (dotted)
are shown.  Singlet absorption is shown in (b) for $\psi$
with $\sigma_{\psi N}^{\rm s} = 5$ 
mb (solid) and 10 mb (dashed) as well as $\psi'$ with 
$\sigma_{\psi' N}^{\rm s} = 15$ 
mb (dot-dashed) and 20 mb (dotted).
A combination of octet and singlet production is assumed in (c).  The curves
represent: $\psi$ absorption with $\sigma_{\rm abs}^{\rm octet} = 1$ mb
and $\sigma_{\rm abs}^{\rm singlet} = 1$ mb (solid) and $\sigma_{\rm abs}^{\rm
octet} = 3$ mb
and $\sigma_{\rm abs}^{\rm singlet} = 5$ mb (dot-dashed);
$\psi'$ absorption with  $\sigma_{\rm abs}^{\rm octet} = 1$ mb
and $\sigma_{\rm abs}^{\rm singlet} = 3.7$ mb (dashed) and  
$\sigma_{\rm abs}^{\rm octet} = 3$ mb
and $\sigma_{\rm abs}^{\rm singlet} = 19$ mb (dotted).  In (d), comover
interactions are shown for $\sigma_{\psi {\rm co}} = 0.67$ mb (solid) 
and $\sigma_{\psi' {\rm co}} = 3.7 \sigma_{\psi {\rm co}}$ (dot-dashed).}
\label{absco}
\end{figure}

The first case, pure octet production, assumes that all charmonium states are
initially produced as $|c \overline c g \rangle$ states with the same 
absorption cross sections, leading to $\sigma_{\rm abs} = 
\sigma_{\psi N}^{\rm o} = \sigma_{\psi' N}^{\rm o} = \sigma_{\chi_{cJ} N}^{\rm
o}$.  Therefore the survival probabilities, Eq.~(\ref{nsurv}), are identical
for all states.
Thus the
feeddown contributions to the $\psi$ in Eq.~(\ref{psisurv})
do not affect the absorption.
After $\sim 0.3$ fm/$c$, the remaining $|c \overline c g \rangle$ states are
expected to
hadronize.  Since
the absorption cross section is established at the production of the state,
the octet cross section is independent of the position $z$ and thus $x_F$
and projectile energy.  In this model we treat absorption as if only the $|c
\overline c g \rangle$ interacts with nucleons, not the final charmonium
states.   
In Fig.~\ref{absco}(a), $\alpha$ is given for several values of the $|c
\overline c g \rangle$ cross section: $\sigma_{\rm abs}=1$, 3, 5, and 7 mb 
corresponding to 
$\alpha = 0.98$, 0.95, 0.92, and 0.90 respectively.  It is obvious that octet
production alone will not modify the shape of $\alpha$ as a function of $x_F$.

We now discuss absorption when all $c \overline c$ pairs are produced as color
singlets.  If the \dd{c}{c} pair is produced as a color singlet, it is 
initially small with a spatial extent on the
order of its production time, $\tau \propto m_c^{-1}$, ignored in the
calculation.   
The proper time required for the formation of
the final charmonium bound state obtained from potential models \cite{KMS}, 
$\tau_\psi \sim 1-2$ fm, 
is considerably longer.  The \dd{c}{c}--$N$ absorption
cross section may be expected to grow as a function of proper time until
$\tau_{\psi_i}$ 
when it saturates at the asymptotic value $\sigma_{{\psi_i} N}^{\rm s}$.   
We simulate the growth of the absorption cross section by \cite{GV,Blol3} \be
\sigma_{\rm abs}(z^\prime -z) = \left\{ \begin{array}{ll} \sigma_{{\psi_i} 
N}^{\rm s} 
\bigg( {\displaystyle \frac{\tau}{\tau_{\psi_i}}} \bigg)^{\kappa} \,\, \, \, \,
\, \, & \mbox{if $\tau<\tau_{\psi}$} \\ \sigma_{{\psi_i} N}^{\rm s} 
& \mbox{otherwise}
\end{array}     \right. \, \, . \label{sigsing}
\ee The exponent $\kappa$ determines 
the increase
of $\sigma_{\rm abs}$ during hadronization of the \dd{c}{c} pair.  If
$\sigma_{\rm abs}$ is proportional to the geometric cross section,
then we expect $\kappa \sim 2$.  (See also \cite{lars} for predictions of
$\sigma_{\psi N}^{\rm s}$ if $\kappa = 1$.)
The proper time $\tau$ is related to the path length
traversed by the \dd{c}{c} pair through nuclear matter, $\tau = (z^\prime - z)/
\gamma v$. The $\gamma$ factor introduces $x_F$ and energy
dependencies in the growth of
the cross section.  Depending on the initial energy of the projectile and
the size of the target, the \dd{c}{c} pair may form a $\psi$ 
inside or outside of the target.

In Fig.~\ref{absco}(b), examples of $\alpha$ are 
given for direct $\psi$ and $\psi'$ absorption at 800 GeV.
The solid and dashed curves assume $\sigma_{\psi N}^{\rm s} = 5$ and 10 mb 
respectively
while the dot-dashed and dotted curves are for $\sigma_{\psi' N}^{\rm s} 
= 15$ and 20
mb.  In our calculations, we assume that the asymptotic
absorption cross sections scale in
proportion to the squares of the
charmonium radii \cite{hpov},  $\sigma_{\psi' N}^{\rm s} \approx
3.7 \sigma_{\psi N}^{\rm s}$ and   $\sigma_{\chi_{cJ} N}^{\rm s} \approx
2.4 \sigma_{\psi N}^{\rm s}$.  Thus each contribution to Eq.~(\ref{psisurv})
has a different $A$ dependence.
The $\psi$ and $\psi'$ formation times are 
different,
$\tau_\psi = 0.92$ fm and $\tau_{\psi'} = 1.5$ fm \cite{KMS}.
The $\psi$ and $\psi'$ results at $x_F < 0$ in Fig.~\ref{absco}(b)
reflect the differences in formation times as well as the gamma shift due to
their masses.  At 800 GeV, by $x_F =0$ the
final-state meson is produced outside the target so that $\alpha \approx 1$ for
$x_F >0$.  Therefore the $A$ dependence of color singlet production is
virtually independent of $\sigma_{{\psi_i} 
N}^{\rm s}$ for $x_F > 0$ at 800 GeV.
At 120 GeV, both states can be produced inside the target at $x_F >
0$ and influence the $A$ dependence at forward $x_F$ as well.

More realistically, $\psi$ production is a combination of octet and singlet
states, as in NRQCD.  
The ratio of octet
to singlet production is energy and $x_F$ dependent \cite{benrot} so that the
relative absorption of each state depends on $x_F$ 
since the octet and singlet absorption cross sections
are expected to be different \cite{Zhang}.  Because the $\psi'$ is directly
produced, the $x_F$
dependence of absorption is straightforward,
\be \frac{d\sigma_{pA}^{\psi'}}{dx_F} = \frac{d\sigma_{pp}^{\psi', \, \rm 
oct}}{dx_F} \int d^2b \, T_A^{\rm eff\, (oct)}(b) +
\frac{d\sigma_{pp}^{\psi', \, \rm sing}}{dx_F} \int d^2b \, T_A^{\rm eff\, 
(sing)}(b) \, \, , \label{psipcombo} \ee
where $\sigma_{\rm abs}^{\rm oct}$ and $\sigma_{\rm abs}^{\rm sing}$
replace $\sigma_{\rm abs}$ in Eq.~(\ref{taeff}).  The $\psi$ distribution
is more complicated since we must account for the fact 
that the octet absorption 
cross section is independent of the charmonium state while the singlet cross
sections are not.  In the octet case, 
the same $ T_A^{\rm eff\, (oct)}(b)$ can be
applied to all states feeding the $\psi$ while singlet absorption is different
for each individual state.  Then,
\be \lefteqn{\frac{d \sigma_{pA}^{\psi, \,{\rm tot}}}{dx_F} = \left[
\frac{d \sigma_{pp}^{\psi, \, {\rm dir, \, oct}}}{dx_F}
+ \sum_{J=0}^2 B(\chi_{cJ} \rightarrow \psi X) \frac{d 
\sigma_{pp}^{\chi_{cJ}, \, {\rm oct}}}{dx_F} \right. } \label{psicomb} \\
& & \left. \mbox{} + B(\psi' \rightarrow \psi X) \frac{d\sigma_{pp}^{\psi', \, 
{\rm oct}}}{dx_F}
\right] \int d^2b T_A^{\rm eff\, (oct)}(b) \nonumber \\ & & \mbox{} + \int d^2b
\left[ \frac{d \sigma_{pp}^{\psi, \, {\rm dir, \, sing}}}{dx_F} T_A^{\psi,
 {\rm dir}, {\rm eff\, (sing)}}(b)
+ \sum_{J=0}^2 B(\chi_{cJ} \rightarrow \psi X) \frac{d 
\sigma_{pp}^{\chi_{cJ}, \, {\rm sing}}}{dx_F} T_A^{\chi_{cJ},\,
 {\rm eff\, (sing)}}(b) \right. \nonumber \\ & & \left. \mbox{}
+ B(\psi' \rightarrow \psi X) \frac{d\sigma_{pp}^{\psi', \, {\rm sing}}}{dx_F}
 T_A^{\psi', \, {\rm eff\, (sing)}}(b)
\right] \nonumber
\, \, . \ee
In Ref.~\cite{Zhang}, the
singlet cross section was assumed to be negligible.  Therefore, $\sigma_{\rm 
abs}^{\rm oct} = 11$ mb was needed to produce an effective $\alpha$
equivalent to the assumption of pure octet production 
with $\sigma_{\rm abs} = 7.3$ mb obtained in 
\cite{klns}.  We use Eq.~(\ref{sigsing})
and $\sigma_{\psi N}^{\rm s} \neq 0$.  Also, in
Ref.~\cite{Zhang}, the authors only calculated 
the $x_F$-integrated cross sections.  Here we use
Eqs.~(\ref{psicomb}) and (\ref{psipcombo}) 
with the full $x_F$ dependence to calculate $\alpha(x_F)$ for $\psi$
and $\psi'$ absorption,
shown in Fig.~\ref{absco}(c) at 800 GeV.
The differences between the $\psi$ and $\psi'$ results arise from their
distinct $x_F$ dependencies in the NRQCD model, predominantly from the
$\chi_{cJ}$ contributions to $\psi$ production.

We have illustrated several different combinations of $\psi'$
and $\psi$ absorption cross sections in Fig.~\ref{absco}(c).
We choose $\sigma_{\psi' N}^{\rm octet} =
\sigma_{\psi N}^{\rm octet}$ as in pure octet production, Fig.~\ref{absco}(a),
and $\sigma_{\psi' N}^{\rm singlet} \approx
3.7 \sigma_{\psi N}^{\rm singlet}$, as in the pure singlet case shown in
Fig.~\ref{absco}(b).  The differences in $\alpha(x_F)$ for $\psi$ and
$\psi'$ are small,
especially between the solid and dashed curves with $\sigma_{\rm abs}^{\rm
octet} = 1$ mb.  The only obvious differences are at $x_F < 0$ when the
$\psi'$ singlet contribution is larger and the $\psi'$ is still produced
inside the target.  However, at large
$|x_F|$, $q \overline q$ annihilation, an octet contribution to $\psi'$, 
begins to 
become more important, causing the change in slope of $\alpha(x_F)$ here.  
This effect is not as strong for the $\psi$ because the $q \overline
q$ contribution does not overtake the $gg$ until larger $x_F$.
It is interesting to note that the effective $\alpha$ in the solid curve 
($\sigma_{\rm abs}^{\rm octet} = 1$ mb and $\sigma_{\rm abs}^{\rm
  singlet} = 1$ mb) is similar to the $\sigma_{\rm abs}^{\rm o} = 1$ result 
shown in Fig.~\ref{absco}(a) for pure octet absorption.  Assuming a 3 mb octet
absorption cross section 
results in a similar
effective $\alpha$ at $x_F<0$ in Figs.~\ref{absco}(a) and (c)---compare the
dashed curve in (a) with the dot-dashed curve in (c).  However, at forward
$x_F$, the effective $\alpha$ in (c) is larger since the color singlet
components escape
without absorption, at least until the growing $q \overline q$ contribution
causes the octet mechanism to dominate $\psi$ absorption once again at 
$x_F > 0.7$.  The effect of singlet absorption would be even weaker at 120 
GeV because the octet contributions make up a larger fraction of the 
production cross section at this energy.

The results with the absorption cross sections shown in Fig.~\ref{absco} 
are only examples of the magnitude of the
effects.  It is clear, both from the data in Fig.~\ref{data} and from the
initial state effects discussed in the following sections, that the model
absorption cross sections must be smaller than those used previously when no
initial state effects were included \cite{RV,klns}.  

\section{Hadronic Comovers in $pA$ Interactions}

Comoving secondaries, formed after $\tau_0 \sim 1-2$ fm, 
may also scatter with the
\dd{c}{c} pair or the $\psi$.  Because $\tau_\psi \stackrel{<}{\sim} \tau_0$, 
the final-state
charmonium is assumed to interact with the comovers.
A spectator hadron moving with a velocity close to that of the charmonium state
enhances the dissociation probability. 

The $A$ dependence of $\psi$ production due to
comovers alone is determined from
\be \sigma_{hA} =
\sigma_{hN} \int d^2b \, S^{\rm co}(b) \, \, , 
\label{survcomo} \ee where the total probability that the $\psi$ survives its 
interactions with comovers is
\be S_\psi^{\rm co} = 0.58 S_{\psi, \, {\rm dir}}^{\rm co} + 0.3
S_{\chi_{cJ}}^{\rm co} + 0.12 S_{\psi'}^{\rm co} \, \, . \label{psicosurv} \ee
The direct $\psi$-comover survival probability is \cite{GV} 
\be S_{\psi, \, {\rm dir}}^{\rm co}(b)  \approx \exp \left\{ -
\int\!d\tau\,{\langle\sigma_{\psi {\rm co}} v\rangle}n(\tau, b)\right\} \, \, .
\label{comosurv} \ee  The other survival probabilities for comover
interactions with charmonium states are similar.
The parameters are the charmonium--comover 
absorption cross sections, the velocity of the $\psi$ relative to the 
comovers, $v \sim 0.6$, and
$n(\tau,b)$, the density of comovers at time $\tau$ and impact parameter $b$.
We take $\sigma_{\psi {\rm co}} = 0.67$ mb from a study of $\psi$ suppression
in nucleus-nucleus data \cite{RV} with $\sigma_{\psi' {\rm co}} \approx 3.7
\sigma_{\psi {\rm co}}$ and $\sigma_{\chi_{cJ} {\rm co}} \approx 2.4
\sigma_{\psi {\rm co}}$ \cite{hpov}, assuming that the asymptotic
charmonium states interact with the comovers.  Integrating 
Eq.~(\ref{comosurv}) 
over $\tau$ and relating the initial density of the system to
the final hadron rapidity density, $n_0 \tau_0 = (\pi R^2)^{-1} (dN/dy)$
\cite{Bjorken}, one finds \cite{GV} \be \int d\tau \, n(\tau, b) \approx
\frac{1}{\pi R^2} \ln\left( \frac{\tau_I}{\tau_0} \right)\frac{dN}{dy}
\sigma_{hN} T_A(b) \label{comofull}
\ee where the effective proper lifetime $\tau_I$ over which the comovers
interact with the $\psi$ is $\tau_I
\sim r_p/v$ and $r_p \sim 0.8$ fm.  This scaling assumption is rather strong
for $pA$ since no collective motion is expected.  However, perhaps within the
interaction tube carved out by the incident proton, scaling may hold.  There is
evidence of rapidity scaling for produced particles in the central region of
$pp$ collisions \cite{Satz91}.

The rapidity density grows with center
of mass energy \cite{Satz91}.  The shape of the produced particle rapidity
density with inclusive $\psi$ production is unknown.  
We assume that the multiplicity slope is the same
on both sides of midrapidity 
\be \frac{dN}{dy} = \frac{dN}{dy}|_{y=0} - a y \, \, , \label{dndyco} \ee
where we take $dN/dy|_{y=0} = 1.07$ in $pp$ interactions at 800 GeV and $a =
0.108$ \cite{Satz91}. The
comover density is depleted at forward rapidities but enhanced
close to the target. 

Since the transverse area over which the comovers are produced, $\pi R^2$, 
is approximately 
equal to $\sigma_{hN}$, the direct $\psi$ survival probability in $hA$ 
collisions may
be recast as \be S_{\psi, \, \rm dir}^{\rm co}(b) 
\approx \exp \left\{ - \langle \sigma_{\psi {\rm co}} 
v \rangle
\frac{dN}{dy} \ln \left( \frac{\tau_I}{\tau_0} \right) T_A(b) \right\}
\, \, . \label{psihaco} \ee The similarity between Eqs.~(\ref{sigfull}) and
(\ref{psihaco}) suggests that $\psi$--comover interactions do not introduce 
any unusual $A$ dependence.  Thus comover contributions to $pA$ interactions,
while small, are difficult to rule out entirely.

\section{Nuclear Shadowing}

Measurements of the nuclear 
charged parton distributions by deep-inelastic scattering off both a large 
nuclear target and a deuterium target show that the ratio $R_{F_2} =
F_2^A/F_2^D$
has a characteristic shape as a function of $x$ \cite{Arn}.  The region  
below $x\sim 0.1$ is referred to as the shadowing region and the range
$0.3<x< 0.7$ is known as the EMC region.  
In both regions, the parton density is depleted in
the heavy nucleus relative to deuterium, {\it i.e.}\ $R_{F_2}<1$.  
At very low $x$, $x \approx 0.001$, $R_{F_2}$ appears to
saturate \cite{E6652}.  Between the shadowing and EMC 
regions, an enhancement, antishadowing, is seen where
$R_{F_2}>1$.  There is also an enhancement as $x \rightarrow 1$, assumed to be
due to nucleonic Fermi motion.  The general behavior of $R_{F_2}$ 
as a function of $x$ is often
referred to as shadowing.  Although this behavior is 
not well understood for all $x$, 
the shadowing effect can be modeled by an $A$ dependent fit to the nuclear
deep-inelastic scattering data. 

We have assumed
that the nuclear parton distributions factorize into 
the nucleon parton distributions, independent of $A$, and a shadowing 
function that parameterizes the 
modifications of the nucleon parton densities
in the nucleus, dependent on $A$, $x$, and $Q^2$:
\be
f_i^A(x,Q^2,A) & = & S^i(A,x,Q^2) f_i^p(x,Q^2) \nonumber \,\, .
\ee
While the location of the parton in the target could influence $S^i$
\cite{ekkv}, the impact parameter is difficult to resolve in $pA$ collisions.
We use three different parameterizations of 
the shadowing function, $S^i(A,x,Q^2)$.

\begin{figure}[htbp]
\setlength{\epsfxsize=\textwidth}
\setlength{\epsfysize=0.6\textheight}
\centerline{\epsffile{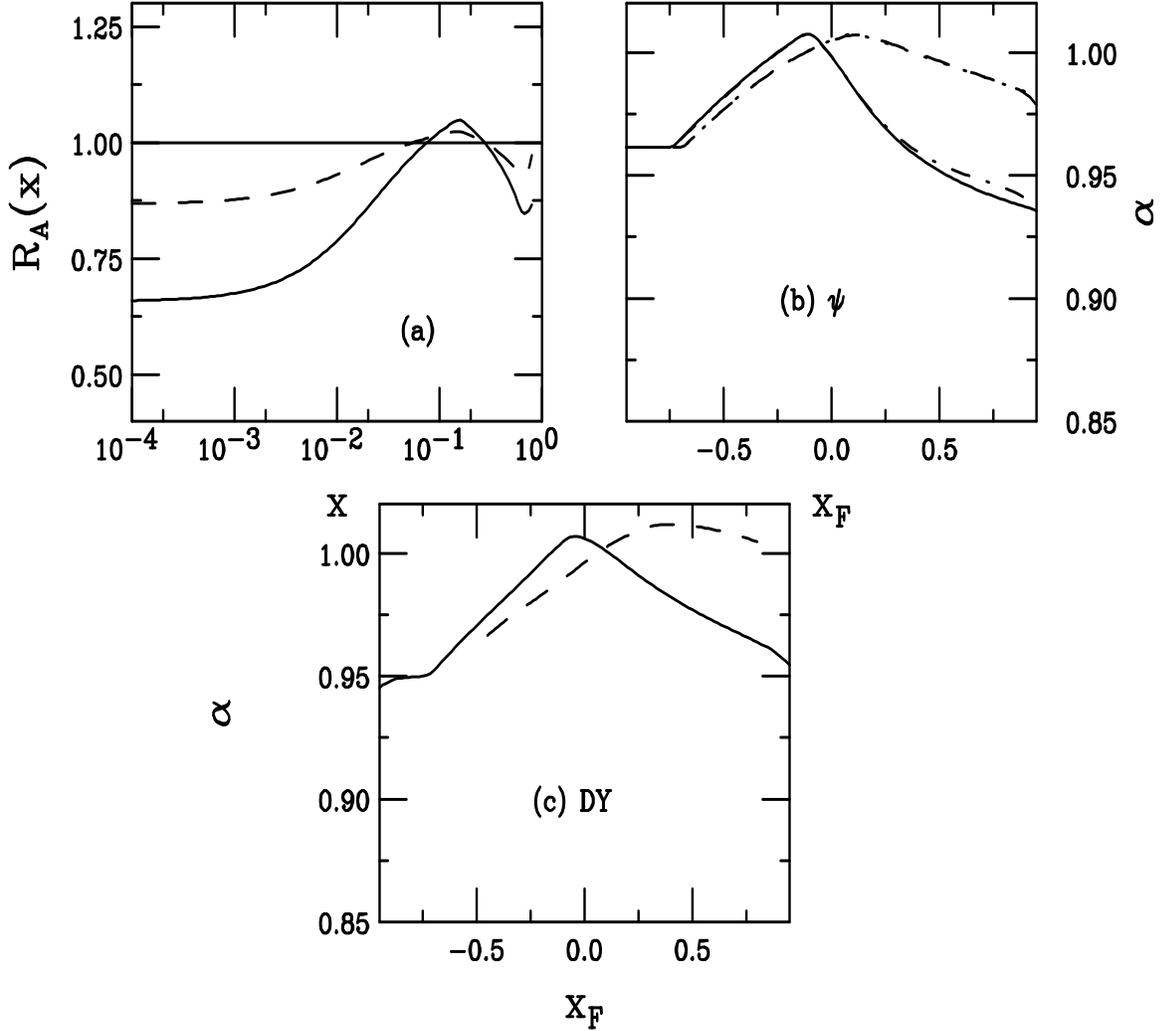}}
\caption[]{(a) The $S_1$ shadowing parameterization for W (solid) and Be 
(dashed)
targets as a function of $x$.   The resulting $A$ dependence for (b) $\psi$ 
production in the CEM and (c) Drell-Yan
production is given at 800 GeV (solid) and 120 GeV (dashed). The NRQCD $\psi$
results are shown in (b) at 800 GeV (dot-dashed) and 120 GeV (dotted).} 
\label{shad1}
\end{figure}

The first parameterization
is a fit to nuclear 
deep-inelastic scattering
data which does not differentiate between quark, antiquark, and gluon
modifications and does not include evolution in $Q^2$.  Therefore
it is not designed to conserve baryon number or momentum.  We define
$R_{F_2} = S_1(A,x)$ \cite{EQC} with \be S_1(A,x) = \left\{
\begin{array}{ll} R_s {\displaystyle \frac{1 + 0.0134 (1/x -1/x_{\rm sh})}{1
+ 0.0127A^{0.1} (1/x - 1/x_{\rm sh})}} & \mbox{$x<x_{\rm sh}$} \\
a_{\rm emc} - b_{\rm emc}x  & \mbox{$x_{\rm sh} <x< x_{\rm fermi}$} \\
R_f \bigg( {\displaystyle \frac{1-x_{\rm fermi}}{1-x}} \bigg)^{0.321} &
\mbox{$x_{\rm fermi} <x< 1$} \end{array} \right. \, \, , \ee
where $R_s = a_{\rm emc} - b_{\rm emc} x_{\rm sh}$, $R_f = a_{\rm emc} -
b_{\rm emc} x_{\rm fermi}$, $b_{\rm emc} = 0.525(1 - A^{-1/3} - 1.145A^{-2/3} +
0.93A^{-1} + 0.88A^{-4/3} - 0.59A^{-5/3})$, and $a_{\rm emc} = 1 + b_{\rm emc}
x_{\rm emc}$.   The fit fixes the $x$ values at the boundaries of the $x$
regions, $x_{\rm sh}=0.15$, 
$x_{\rm emc}=0.275$, and $x_{\rm fermi}=0.742$.
Thus, the nuclear parton distributions are modified so
that \be f_i^A(x,Q^2) & = & S_1(A,x) f_i^p(x,Q^2) . \ee
The parameterization is available for all $A$ and is designed so that $S_1 
\equiv 1$
when $A=1$.  Figure \ref{shad1}(a) shows the parameterization for $A = 184$ and
$A = 9$.  Note that the antishadowing region is rather narrow and saturation
appears at $x<10^{-3}$.  
Figures \ref{shad1}(b) and (c) give
$\alpha(x_F)$ for $\psi$ and Drell-Yan production respectively.  
Since this parameterization affects
all partons equally, the results are independent of the
chosen parton distribution function.  They are also virtually independent of
the charmonium production mechanism although there is a slight model dependence
because the CEM involves an integral over $2m_c/\sqrt{s}
< x < 2m_D/\sqrt{s}$, Eq.~(\ref{cevap}), while in NRQCD $x_1$ and $x_2$ are
either both fixed, as in Eqs.~(\ref{psiqq}), (\ref{chigg0}), (\ref{chigg2}),
(\ref{chiqq0}), and (\ref{chiqq12}) or $x_1$ is fixed by the delta function in
Eq.~(\ref{signrqcd}) while $4m_c^2/x_1s < x_2 < 1$.

The second parameterization, $S_2^i(A,x,Q^2)$, modifies the
valence quark, sea quark and gluon distributions separately and also includes 
$Q^2$ evolution \cite{KJE}, beginning at $Q = Q_0 = 2$ GeV and continuing
up to $Q = 10$ GeV.
It is based on a fit to the data using the
Duke-Owens parton densities \cite{DO}.  
In this case, the nuclear parton densities are modified so
that \be f_V^A(x,Q^2) & = & S_2^V(A,x,Q^2) f_V^p(x,Q^2) \, \, , \\  
f_S^A(x,Q^2) & = & S_2^S(A,x,Q^2) f_S^p(x,Q^2) \\ f_G^A(x,Q^2) & =
& S_2^G(A,x,Q^2) f_G^p(x,Q^2) \, \, , \ee where $f_V = u_V + d_V$ is 
the valence 
quark density and $f_S = 2(\overline u + \overline d + \overline s)$ is
the total
sea quark density.  It is assumed that $S_2^V$ and $S_2^S$ are the same for
all valence and sea
quarks, consistent with the symmetric sea of the Duke-Owens parton
distributions.   
These modifications conserve baryon number,
$\int_0^1 dx \, f_V^p(x,Q^2) =\int_0^1 dx \, f_V^A(x,Q^2) $, and the parton
momentum sum,
$\sum_P \int_0^1 dx \, x f_P^p(x,Q^2) = \sum_P \int_0^1 dx \, x
f_P^A(x,Q^2)$ where $P = V$, $S$, and $G$,
at all $Q^2$.  Using parton densities other than Duke-Owens may lead to small
deviations in the conservation rules.  

\begin{figure}[htbp]
\setlength{\epsfxsize=\textwidth}
\setlength{\epsfysize=0.6\textheight}
\centerline{\epsffile{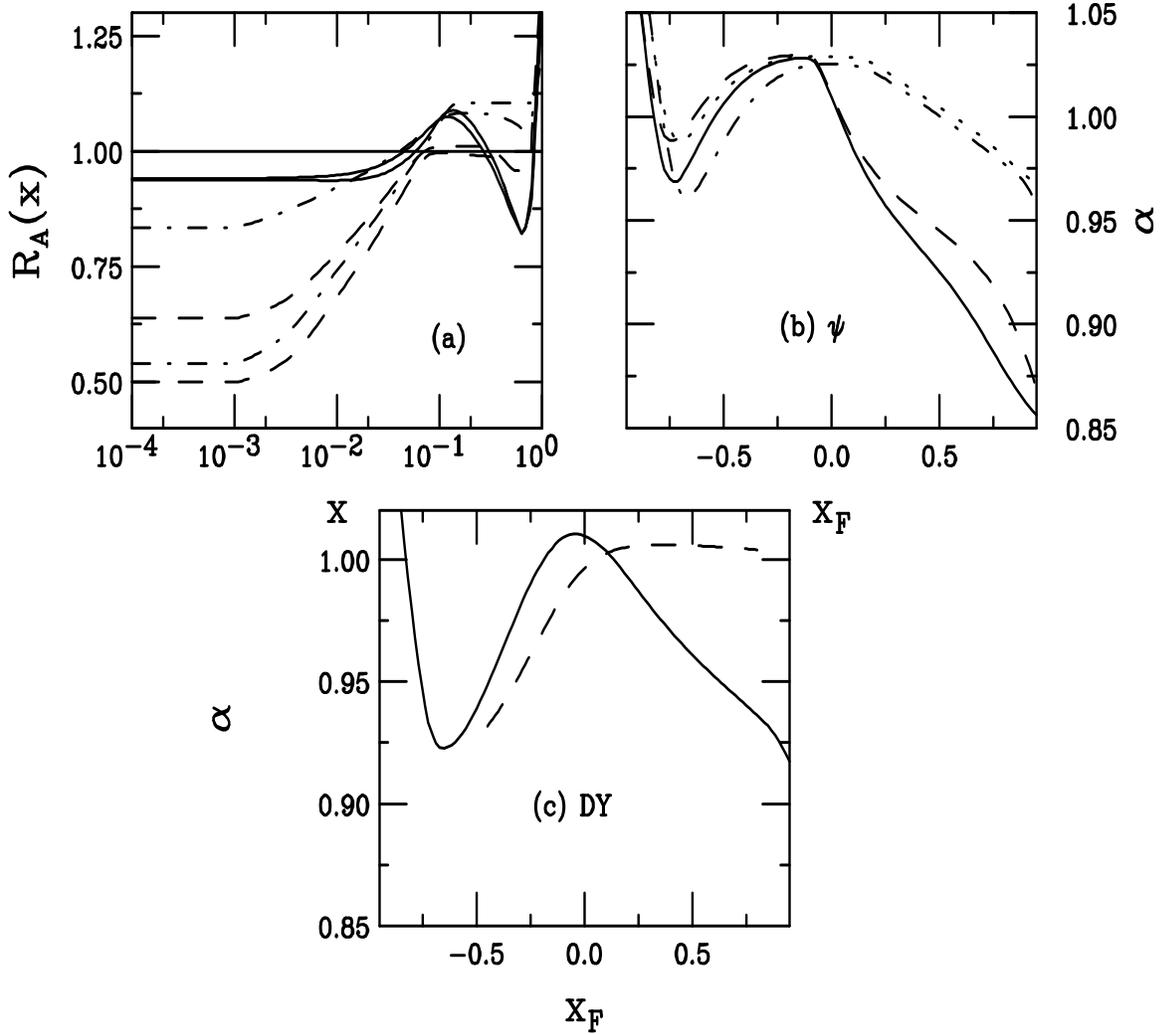}}
\caption[]{(a) The $S_2$ shadowing parameterization for $A= 200$ as a 
function of $x$. The valence ratios, $R_V$, are given by the
solid curves, the sea quark ratios, $R_S$, by the dashed curves and the gluon
ratios, $R_G$, are given by the dot-dashed curves.  At small $x$, the
lower curves are at $Q = 2$ GeV and the upper are at $Q=10$ GeV.
In (b) the $\psi$ $A$ dependence is illustrated for the
CEM with
MRST LO distributions at 800 GeV (solid) and 120 GeV (dot-dashed).
The NRQCD results with the CTEQ 3L densities
are also shown at 800 GeV (dashed)
and 120 GeV (dotted). 
In (c) the Drell-Yan $A$ dependence is given for the MRST LO
distributions at 800 GeV (solid) and 120 GeV (dashed).} 
\label{shad2}
\end{figure}

The parameterization is only available for $A = 32$ and 200. It is thus applied
only to the tungsten target and the beryllium densities are left unmodified.
Figure \ref{shad2}(a) shows the ratios $S_2^V$,
$S_2^S$ and $S_2^G$ at $Q = Q_0$ and $Q = 10$ GeV.  
At $Q_0$ the sea quarks are
shadowed more strongly 
at low $x$ than the gluons.  Both the valence quarks and gluons
are antishadowed while the sea quarks are not.  The effects of evolution are
weakest for the valence quarks and strongest for the gluons.  Figures
\ref{shad2}(b) and (c) show $\alpha(x_F)$ for $\psi$ and Drell-Yan production
respectively.  The $\psi$ results in the CEM
are given for the MRST LO \cite{mrsg} distributions and in the NRQCD approach 
with the CTEQ 3L \cite{cteq3} parton distributions.  
The main differences in the
production models appear at negative $x_F$, corresponding to the EMC dip at
large $x$ and appears because of the evolution of the gluon distributions at
large $x$.  The two calculations evolve differently because $m_c = 1.2$ GeV in
the CEM and 1.5 GeV in NRQCD.  The larger scale causes a smaller EMC dip in the
shadowing ratio for the NRQCD calculation.
The differences between
production models at large $x_F$ are due in part to the $gq$ scattering
contribution.  Since this component is virtually negligible at 120 GeV, the
model dependence is then small for $S_2$.
Choosing other parton distribution functions for CEM $\psi$ and Drell-Yan
production results in very similar ratios as for MRST LO. 

A more recent shadowing
parameterization, $S_3^i(A,x,Q^2)$, based on the GRV LO parton distributions
\cite{GRV92}, is now available \cite{EKRS3,EKRparam}.  The initial scale was
chosen to equal the charm quark mass in the GRV LO distributions, $Q = Q_0 =
1.5$ GeV.  At this scale all sea quark ratios are assumed to be equal, as are
both the valence ratios.  The parameters are constrained by nuclear
deep-inelastic scattering and Drell-Yan data.  The gluon ratio is then fixed by
the momentum sum rule as well as $\psi$ electroproduction data.  
Above $Q_0$, the individual quark and gluon
distributions are evolved separately.  The gluon distribution has a larger
antishadowing peak in this parameterization while the sea quarks are 
shadowed in the
same region, a significant difference from $S_2$.  
The Drell-Yan data on the violation of the
Gottfried sum rule \cite{NA51} is taken to account, thus $S_3^{\overline u} 
\neq S_3^{\overline d}$ above $Q_0$.  
Evolution is taken up to $Q = 100$ GeV and the
parameterization is generalized to all $A$, both improvements over $S_2$.
Again however, using other parton densities besides GRV LO could lead to small
deviations from the conservation rules.

\begin{figure}[htbp]
\setlength{\epsfxsize=\textwidth}
\setlength{\epsfysize=0.6\textheight}
\centerline{\epsffile{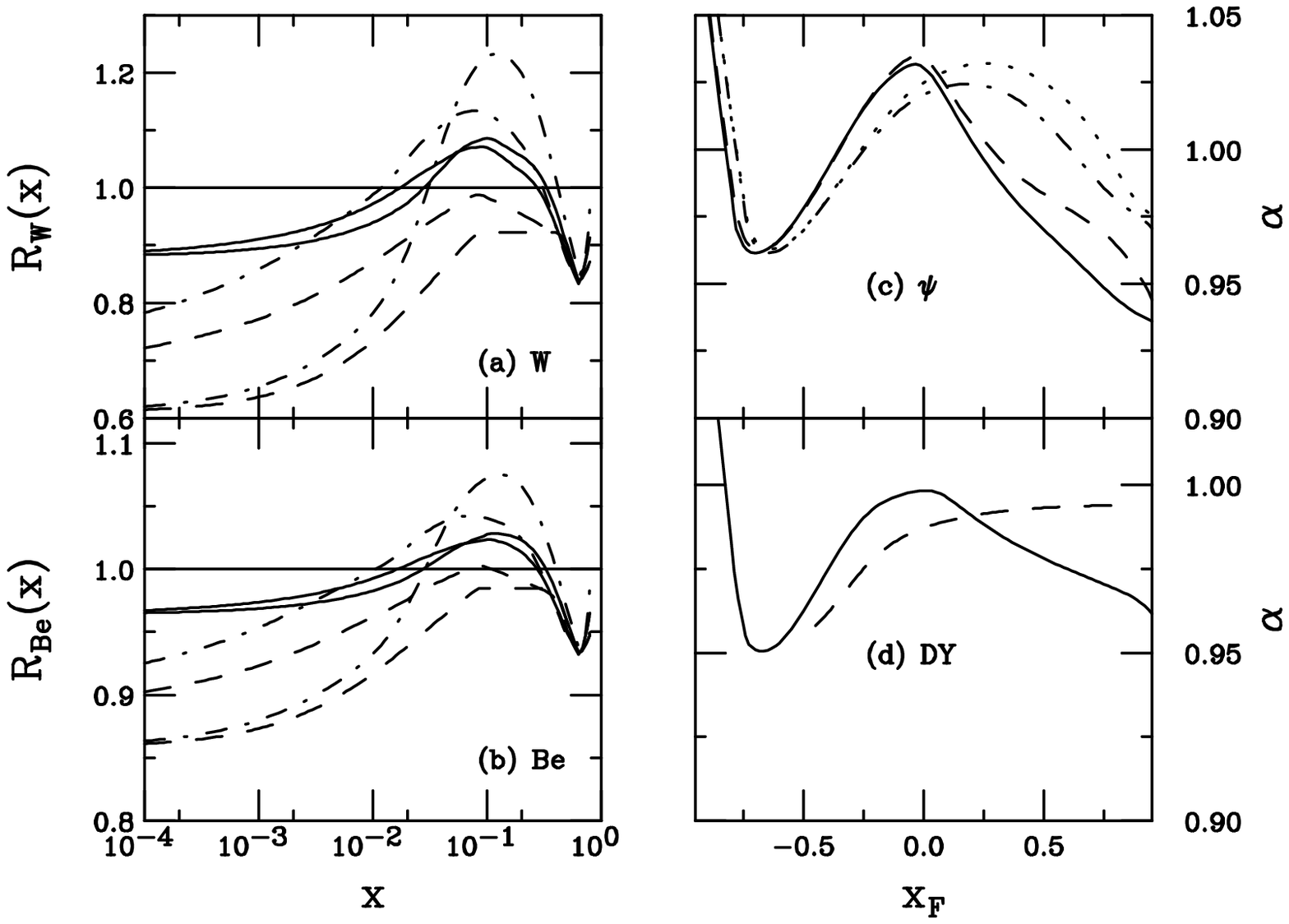}}
\caption[]{The $S_3$ shadowing parameterization as a function of $x$ for (a) W
and (b) Be targets. The valence up ratios, $R_{u_V}$, are given by the
solid curves, the $\overline u$ ratios, $R_{\overline u}$, 
by the dashed curves and the gluon
ratios, $R_G$, are given by the dot-dashed curves.  At small $x$, the
lower curves are at $Q = 1.5$ GeV while the upper curves are at $Q=10$ GeV.
The resulting $A$ dependence for (c) $\psi$ and (d) Drell-Yan
production is given.  In (c) the $\psi$ $A$ dependence is illustrated for the
CEM with
MRST LO distributions at 800 GeV (solid) and 120 GeV (dot-dashed).
The NRQCD results with the CTEQ 3L densities
are also shown at 800 GeV (dashed)
and 120 GeV (dotted). 
In (d) the Drell-Yan $A$ dependence is given for the MRST LO
distributions at 800 GeV (solid) and 120 GeV (dashed). } 
\label{shad3}
\end{figure}

In Fig.~\ref{shad3}(a) and (b), we show ratios for the $u_V$, $\overline u$ and
$g$ densities
at $Q = Q_0$ and 10 GeV for W and Be respectively.  
Figures \ref{shad3}(c)
and (d) give the corresponding $\alpha(x_F)$ for $\psi$ and Drell-Yan
production.  The $\psi$ results in the CEM are given for the MRST LO
distributions.  Since gluon shadowing is 
not as strong as in
the $S_2$ parameterization at low $x$, the effective $\alpha$
is larger at large $x_F$ than in Fig.~\ref{shad2}(b).  We have checked the CEM
results with other parton distributions and found that the differences between
the parton distributions are also more pronounced at large $x_F$
since the individual quark
and antiquark distributions evolve separately while with $S_2$, the valence and
sea quarks respectively were considered together.  In NRQCD, the effective
$\alpha$ with the CTEQ 3L densities at 800 GeV
is similar to the CEM results except at
larger $x_F$ due to the $gq$ scattering contribution, as with the $S_2$
parameterization.  However, at 120 GeV, the
model dependence is more pronounced than with $S_2$, due to the larger relative
importance of $q \overline q$ annihilation in the CEM than in NRQCD. This
difference is again less important with the $S_2$ parameterization because it
does not distinguish between the individual quark and antiquark distributions.
At negative $x_F$ there is no difference due 
to evolution at the EMC dip because
the ratios at large $x$ in Figs.~\ref{shad3} (a) and (b) are essentially
independent of $Q^2$. The Drell-Yan results are only shown for the MRST LO 
distributions.  The reduced
antiquark shadowing at low $x$ results in a larger $\alpha$ than with the $S_2$
parameterization.

To summarize, we note that the shape of $\alpha(x_F)$ is fixed by each
parameterization.  It is clear from the results in
Figs.~\ref{shad1}-\ref{shad3} that shadowing alone is insufficient to describe
the preliminary E866 $\psi$ data as a function of $x_F$.  This fact has been
known for some time since the NA3 \cite{NA3} and E772 \cite{E772} $\psi$ $A$
dependence was similar as a function of $x_F$ but not as a
function of $x_2$ as would be expected if the nuclear dependence was dominated
by shadowing.

\section{Effects of Energy Loss}

Partons are expected to lose energy when traversing matter.  This effect
has been
discussed primarily in the context of jet quenching \cite{elossrefs,bdmps1}.  
Since the projectile parton is typically expected to feel the effects of
energy loss, the scaling of the $A$ dependence at different energies with
$x_F$ or $x_1$ suggested that energy loss could be the cause.
We will introduce three models of energy
loss that have been applied earlier to $\psi$ production and discuss their 
influence in the context of the E866 data.

\subsection{Initial State Loss} 

Initial state energy loss, as studied by Gavin and Milana \cite{GM} and
subsequently developed by Brodsky and Hoyer \cite{BH}, takes a 
multiple scattering approach that essentially depletes the 
projectile parton momentum fraction, $x_1$, as the parton moves through the
nucleus.  Both the quarks and gluons can scatter elastically and lose energy
before the hard scattering.  This loss produces a similar effect for Drell-Yan
and $\psi$ production.  The motivation for this model stemmed from 
the fact that the $A$ dependence of $\psi$ production at 200 and 800 GeV
seemed to scale with $x_F$ (or $x_1$) and not $x_2$ \cite{NA3,E772}.  The
projectile parton momentum fraction involved in the hard scattering is then
$x_1' = x_1 - \Delta x_1$ where $x_1$ is the original projectile parton
momentum fraction when the parton first entered the target
and $\Delta x_1$ represents the loss in $x_1$ due to multiple
scatterings.  Thus the shifted value, $x_1'$, enters the partonic cross
sections but the parton distributions must be evaluated at the initial $x_1$.
An additional delta function is added to Eqs.~(\ref{cemdef}) and
(\ref{signrqcd}) with the corresponding integral over $x_1'$ so that 
Eq.~(\ref{cemdef}) becomes
\be      \frac{d \sigma^{c \overline c}}{dx_F dm^2} & = & \frac{1}{s}   
\int_0^1 dx_1' dx_1 dx_2 \, \delta(x_1' - x_1 + \Delta x_1) \nonumber \\
& & \mbox{} \times \delta ( x_F - x_1' + x_2 )\, \delta(x_1' x_2 s - m^2) \,
 H_{AB}(x_1,x_1',x_2;m^2)  \, \,  \label{cemloss} \ee
while Eq.~(\ref{signrqcd}) is then
\be      \frac{d \sigma^C}{dx_F} & = & \sum_{i,j} 
\int_0^1 dx_1' dx_1 dx_2 \, \delta(x_1' - x_1 + \Delta x_1) \nonumber \\
& & \mbox{} \times \delta ( x_F - x_1' + x_2 ) f_i^A(x_1,\mu^2)f_j^B(x_2,\mu^2)
\widehat{\sigma}(ij \rightarrow C) \, \, . \label{nrqcdloss} \ee
We first discuss the model by Gavin and Milana \cite{GM} and then the
modifications suggested by Brodsky and Hoyer \cite{BH} with later refinements
by Baier {\it et al.} \cite{bdmps1}.

The first model of initial-state energy loss applied to $J/\psi$ production
was proposed by Gavin and Milana \cite{GM}, referred to as GM hereafter.  In
their model, they assumed that
\be \Delta x_1 = \epsilon_i x_1 A^{1/3} \left( \frac{Q}{Q_0} \right)^{2n} 
\label{gmdelx1} \ee  with $n=1$.  We do not
include the $Q^2$ dependence in our calculations so that here $n=0$.  
The energy loss
depends on the parton identity in this formulation.  The initial $x_1$ is
\be
x_1 = \frac{x_1'}{1 - \epsilon_i A^{1/3}} \, \,  \ee
where $i=q$ or $g$ with $\epsilon_q = 0.00412$ and $\epsilon_g = 9
\epsilon_q/4$ 
due to the difference in the color factors.  When $n=0$,
Eq.~(\ref{gmdelx1}) corresponds to $-dE/dz|_q
\sim 1.5$ GeV/fm and $-dE/dz|_g \sim 3.4$ GeV/fm \cite{GM}.  
In our calculations, we assume only initial
state elastic scattering of the quarks and gluons.
Final state effects on the $\psi$ included in Ref.~\cite{GM}
are left out here under the assumption that final-state absorption provides a
compensatory effect.

\begin{figure}[htbp]
\setlength{\epsfxsize=\textwidth}
\setlength{\epsfysize=0.6\textheight}
\centerline{\epsffile{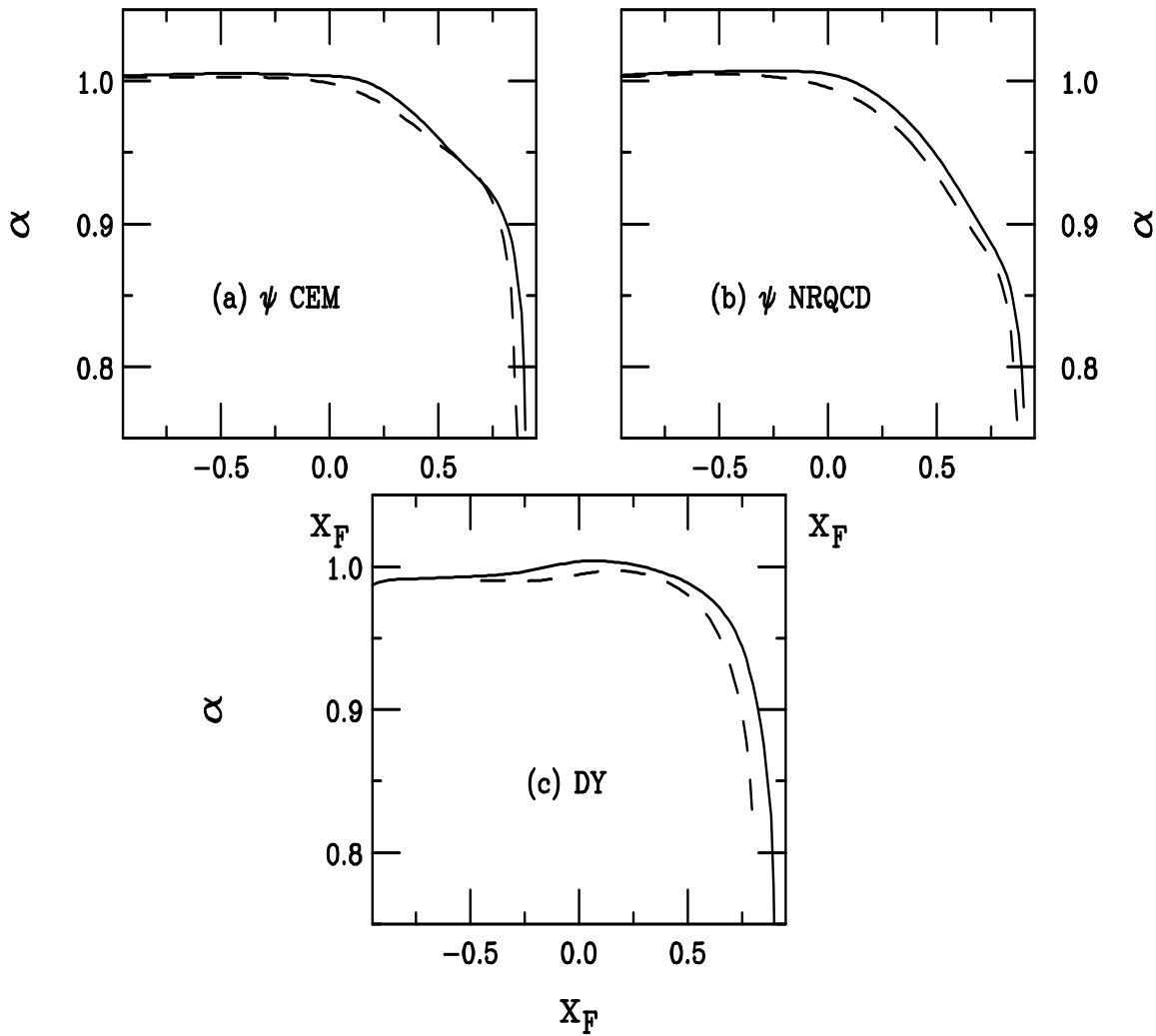}}
\caption[]{The $A$ dependence assuming GM loss
for $\psi$ production in (a) the CEM and in (b)
NRQCD and (c) Drell-Yan production.  In (a) the $\psi$ $A$ 
dependence is illustrated for the
MRST LO distributions at 800 GeV (solid) and 120 GeV (dashed).
The NRQCD results with CTEQ 3L
are shown in (b) at 800 GeV (solid)
and 120 GeV (dashed). In (c) the Drell-Yan $A$ dependence 
is given for the MRST LO
distributions at 800 GeV (solid) and 120 GeV (dashed). } 
\label{loss1}
\end{figure}

In Fig.~\ref{loss1} we show the results for this mechanism alone on the $A$
dependence of $\psi$ and Drell-Yan production.  
The $A$ dependence is weak at negative $x_F$ where $x_1$ is already small so 
that further reduction does not
significantly change the quark and gluon distributions.  This is true even for
parton distributions that increase as $x_1^{-a}$ when $x_1$ is small
and $a = 0.3-0.5$.
As $x_F$ increases,
$x_1$ grows larger and if the parton densities behave as $\sim (1-x_1)^{n_P}$
as $x_1 \rightarrow 1$, a slight decrease in $x_1$ is magnified.  
The effect should be stronger for $\psi$ than Drell-Yan production
because $n_g \sim 5 > n_{q_V} \sim 3$ 
in simple spectator counting models \cite{jgunion} and the valence quark 
distributions are most important for Drell-Yan production
at large $x_F$ (and $x_1$).  The choice of parton densities does not 
change the shape of $\alpha(x_F)$.  The energy dependence is also rather weak.
A comparison of Figs.~\ref{loss1}(a) and (b) shows that the behavior of
$\alpha(x_F)$ does not depend strongly on the $\psi$ production model although
there is evidence that the effect begins to be nonnegligible at a lower $x_F$
in the NRQCD approach.  

\begin{figure}[htbp]
\setlength{\epsfxsize=\textwidth}
\setlength{\epsfysize=0.6\textheight}
\centerline{\epsffile{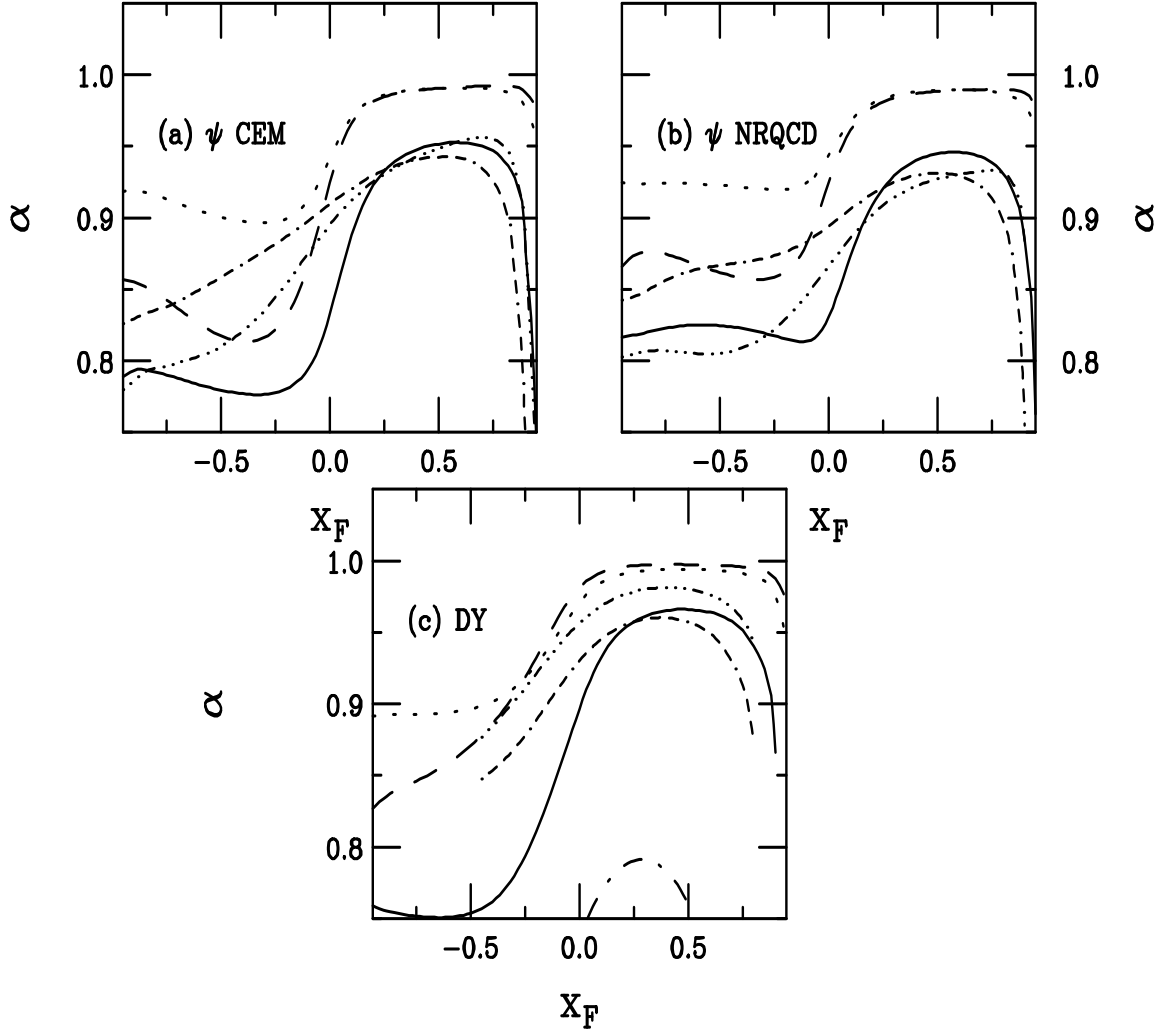}}
\caption[]{The $A$ dependence assuming BH loss
for $\psi$ in (a) the CEM and in (b)
NRQCD and (c) Drell-Yan production.  The MRST LO distributions are used for
CEM $\psi$ and Drell-Yan production while CTEQ 3L densities are used with NRQCD
$\psi$ production. The maximum BH loss, Eq.~(\protect\ref{bhmax}), 
is shown in the
solid curves at 800 GeV and in the dot-dashed curves at 120 GeV. The minimum 
BH loss, Eqs.~(\protect\ref{bhminq}) and (\protect\ref{bhming}), 
is shown in the
dashed curves at 800 GeV and in the dot-dot-dot-dashed curves at 120 GeV.  
The original BH loss, Eq.~(\protect\ref{bhdelx}), is shown in the
dotted curves at 800 GeV and in the dot-dash-dash-dashed curves at 120 GeV.} 
\label{loss3}
\end{figure}

Later, Brodsky and Hoyer \cite{BH}, BH, argued that the energy loss in the
Gavin and Milana model was too large because there is not enough time
after the initial QCD bremsstrahlung for the color field of the parton to be
regenerated.  Therefore, the subsequent interactions of the parton in the
target do not lead to a large increase in energy loss \cite{LPM}.  From the
uncertainty principle they deduced that the loss should be independent of
parton type and the change in $\Delta x_1$ should be bound so that
\be \Delta x_1 < \frac{\langle k_\perp^2 \rangle L_A}{2E}
\label{bhlim} \ee 
where $L_A$ is the path length through the medium and $\langle
k_\perp^2 \rangle$ is the average transverse momentum of gluons radiated by
the incoming parton.  If $E = x_1s/2m_p$ and $L_A \sim R_A \propto A^{1/3}$,
then \be \Delta x_1 \leq \frac{\kappa}{x_1s} A^{1/3} \label{bhdelx} \ee 
where $\kappa \propto m_p \langle k_\perp^2 \rangle$.  The
average radiative loss is thus expected
to be $-dE/dz \sim 0.25$ GeV/fm with another 0.25 GeV/fm loss expected to
arise from elastic scattering.  In this case, when $\Delta x_1 \propto
c/x_1$, $x_1 = 0.5(x_1' + \sqrt{(x_1')^2 + 4c})$.  The $x_F$ dependence of
$\alpha$ when $c = \kappa A^{1/3}/s$, referred to henceforth as ``original BH
loss'', is given by the dotted and dot-dash-dashed
curves in Fig.~\ref{loss3} for $\psi$ and Drell-Yan production.

Subsequently, the bound on $dE/dz$ was refined through the work of Baier {\it
et al.}\ \cite{bdmps1,bdms} where they determined \be -\frac{dE}{dz} = \frac{3
\alpha_s}{4} \langle p_{\perp W}^2 \rangle \label{bdmpsdedz} \ee
with $\langle p_{\perp W}^2 \rangle$ the characteristic squared transverse
momentum of the parton\footnote{After the 
inclusion of other diagrams suggested by
Zakharov \cite{bgz}, Baier {\it et al.}\ concluded that the loss derived in
Ref.~\cite{bdmps1} was, in fact, a factor of two larger \cite{bdms}.  This
difference is reflected in Eq.~(\ref{bdmpsdedz}).}.
The value of
the radiative loss is independent of the details of the scattering process
as long as $L_A$ is large.  In this description, $\Delta x_1$ is
then \be \Delta x_1 = \frac{3 \alpha_s}{2} \frac{m_p}{x_1 s} L_A
\langle p_{\perp W}^2 \rangle \label{bdmpsdx} \ee
where the average transverse momentum $\langle p_{\perp W}^2 \rangle$ is
proportional to $A^{1/3}$ \cite{bdmps1}. Since $\langle p_{\perp W}^2 \rangle
\propto A^{1/3}$,
$\Delta x_1 \propto A^{2/3}$ in Eq.~(\ref{bdmpsdx})
rather than $A^{1/3}$ as postulated by Brodsky
and Hoyer \cite{BH}, Eq.~(\ref{bhdelx}), 
because they assumed that $\langle k_\perp^2 \rangle$ was
independent of $A$.  

Two estimates of $\langle p_{\perp W}^2 \rangle$ were provided in
Ref.~\cite{bdmps1}.  The larger value, used as an upper limit, comes
from a single nuclear rescattering of photoproduced dijets \cite{LQS}, 
\be  \langle p_{\perp W}^2 \rangle = \pi^2 \alpha_s \lambda_{\rm
  LQS}^2 A^{1/3} \frac{C_A \sigma^{\gamma A}_g + C_F \sigma^{\gamma
  A}_q}{\sigma^{\gamma A}} \, \, .
\label{bhlqs} \ee  They obtained $0.05 < \lambda^2_{\rm LQS} < 0.1$ GeV$^2$ by
assuming that dijet production is dominated either by quarks or gluons using
the measured $p_T$ broadening as a function of $A$.  With the 
lower bound on $\lambda^2_{\rm LQS}$,
\be  \langle p_{\perp W}^2 \rangle \simeq 0.658 \, \alpha_s \, A^{1/3} 
\, {\rm GeV}^2 \, \, . \label{bhmax} \ee  
Since the initial states could not be explicitly identified, we
assume that $\langle p_{\perp W}^2 \rangle$ is identical 
for quarks and gluons.  Then
when $\alpha_s \sim 0.3$ and $A=184$, we find $-dE/dz
\simeq 1.28$ GeV/fm with Eq.~(\ref{bhmax}). 
We refer to this as ``maximum BH loss''
in the remainder of the discussion even though $dE/dz$ is actually smaller
than the proposed GM loss.  (The difference in the subsequent shapes of
$\alpha(x_F)$ lies in the form of $\Delta x_1$.) The second estimate 
depends on the nucleon gluon distribution and contains explicit color factors
so that \be \langle p_{\perp W}^2 \rangle_q & = & 
\frac{2 \pi^2 \alpha_s}{3} \rho_A xG(x,Q^2) L_A \simeq 0.07 \, \alpha_s \,
A^{1/3} \, {\rm GeV}^2 \label{bhminq} \\ 
\langle p_{\perp W}^2 \rangle_g & = & 
\frac{9}{4} \langle p_{\perp W}^2 \rangle_q \simeq 0.15 \, \alpha_s \,
A^{1/3} \, {\rm GeV}^2 \label{bhming} \ee
where $xG(x) \sim 1-2$ for the $x_1$ range of E866. This lower estimate is
referred to subsequently as ``minimum BH loss''.  
Now when $\alpha_s \sim 0.3$ and $A=184$, $-dE_q/dz \simeq 0.12$ 
GeV/fm and $-dE_g/dz \simeq 0.28$ GeV/fm.  

The resulting $x_F$ dependence of $\alpha$ is shown in
Fig.~\ref{loss3} at 800 and 120 GeV.  
At negative $x_F$, $x_1$
can be considerably larger than $x_1'$, up to an order of magnitude as $x_F
\rightarrow -1$ at 800 GeV.  The difference in shapes at negative $x_F$ between
the CEM and NRQCD arise from the relative importance of $q \overline q$
annihilation and $gg$ fusion (as well as $qg$ scattering in NRQCD).  Even
though the energy loss is the same for quarks and gluons with the original and
maximum BH loss estimates,
the relative change is larger for the gluon than the sea quark distributions
when $x_1$ is small.  At large negative $x_F$, the $q \overline q$ contribution
is dominant with $\Sigma_q \, q(x_1) \overline q(x_2)
\approx \overline u(x_1) u(x_2) + \overline d(x_1) d(x_2)$ 
and the change in the projectile sea quark distribution is less than that of
the gluon distribution.   When $gg$ fusion
dominates, $\alpha(x_F)$ decreases.  In the NRQCD model, the $qg$
contribution tends to balance this difference, leading to the flatter
$\alpha(x_F)$ for $x_F <0$, particularly at 800 GeV.  As $x_F$ approaches zero,
the change in all the distributions becomes smaller.  The change in $x_1$ due
to the minimum BH loss is only $\sim 20$\% at $x_F \approx 0$, decreasing to 
less than 8\% at $x_F = 0.1$.  Note also that the
predicted minimum loss for gluons, Eq.~(\ref{bhming}), and the 
original BH loss, Eq.~(\ref{bhdelx}), result in a similar $\alpha(x_F)$ at
forward $x_F$ even
though the $A$ dependencies of the two are different. That is because the
shift in $x_1$ is reduced at large $x_F$, the change in all parton densities
is small when $x_1 \approx 0.1-0.9$ and the original BH model $dE/dz$ is very 
similar to $dE_g/dz$ for the minimum BH loss.
At 800 GeV, the drop at large $x_F$ is due to loss by 
valence quarks since at large $x_F$, $\Sigma_q q(x_1) \overline 
q(x_2) \approx \overline u(x_2) u(x_1) + \overline d(x_2) d(x_1)$.  
At 120 GeV, the
effect of the loss is larger since $\Delta x_1 \propto 1/s$.  The
correspondingly higher $x_1'$ values at 120 GeV reduce the gluon
distribution shift relative to $q \overline q$.  This, as well as the greater
importance of $q \overline q$ annihilation, 
results in the different shapes of $\alpha(x_F)$ for the
two energies at negative $x_F$.  

The Drell-Yan results are similar
to the calculated $\psi$ results except that the Drell-Yan loss is weaker at
larger $x_F$.  Part of the difference is because $x_1$ and the scale $M$ at
which the parton densities are evaluated are both greater than for the $\psi$.
We can also see that the
maximum BH loss is almost certainly too large to explain the current Drell-Yan
results.  Indeed, at 120 GeV,
$\alpha(x_F)$ barely appears on the plot.  The Drell-Yan results are similar
to the calculated $\psi$ results except 

The large change in $x_1$, appearing as large $\Delta x_1$, suggests that the
calculation may not be applicable for $\Delta x_1 > x_1$.  At 800 GeV, $\Delta
x_1 = x_1$ occurs when $0.03 \leq x_1 \leq 0.09$, depending on the loss
estimate, corresponding to a minimum $x_F$ of $-0.1 \leq x_F \leq 0.02$.  At
120 GeV, the $x_1$ values are larger, $x_1 \approx 0.1-0.3$ for the original
and minimum BH loss estimates and 0.6 for the maximum BH estimate,
corresponding to $x_F \approx -0.2-0$ and 0.5 respectively.  In our
calculations, we will apply the model over all $x_F$.

We finally note that neither of these initial state models of energy
loss alone can reproduce the data.  GM loss does not have the same curvature
of the data at large $x_F$ while BH loss is too weak at large $x_F$ and too
strong at low $x_F$.

\subsection{Final State Loss} 

The second model of energy loss we consider is 
applicable only to the quarkonium system and not to Drell-Yan production
which does not involve color confinement in the
final state \cite{KS}.  
When a $c \overline c$ pair is produced in a color octet state, it 
has to emit a soft gluon in order to produce the final-state $\psi$ or 
$\psi^\prime$.  This $c \overline c$ can propagate some distance, essentially
longer than its path through the nucleus, before the soft gluon is finally
emitted.  This is because the Landau-Pomeranchuk-Migdal effect \cite{LPM}
in QCD causes a delay
in the emission of the third gluon to neutralize the color of the $c \overline
c$ state due to successive interactions of the colored $c \overline c$ pair in
the medium.  However, each successive interaction of the $c \overline c$ pair
degrades its momentum.

This final-state loss model, developed by Kharzeev and Satz \cite{KS} and
referred to as KS loss here, is
applicable only when the $c \overline c$ pair interacts in the color octet
state, essentially for $x_F \geq 0$.  After $n$ interactions along its path
length before leaving
the target, the pair's momentum is reduced by $\sim \kappa L_A$ where $\kappa$,
the hadronic string tension, is determined from lattice studies of confinement
between colored objects, $\kappa \sim (9/4)$ GeV/fm \cite{KS}, and $L_A$ is the
distance the pair has traveled through the target, calculated for the nuclear
shape
distributions in Ref.~\cite{JVV}.  A $\psi$ state observed at a given $x_F$
has actually been produced with a higher value, $x_F/\delta$, where
$\delta \approx 1 - \kappa L_A/P_\psi$
and $P_\psi$ is the $\psi$ momentum in the center of mass frame.

The $x_F$ distribution $G_A(x_F)$ then has two parts \cite{KS},
\be G_A(x_F) \propto S_A G_p(x_F) + (1 - S_A) \frac{G_p(x_F/\delta)}{\delta}
\theta(1 - x_F/\delta) \, \, , \label{kseq} \ee
where $G_p(x_F)$ is the $x_F$ distribution in $pp$ interactions and $S_A$ is
the survival probability for the $c \overline c$ pair not to break up on its
way out of the target, calculated in Eq.~(\ref{sigfull}) for pure octet
production.  The second term includes the scatterings in the target
that cause the shift in $x_F$.  The effect of Eq.~(\ref{kseq}) does not
produce an integrated $\psi$ suppression:  the integrated
$\alpha$ in Eq.~(\ref{alfint})
is unchanged with this mechanism, only $\alpha(x_F)$ changes due to the
shift in $x_F$.

\begin{figure}[htbp]
\setlength{\epsfxsize=0.5\textwidth}
\setlength{\epsfysize=0.3\textheight}
\centerline{\epsffile{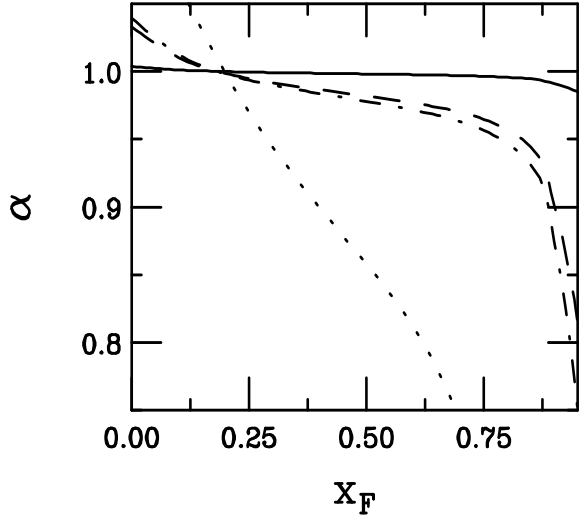}}
\caption[]{The $A$ dependence of $\psi$ production assuming KS
loss for $x_F>0$.  Octet cross sections of 1 mb (solid), 20 mb
(dashed) and 40 mb (dot-dashed) are calculated with the MRST LO parton
densities at 800 GeV.  At 120 GeV, a 40 mb octet cross section is assumed
(dotted). } 
\label{loss2}
\end{figure}

The resulting $x_F$ dependence is shown in Fig.~\ref{loss2} for
$x_F \geq 0$.  
We have illustrated the effect for the MRST LO parton
densities at 800 GeV with three different color octet cross sections: 1 mb, 20
mb and 40 mb.  A 20 mb cross section was chosen originally \cite{KS} to be 
as large as a typical meson-nucleon inelastic
cross section at the same energy, $\sigma_{\rm abs} \approx \sigma_{\pi N}^{\rm
inel} \approx 20$ mb.  The 1 mb cross section shows a minimal effect while
$\sigma_{\rm abs} \approx 40$ mb sets the scale for the maximum effect since
then $\sigma_{\rm abs} > \sigma_{pp}^{\rm inel}$.
While a large octet cross section is needed to produce a strong effect at $x_F
> 0.5$, the normalization amplifies 
$\alpha(x_F \sim 0)$ so that the shape of the
dependence is significantly different from the behavior of the E866 data at low
$x_F$.  Including shadowing will further increase $\alpha$ for $x_F \sim 0$, as
we will see.  Due to its nature, this 
model is limited to the case where all $\psi$'s are assumed to be produced 
in pure color octet states.

It is clear that KS loss alone cannot account for the shape
of the $\psi$ and $\psi'$ data in Fig.~\ref{data}.  A combination of effects
is needed.

\section{Intrinsic Charm}

The wavefunction of a proton in QCD can be represented as a
superposition of Fock state fluctuations, {\it e.g.}\ $\vert uudg
\rangle$, $\vert uud q \overline q \rangle$, $\vert uud Q \overline Q \rangle$,
\ldots of the $\vert uud \rangle$ state.
When the projectile scatters in the target, the
coherence of the Fock components is broken and the fluctuations can
hadronize \cite{intc,BHMT}.  These
intrinsic $Q
\overline Q$ Fock states are dominated by configurations with
equal rapidity constituents so that, unlike sea quarks generated
from a single parton, the intrinsic heavy quarks carry a large
fraction of the parent momentum \cite{intc}.
 
The frame-independent probability distribution of a $5$--particle
$c \overline c$ Fock state in the proton is
\be
\frac{dP^5_{\rm ic}}{dx_i \cdots dx_5} = N_5 \alpha_s^4(m) 
\frac{\delta(1-\sum_{i=1}^5 x_i)}{(m_p^2 - \sum_{i=1}^5
(\widehat{m}_i^2/x_i) )^2} \, \, ,
\label{icdenom}
\ee
where $N_5$ normalizes the $|uud c
\overline c \rangle$ probability, $P^5_{\rm ic}$.  The delta function
conserves longitudinal momentum.  The denominator is
minimized when the heaviest constituents carry the largest fraction of the
longitudinal momentum, $\langle x_Q \rangle > \langle x_q \rangle$, maximizing
$P_{\rm ic}^5$.  We choose
$\widehat{m}_q = 0.45$ GeV and $\widehat{m}_c = 1.8$ GeV \cite{VBH1}.
 
The intrinsic charm production cross section from the $5$-particle state
can be related to
$P^5_{\rm ic}$ and the inelastic $pN$ cross section by
\be
\sigma^5_{\rm ic}(pN) = P^5_{\rm ic} \sigma_{p N}^{\rm in}
\frac{\mu^2}{4 \widehat{m}_c^2} \, \, .
\label{icsign}
\ee
The factor of $\mu^2/4 \widehat{m}_c^2$ arises from the soft
interaction which breaks the coherence of the Fock state. We 
assume that the NA3 diffractive $\psi$  cross
section \cite{NA3}, the second term in Eq.~(\ref{twocom}) proportional to
$A^\beta$, 
can be attributed to intrinsic charm and find $\mu^2 \sim 0.1$ GeV$^2$.  

While the total intrinsic charm cross section is relatively easy to define,
there are some uncertainties in the relative weights of
open charm and $\psi$ production from
an intrinsic charm state.  In general, the 
$\psi$ production cross section is significantly
smaller than the open charm production cross section.  There are several
factors that can suppress $\psi$ production relative to open charm in standard
charmonium production models such as the CEM and NRQCD as well as in the
intrinsic charm model.  As in the CEM, the probability to produce a $\psi$ from
an intrinsic $c \overline c$ state is proportional to the fraction of 
intrinsic $c \overline c$ production
below the $D \overline D$ threshold. The fraction of $c \overline c$
pairs with $2m_c < m < 2m_D$ is
\be
f_{c \overline c/h} = \int_{4 m^2_c}^{4
m_D^2} dm^2 \ \frac{dP_{\rm ic}}{dm^2}
\ \Bigg/ \int_{4 m^2_c}^s 
dm^2 \ \frac{dP_{\rm ic}}{dm^2}
\, \, ,
\ee
typically smaller than that obtained in
the CEM \cite{HPC}.  
However, as discussed in Section 2.1, not all $c \overline c$ pairs
below the $D \overline D$ threshold will produce a final state $\psi$.  The
fraction that actually become $\psi$'s is rather small, on the order of 2.5\%
in the CEM \cite{HPC}.  Since the additional suppression factors involved in
the intrinsic charm 
model are not completely fixed \cite{vbpsi2}, rather than discuss all
the uncertainties here, we will use an effective intrinsic charm
probability, $P_{\rm ic}^{\rm eff}$.  
The EMC charm structure function data is consistent with 
$P^5_{\rm ic} = 0.31$\%  for low energy virtual photons but $P^5_{\rm ic}$
could be as large as 1\% for the highest virtual photon energies
\cite{EMCic,hsv}.  Typically the more conservative result is used but in this
paper, we will use the larger value in most of our calculations and show the
effect of reducing and/or eliminating the intrinsic charm component.

\begin{figure}[htbp]
\setlength{\epsfxsize=\textwidth}
\setlength{\epsfysize=0.6\textheight}
\centerline{\epsffile{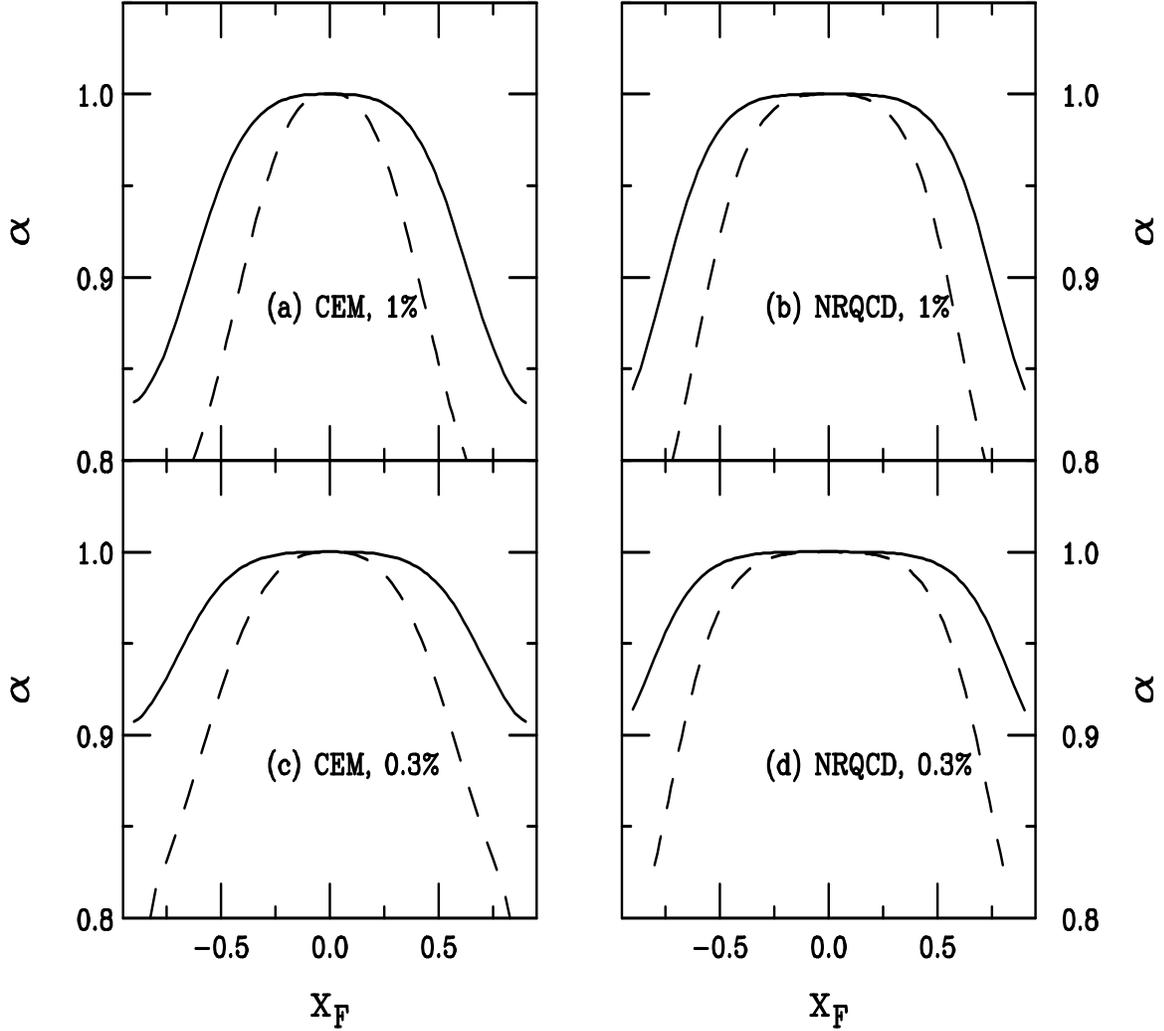}}
\caption[]{The $A$ dependence of intrinsic charm at 800 GeV (solid) and 120 GeV
(dashed).  In (a) and (c) an effective production probability of 1\% is 
assumed in the CEM 
and in NRQCD respectively while in (c) and (d) 
$P_{\rm ic}^{\rm eff} = 0.31$\% is assumed in the
CEM and in NRQCD. } 
\label{alfic}
\end{figure}

Including a delta function to combine the $x_c$ and $x_{\overline c}$ in the
$\psi$ state, the $\psi$ $x_F$ distribution from intrinsic charm is
\be \frac{d\sigma_d}{dx_F} =  \sigma_{p N}^{\rm in}
\frac{\mu^2}{4 \widehat{m}_c^2} \int \prod_{i=1}^5 dx_i
\, \frac{dP^{\rm eff}_{\rm ic}}{dx_1 \ldots dx_5} 
\, \delta(x_F - x_c - x_{\overline c}) \,
\, . \label{icdist} \ee Only the 5 particle Fock state is considered.
The intrinsic charm contribution is included as in Eq.~(\ref{twocom}) with
$\beta = 0.71$.  The total $A$ dependence for intrinsic charm alone is shown
in Fig.~\ref{alfic} assuming $P_{\rm ic}^{\rm eff} = 1$\% and 0.31\% with both
charmonium production models.  The contribution is symmetric around $x_F = 0$
since the projectile and target fragmentation regions are treated equally. 
Figure~\ref{alfic}(a) and (c) show
$\alpha(x_F)$ in the CEM.  
A larger effect is seen at high $x_F$
with the NRQCD model, Fig.~\ref{alfic}(b), because the NRQCD
$x_F$ distribution is narrower.  Since charmonium production models outlined in
Section 2 have a larger energy dependence than the intrinsic charm cross
section in Eq.~(\ref{icsign}), intrinsic charm is more important at 120 GeV.
When $P_{\rm ic}^{\rm eff} = 0.31$\%, the intrinsic charm contribution to
the total $A$ dependence is quite small and only significant for the largest
$x_F$ values due to the reduced $A$ dependence of the mechanism.  Assuming a
1\% probability enhances the intrinsic charm effect at large $x_F$ and even
suggests that intrinsic charm can influence $\alpha(x_F)$ at $x_F \sim 0$.

\section{Results and Predictions}

We now have a comprehensive model with which we can confront the nuclear
dependence of $\psi$, $\psi'$, and Drell-Yan production.  The nuclear effects
included in the model are shadowing of the parton
distributions, energy loss,
nuclear absorption, comover interactions, and the `diffractive'
intrinsic charm component.  It is clear from an examination of the individual
nuclear effects shown in Figs.~\ref{absco}-\ref{alfic} that no single mechanism
can correctly predict the shape of the E866 $x_F$ data.  

To compare with the preliminary data, we calculate $\alpha(x_F)$ of
$\psi$ and $\psi'$ 
production with all three models of energy loss and all shadowing
parameterizations.  
For pure octet absorption, we use all three models of
energy loss.  Only GM and BH loss are used with pure singlet absorption and
the combination of singlet and octet production.  The CEM model with the MRST
LO parton distributions is used to
calculate charmonium production for pure octet and pure singlet absorption.
NRQCD is used as the basic production model for the octet/singlet combination,
Eq.~(\ref{psicomb}).  All three shadowing parameterizations are used in each
case.  We use an effective intrinsic charm probability of 1\% but will examine
the relative importance of intrinsic charm to the overall description of the
large $x_F$ E866 data.
The absorption cross sections are chosen so that the
shadowing parameterization gives reasonable agreement with the
magnitude of $\alpha(x_F)$ for both $\psi$ and $\psi'$ production at $x_F > 0$
with GM loss.  We do not actually make detailed fits to the data to obtain the
cross sections.
The resulting absorption cross sections are given in Table~\ref{sigtable} for
GM and BH loss.  The KS model of energy loss is always
calculated with an octet absorption cross section of 40 mb.  The NA3 Pt/$^2$H
ratio as a function of $x_F$ at 200 GeV is also compared to the model
calculations.  We make predictions of the $\psi$ and $\psi'$ $A$ dependence at
120 GeV.  Finally, we show the
combined effects of shadowing and initial-state energy loss on Drell-Yan
production at 800 and 120 GeV.

\begin{figure}[htbp]
\setlength{\epsfxsize=\textwidth}
\setlength{\epsfysize=0.6\textheight}
\centerline{\epsffile{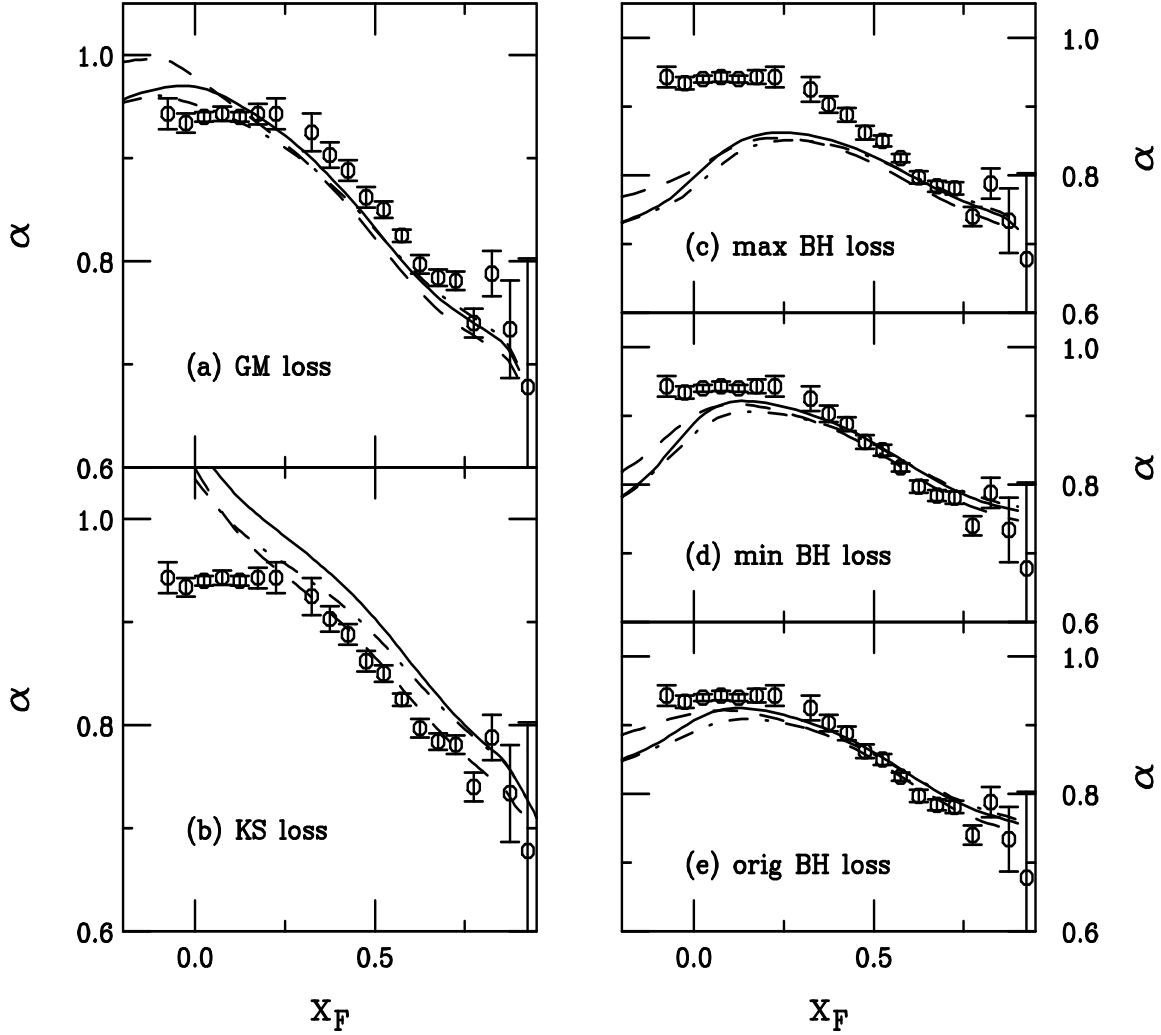}}
\caption[]{All effects are compared with the preliminary E866
$\psi$ data \protect\cite{MJL} assuming pure octet absorption.  
In (a) and (b), GM
and KS loss are assumed. Energy loss effects associated with the
BH bound are shown in (c), (d), and (e) for the estimated maximum and minimum
loss and the original bound respectively.
All calculations are in the CEM with the
MRST LO parton densities.  The curves represent shadowing with the
$S_3$ (solid), $S_2$ (dashed) and $S_1$ (dot-dashed) parameterizations. } 
\label{psioct}
\end{figure}

\begin{figure}[htbp]
\setlength{\epsfxsize=\textwidth}
\setlength{\epsfysize=0.6\textheight}
\centerline{\epsffile{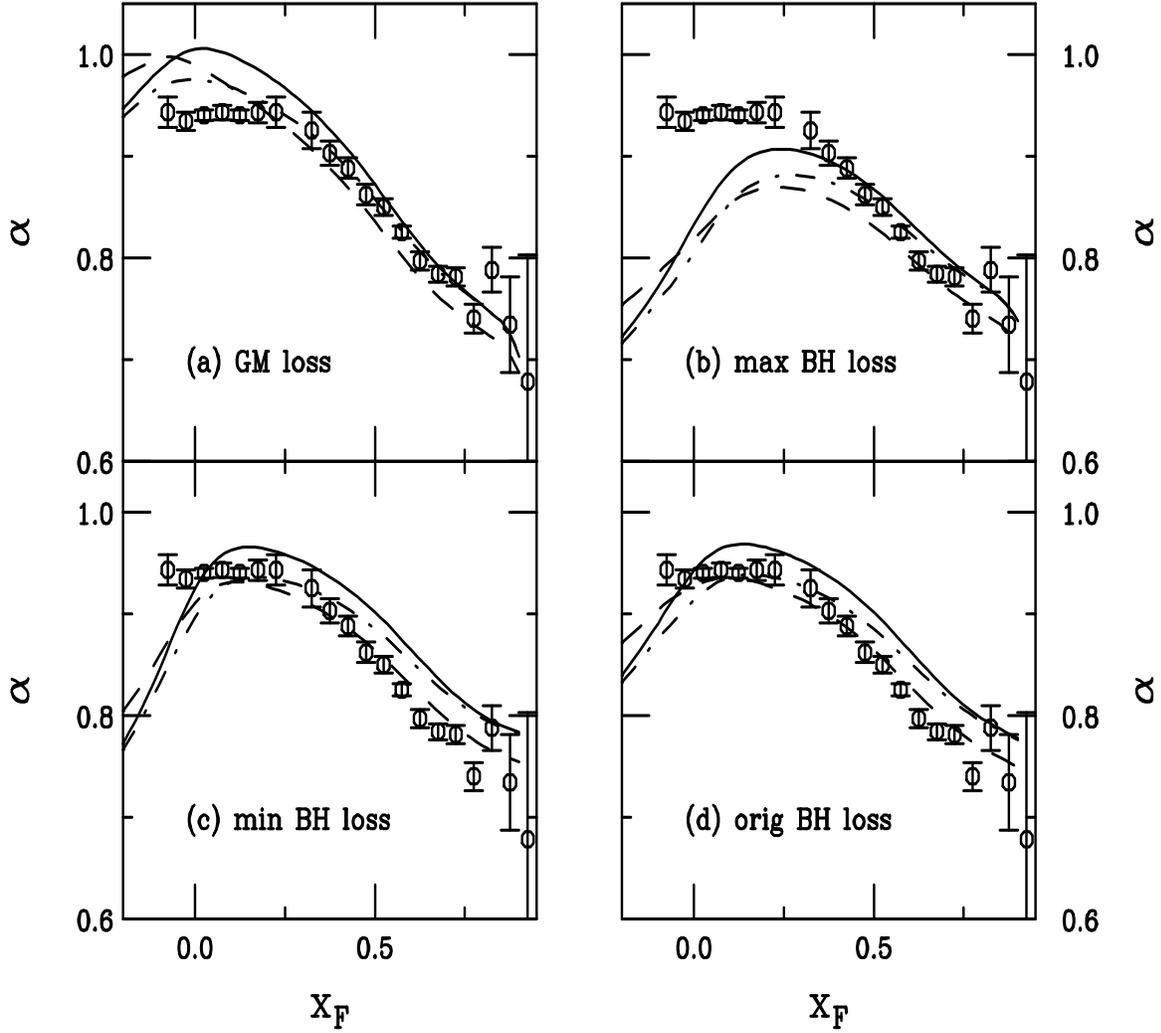}}
\caption[]{All effects are compared with the preliminary E866
$\psi$ data \protect\cite{MJL} assuming pure singlet absorption.  
In (a), GM loss is assumed. Energy loss effects associated with the
BH bound are shown in (b), (c), and (d) for the estimated maximum and minimum
loss and the original bound respectively.
All calculations are in the CEM with the
MRST LO parton densities.  The curves represent shadowing with the
$S_3$ (solid), $S_2$ (dashed) and $S_1$ (dot-dashed) parameterizations. } 
\label{psising}
\end{figure}

\begin{figure}[htbp]
\setlength{\epsfxsize=\textwidth}
\setlength{\epsfysize=0.6\textheight}
\centerline{\epsffile{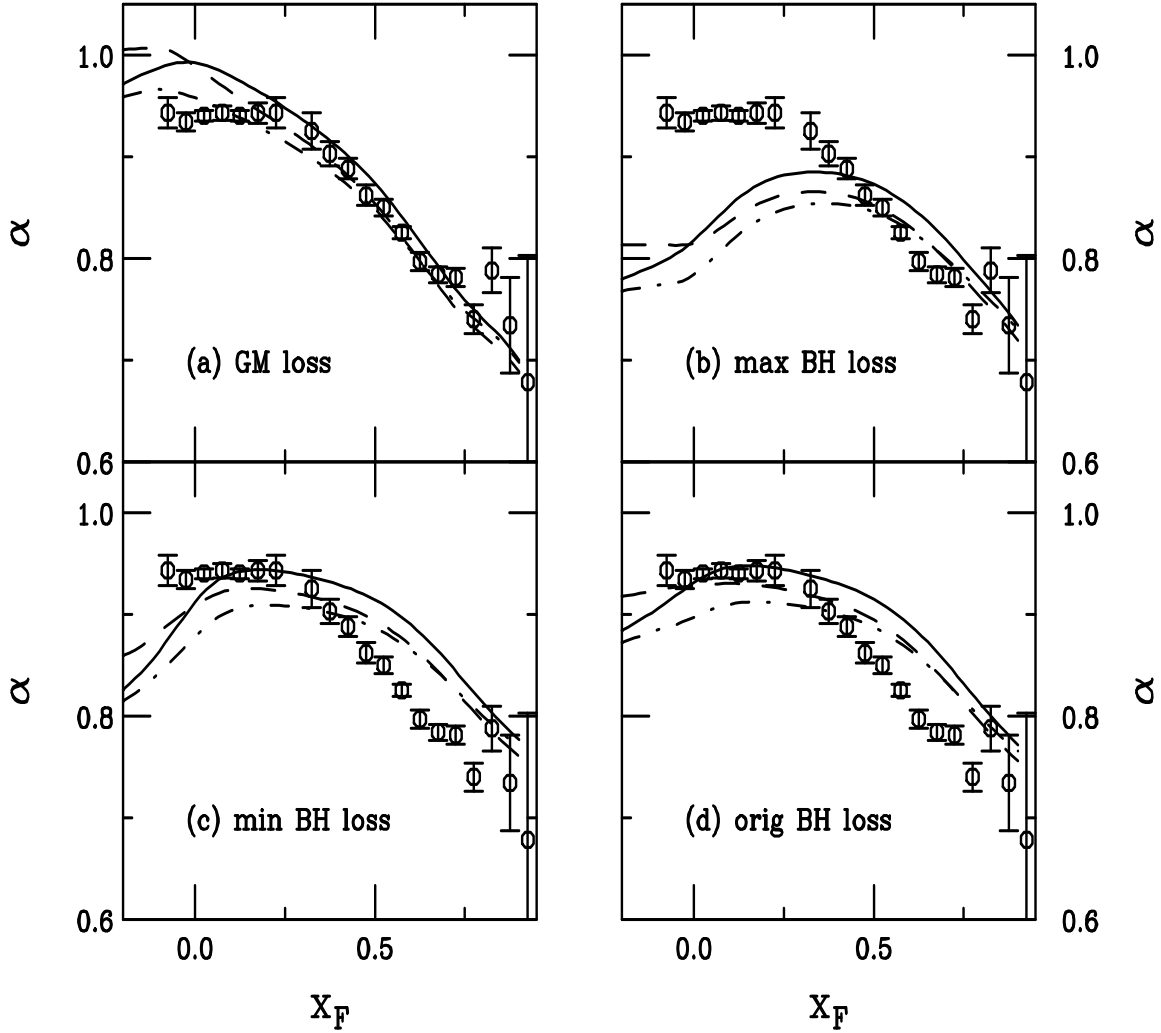}}
\caption[]{All effects are compared with the preliminary E866
$\psi$ data \protect\cite{MJL} 
assuming a combination of octet and singlet production
and absorption.  
In (a), GM loss is assumed. Energy loss effects associated with the
BH bound are shown in (b), (c), and (d) for the estimated maximum and minimum
loss and the original bound respectively.
All calculations are in NRQCD with the
CTEQ 3L parton densities.  The curves represent shadowing with the
$S_3$ (solid), $S_2$ (dashed) and $S_1$ (dot-dashed) parameterizations. } 
\label{psicombin}
\end{figure}

We first compare our full model results with the preliminary 
E866 $\psi$ data in
Figs.~\ref{psioct}, \ref{psising}, and \ref{psicombin}.  
Each figure shows the difference in the 
shadowing mechanisms for each type of energy loss with a particular absorption 
mechanism.
In Fig.~\ref{psioct}, the pure octet
absorption mechanism is shown.  
The results with GM loss, 
Fig.~\ref{psioct}(a), best reproduce the
general trend of the data for $x_F > 0.1$.   
In general the agreement is worse at 
$x_F<0$ because $x_2$ is in the antishadowing region where antishadowing of
gluons enhances $\alpha(x_F)$, see
Figs.~\ref{shad1}-\ref{shad3}.  The KS loss model is
typically above the data, except for the $S_2$ parameterization, and is
inconsistent with the shape of the
preliminary E866 data at $x_F < 0.2$.  
Since the KS model is only applicable for 
$x_F \geq 0$, we make no further calculations with this model.
The calculations of $\alpha(x_F)$ with BH loss,
Fig.~\ref{psioct}(c)-(e), do not match the data very well at low $x_F$, 
particularly for the
maximum BH loss.  While the results 
with the maximum loss, Eq.~(\ref{bhmax}), 
produce the largest reduction at large $x_F$, the
negative $x_F$ region is far off due to the drop in $\alpha$ at negative 
$x_F$.  
The curvature of $\alpha(x_F)$ changes at $x_F \sim 0.1-0.25$, the point
at which the slope of the BH loss flattens in Fig.~\ref{loss3}(a).
Better results are achieved with the lower estimates of the BH loss, the
minimum estimate, Eqs.~(\ref{bhminq}) and (\ref{bhming}), 
and the original suggestion,
Eq.~(\ref{bhdelx}).  The data are somewhat overestimated at $x_F < 0.2$ with
the minimum loss but the overestimate is slight for the original BH loss 
with the $A^{1/3}$ dependence.
Choosing a smaller absorption cross section would improve the agreement with
the data at low $x_F$ although it would worsen the agreement at $x_F > 0.25$.
We also note that none of the absorption cross sections
are greater than 3 mb, already more than a factor of two less than the 7.3 mb
effective absorption cross section found in Ref.~\cite{klns}.

Pure singlet absorption, shown in Fig.~\ref{psising}, results in somewhat
poorer agreement with the data than pure octet absorption because
the $\psi$ is always produced outside the target when $x_F >0$ at 800 GeV.
Therefore changing the absorption cross section would not improve the agreement
with the data.  The choice of parton distribution function 
in the CEM model results in small changes in
the shape of $\alpha(x_F)$ and does not influence the overall agreement with
the data.

\begin{figure}[htbp]
\setlength{\epsfxsize=\textwidth}
\setlength{\epsfysize=0.3\textheight}
\centerline{\epsffile{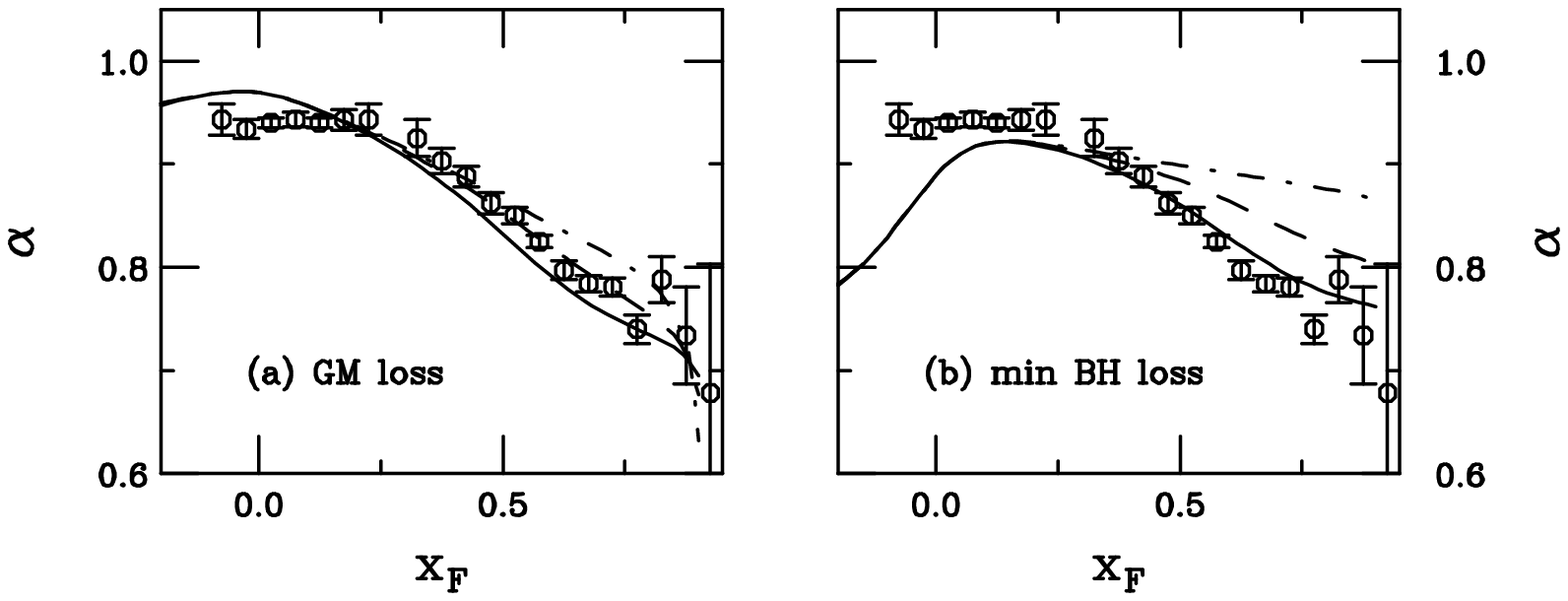}}
\caption[]{The effective probability of intrinsic charm is varied for pure 
octet production with (a) GM loss and (b) the minimum BH loss.
The curves represent an effective intrinsic charm probability of
1\% (solid), 0.31\% (dashed) and 0\% (dot-dashed). } 
\label{psiictest}
\end{figure}

A combination of octet and singlet absorption in the NRQCD $\psi$ production
model produces rather good 
agreement for all shadowing parameterizations when the
GM model is applied, Fig.~\ref{psicombin}(a).  
The difference in curvature at $x_F > 0$ between the calculations with
BH loss and the preliminary data are larger than in the CEM because 
the $A$ dependence of shadowing in NRQCD is weaker at positive $x_F$ than that
of the CEM, see Figs.~\ref{shad1}-\ref{shad3}.  As explained in Section 6, the
difference in the $A$ dependence of the two production models is due to the
chosen $x$ values and the charm quark mass which sets the scale for evolution.

In all cases, the most striking disagreement of the GM loss model 
with the data occurs at $x_F < 0.1$ when the calculated $\alpha$ slightly
overshoots the data due to the antishadowing of the gluon distribution.
The shape of $\alpha(x_F)$ here depends most strongly on the shadowing
parameterization since the other $x_F$ dependent contributions are rather
slowly varying.  
None of these parameterizations produce the same curvature
as the data and, even if they did, the additional absorption required for the
calculations to agree with the data would ruin the agreement of the model
with the data
at forward $x_F$.  Increasing absorption at $x_F < 0$ by artificially
enhancing the comover density would not significantly improve the agreement.

The pure octet and pure singlet calculations with BH loss are in reasonably
good agreement with the data for $x_F > 0.2$
The maximum estimated loss is in clear
disagreement with the data at all $x_F$, both in shape and in magnitude.  This
disagreement would persist, even if final-state absorption of the $\psi$ were
ignored\footnote{S. Gavin has addressed the E772 $\psi$ data with a combination
of BH-type loss and shadowing without final-state effects
and found a similar level of agreement as is seen
here \cite{Gavin}.}.  
It is also then unlikely to produce results consistent with the 
minimal $A$ dependence of Drell-Yan production, especially at lower energies, 
as we discuss later.  The agreement with the minimum and original BH
calculations is reasonable for $x_F > 0.2$.  At lower $x_F$ values, the change
in $\Delta x_1$ due to the energy loss is large.   However, since $\Delta x_1
\sim x_1$ at $x_F \sim 0$, the model is at the limit of applicability and
therefore the magnitude of the disagreement is suspect.

To show the influence of intrinsic charm, we take the $S_3$ shadowing
parameterization with GM and minimum BH loss and vary $P_{\rm ic}^{\rm eff}$
between 0 and 1\% in Fig.~\ref{psiictest}.  We choose pure octet production and
absorption because the agreement with the data seems to be among the best, see
Figs.~\ref{psioct}-\ref{psicombin} .
Since the GM loss mechanism alone causes strong reduction in $\alpha$ at large
$x_F$, see Fig.~\ref{loss1}, including intrinsic charm does not have a large
effect.  It would appear from Fig.~\ref{psiictest}(a) that $P_{\rm ic}^{\rm
eff} = 0.31$\% agrees best with the data although the agreement is reasonable
in 
all three cases.  The same is true for the combination model but pure singlet
absorption would require a larger intrinsic charm probability to agree with the
data.  On the other hand, the relatively good agreement of the minimum BH loss
calculations with the data at large $x_F$ is due to the intrinsic charm
contribution.  Without intrinsic charm with $P_{\rm ic}^{\rm eff} =1$\%, 
the model calculations would not agree with the data.  The
minimum and original BH loss models affect $\alpha$ weakly at positive $x_F$
and shadowing alone can only reduce $\alpha$ to $\sim 0.85$ as $x_F \rightarrow
1$ with the $S_2$ parameterization, see Fig.~\ref{shad2}.  Thus increasing the
relative intrinsic charm contribution is the only way to produce agreement with
the data at large $x_F$.  This is clearly shown in Fig.~\ref{psiictest}(b).
With no intrinsic charm, $\alpha(x_F)$ is relatively flat at large $x_F$.
Similar results are obtained with the pure singlet and combination absorption
models.  Note that for both loss mechanisms, intrinsic charm only affects the
shape of $\alpha(x_F)$ at $x_F > 0.25$.

\begin{figure}[htbp]
\setlength{\epsfxsize=\textwidth}
\setlength{\epsfysize=0.6\textheight}
\centerline{\epsffile{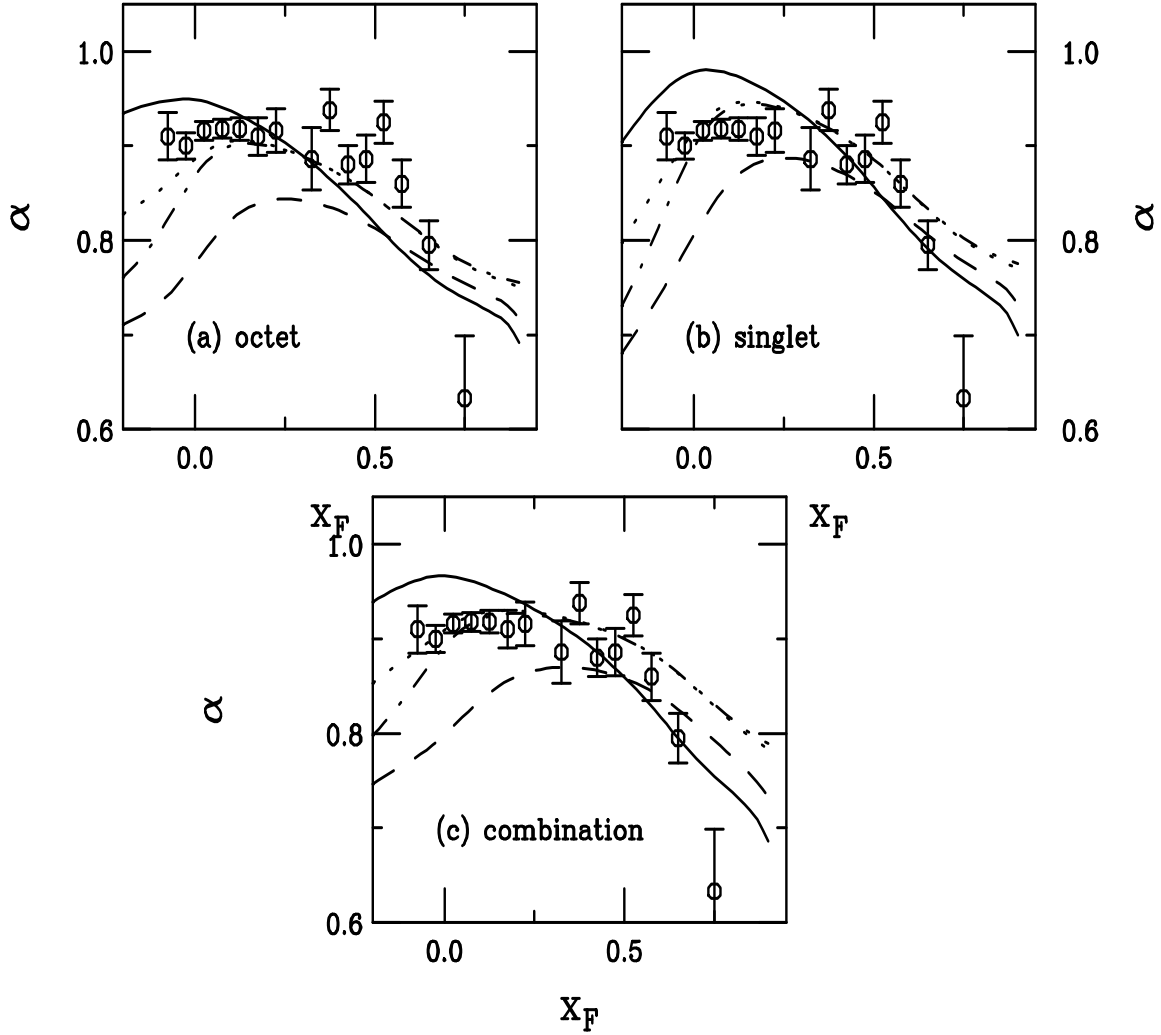}}
\caption[]{All effects are combined and compared with the preliminary E866
$\psi'$ data \protect\cite{MJL}.   
In (a), pure octet absorption is assumed. The results for pure 
singlet absorption 
are given in (b) and
combined octet/singlet absorption in (c). All calculations in (a) and (b) 
are in the CEM with the
MRST LO parton densities while the calculations in (c) are in NRQCD with the
CTEQ 3L parton densities. The $S_3$ parameterization is used for calculations
with different energy loss models.  The GM loss is shown in the solid curve
while the dashed, dot-dashed, and dotted curves are calculations with the
maximum, minimum, and original BH loss estimates respectively. } 
\label{psip800}
\end{figure}

The corresponding $\psi'$ calculations are compared to the data in
Fig.~\ref{psip800} with the $S_3$
shadowing parameterization.  
For each absorption
model, we show the $A$ dependence with the GM loss and the three estimates
of BH loss. 
Because of the greater uncertainties in the data, none of the calculations
are fully incompatible with the data at forward $x_F$.  
The largest discrepancies between model and data are at low to negative $x_F$.
GM loss produces the largest $\alpha$ because of negligible loss effects at
$x_F \approx 0$ combined with antishadowing.  The minimum and original BH loss
models agree relatively well with  all the data for the three absorption
models.  Note that these two loss models coincide at large $x_F$ both 
because of their similar behavior at large $x_F$, see Fig.~\ref{loss3}, and
the intrinsic charm contribution at forward $x_F$.
All the calculations result in a
slightly lower $\alpha$ for the $\psi'$ than the $\psi$ due to the larger
comover cross section,
$\sigma_{\psi' {\rm co}}$.  

The $A$ dependence of $\psi$ and
$\psi'$ production has been shown to be similar in previous measurements
\cite{E772}, albeit not to
high precision.  To compare the two results here, we calculate the integrated
$\alpha$ in the interval $-0.2 \leq x_F \leq 0.8$ for all energy loss models
and all shadowing parameterizations.  The results 
are shown in Tables~\ref{psialfoct}-\ref{psialfcombo}.  
The change in $\alpha$ between $\psi$
and $\psi'$ at 800 GeV is small, typically a $2-3$\% difference.  
One might expect
that for pure octet absorption, the integrated $\alpha$ would be identical
for $\psi$ and $\psi'$.  However, the $pA$ comover interactions are treated
assuming formed $\psi$ and $\psi'$ interact with secondaries and the $\psi'$
comover absorption cross section is larger.  
Thus, even though the comover interaction cross sections are typically
significantly smaller than the corresponding nucleon absorption cross sections,
the difference is large enough to cause the observed 2\% shift in the
integrated $\alpha$ in the Tables.  (See also Fig.~\ref{absco}(d) which
highlights the differences in the assumed comover cross sections.)
Indeed, without this difference in the cross sections, the model calculations
would not agree as well with the data in Fig.~\ref{psip800}. 

We note that we do not expect our values of $\alpha$ 
to agree in detail with the integrated data because
our estimates do not agree with the preliminary
data at all $x_F$.  The GM model always overestimates the data at low to
negative $x_F$.  Thus the GM results can
then be expected to overestimate the integrated $\alpha$ of the data.
Typical $\alpha$ values are between 0.94 and 0.98.
The original and minimum estimates of BH loss typically underestimates the low
and negative $x_F$ data and should therefore underestimate the measured total
$\alpha$.  In this case, $0.87 < \alpha < 0.94$, similar to $\alpha = 0.91$
\cite{E772}.  Even though the nuclear absorption cross sections are small, the
effective absorption can be large, compatible with that obtained assuming
absorption is the only source of the $\psi$ $A$ dependence.

The integrated $\alpha$ 
also depends on the shadowing parameterization.  Typically $\alpha$ with $S_2$
is largest with the octet absorption
because the $S_2$ parameterization is not available for $A=9$ and
thus treats the Be nucleus like a proton.  Therefore, although the $S_3$
parameterization has a larger gluon antishadowing effect than $S_2$, the
calculated value of $\alpha$ is larger for $S_2$ near $x_F \approx 0$.
This difference, along with smaller absorption cross sections used
with $S_2$, see
Table~\ref{sigtable}, results in a larger integrated $\alpha$ with $S_2$ for 
pure octet absorption, Table~\ref{psialfoct}.   Since the
$\psi$ and $\psi'$ are formed after they have left the nucleus in the pure
singlet case at 800 GeV, the $\alpha$ obtained with $S_3$ is larger at this
energy, see Table~\ref{psialfsing}.  In the
combination octet/singlet calculations, Table~\ref{psialfcombo}, $\alpha$ is
very similar for $S_2$ and $S_3$. 

\begin{figure}[htbp]
\setlength{\epsfxsize=\textwidth}
\setlength{\epsfysize=0.6\textheight}
\centerline{\epsffile{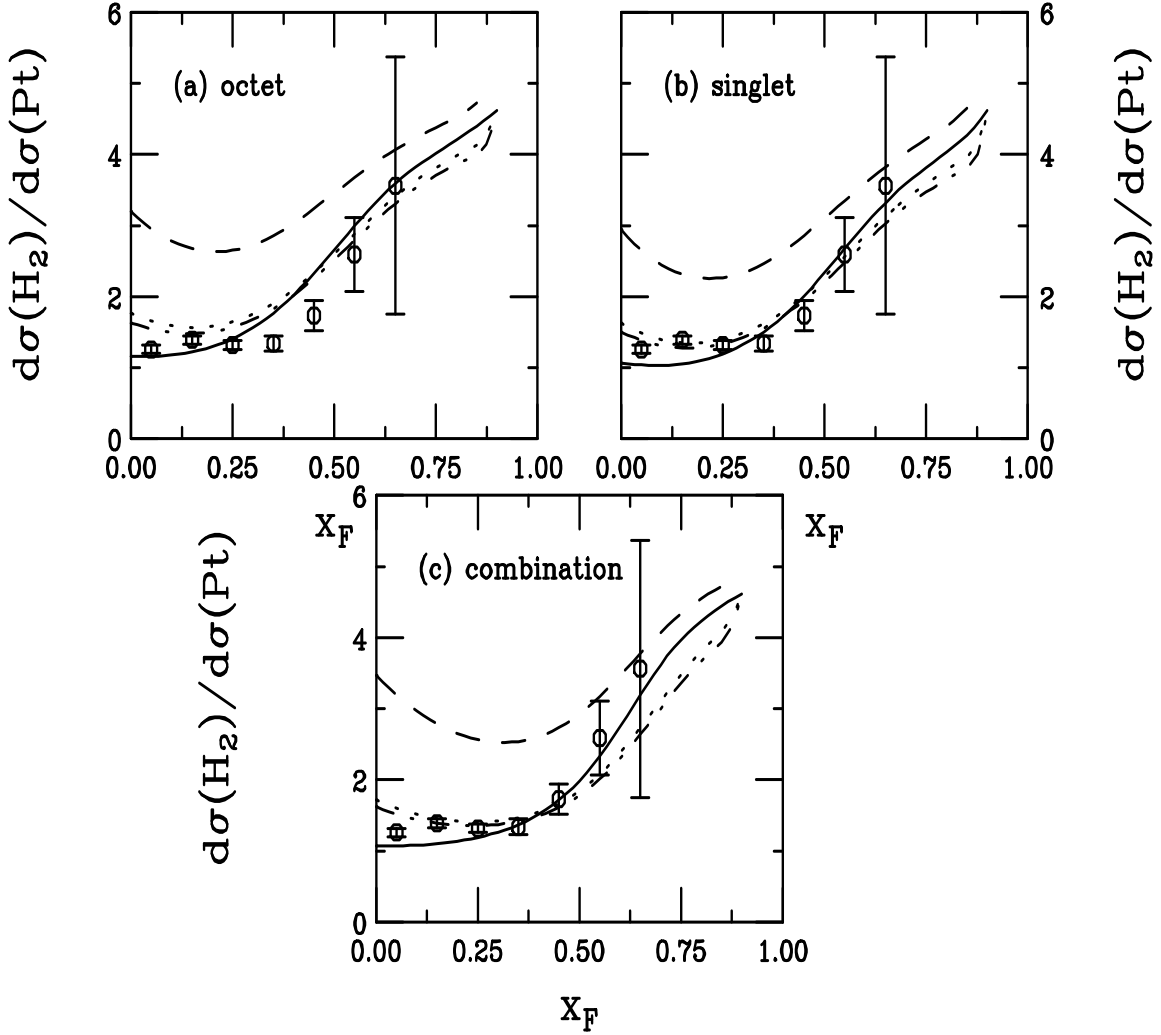}}
\caption[]{All effects are combined and compared with the NA3
$\psi$ data \protect\cite{NA3}.   
In (a), pure octet absorption is assumed. The results for pure 
singlet absorption 
are given in (b) and
combined octet/singlet absorption in (c). All calculations in (a) and (b) 
are in the CEM with the
MRST LO parton densities while the calculations in (c) are in NRQCD with the
CTEQ 3L parton densities. The $S_3$ parameterization is used for calculations
with different energy loss models.  The GM loss is shown in the solid curve
while the dashed, dot-dashed, and dotted curves are calculations with the
maximum, minimum, and original BH loss estimates respectively. } 
\label{na3fig}
\end{figure}

A measurement of the $\psi$ $A$ dependence
at 120 GeV has been proposed.  Such an energy would be the closest to the NA50
Pb+Pb measurement.  The
most precise lower energy data with a proton beam was taken at 200 GeV 
by the NA3 collaboration
\cite{NA3}.  We compare our calculations of the ratio Pt/$^2$H to their 
data in Fig.~\ref{na3fig}.  
The three absorption mechanisms are shown with the data for GM loss and the
three BH loss estimates with the $S_3$ shadowing parameterization.
We see that essentially none of the calculations contradict
the large $x_F$ data, presumably due to the relatively poor statistics for 
$x_F > 0.5$.  The best agreement at all $x_F$
is obtained with minimum and original BH loss with the pure singlet and
combination absorption models.  GM loss 
tends to underestimate the low $x_F$ data.  The 
maximum BH loss estimate produces a higher ratio at low $x_F$, as seen in
Fig.~\ref{na3fig}(a) and (c), far above the data.  On the whole,
the results at 200 GeV concur with those at 800 GeV.
 
\begin{figure}
\setlength{\epsfxsize=\textwidth}
\setlength{\epsfysize=0.6\textheight}
\centerline{\epsffile{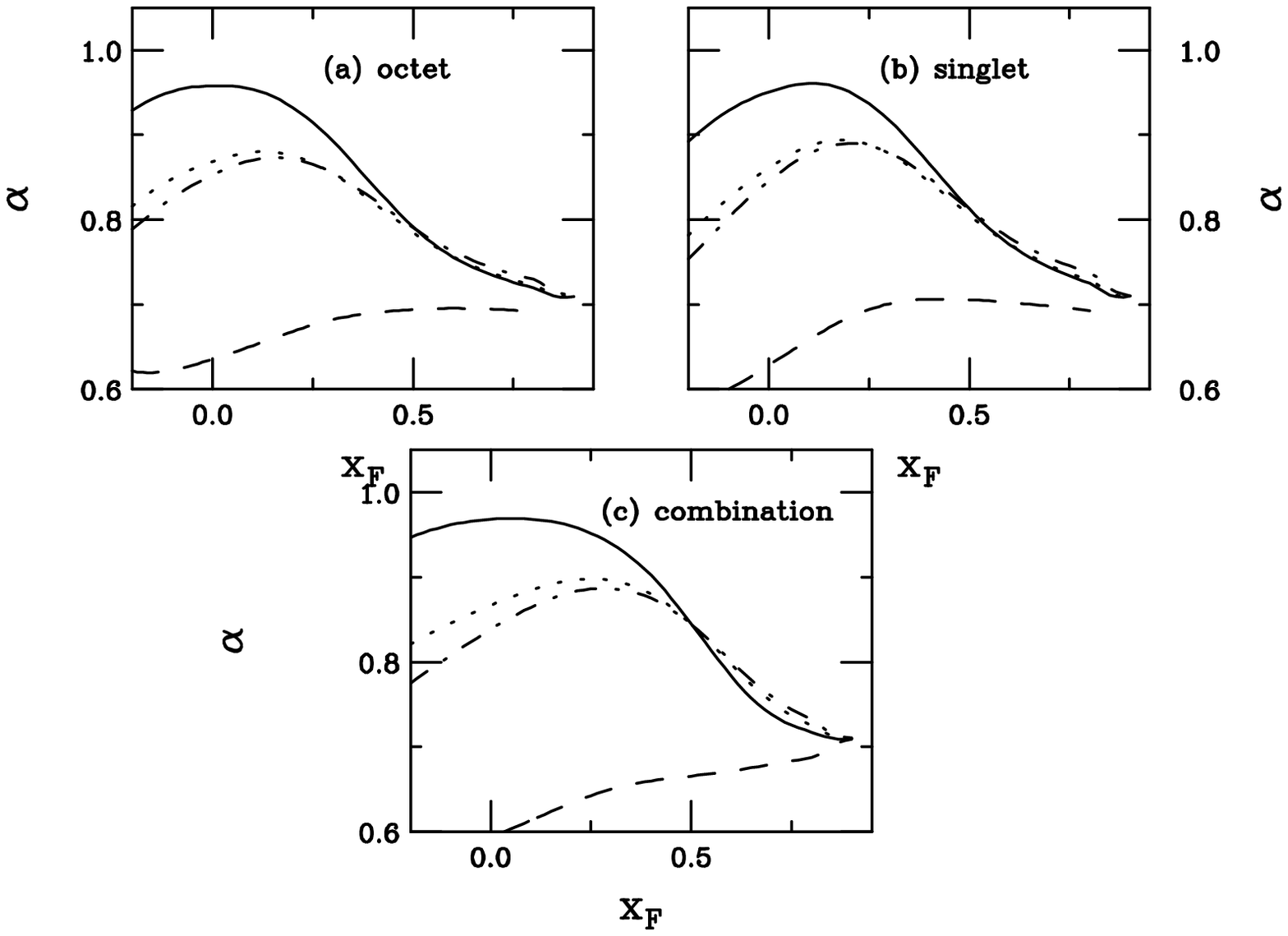}}
\caption[]{Predictions are made for $\psi$ production at 120 GeV.
In (a), pure octet absorption is assumed. The results for pure 
singlet absorption 
are given in (b) and
combined octet/singlet absorption in (c). All calculations in (a) and (b) 
are in the CEM with the
MRST LO parton densities while the calculations in (c) are in NRQCD with the
CTEQ 3L parton densities. The $S_3$ parameterization is used for calculations
with different energy loss models.  The GM loss is shown in the solid curve
while the dashed, dot-dashed, and dotted curves are calculations with the
maximum, minimum, and original BH loss estimates respectively. } 
\label{psilo}
\end{figure}

\begin{figure}
\setlength{\epsfxsize=\textwidth}
\setlength{\epsfysize=0.6\textheight}
\centerline{\epsffile{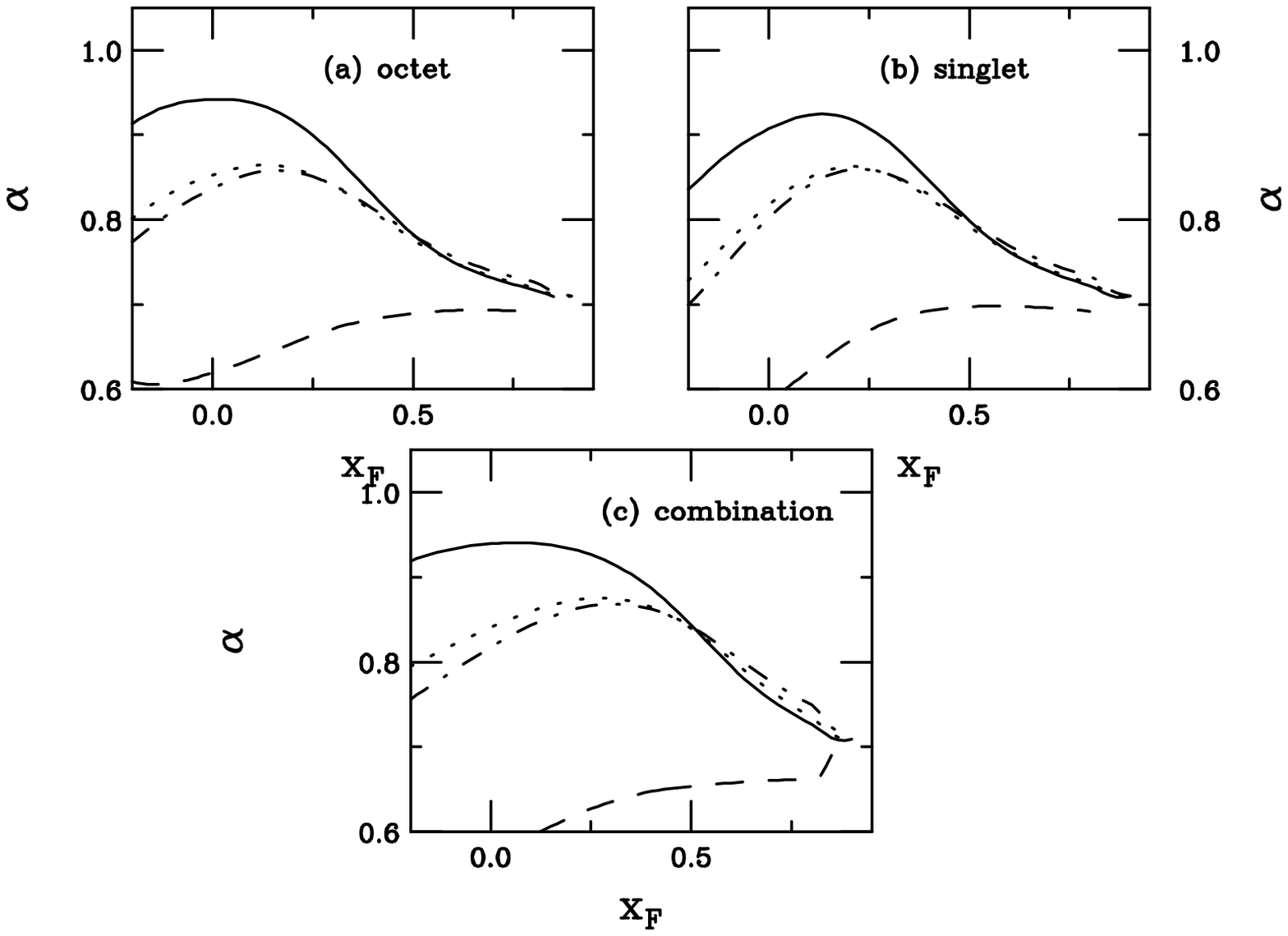}}
\caption[]{Predictions are made for $\psi'$ $A$ dependence at 120 GeV.
In (a), pure octet absorption is assumed. The results for pure 
singlet absorption 
are given in (b) and
combined octet/singlet absorption in (c). All calculations in (a) and (b) 
are in the CEM with the
MRST LO parton densities while the calculations in (c) are in NRQCD with the
CTEQ 3L parton densities. The $S_3$ parameterization is used for calculations
with different energy loss models.  The GM loss is shown in the solid curve
while the dashed, dot-dashed, and dotted curves are calculations with the
maximum, minimum, and original BH loss estimates respectively. } 
\label{psip120}
\end{figure}

Figures~\ref{psilo} and \ref{psip120} show the predictions for the
$\psi$ and $\psi'$ $A$ dependence at 120 GeV.  The calculations with GM
loss show a 
plateau-like behavior at this energy and are very similar to each other
over all $x_F$.  This is
due to the widening of the gluon antishadowing region over the $x_F$ interval,
see Figs.~\ref{shad1}-\ref{shad3}.  In the pure singlet model, since 
the $\psi$ and $\psi'$ may be produced 
inside the target at forward $x_F$, there is a peak in the calculated $\psi$
$\alpha(x_F)$ at $x_F \approx 0.2$, see Fig.~\ref{psilo}(b).
This peak is shifted
slightly forward for $\psi'$ production since the $\psi'$ singlet absorption
cross section is larger, see Fig.~\ref{psip120}(b).  
There is no forward $x_F$ peak when the octet model is considered because the
nucleon absorption is treated identically at all $x_F$.  Since the combination
octet/singlet model includes both types of absorption, there is a rather wide
plateau over $-0.2 < x_F < 0.3$.  The shape of $\alpha(x_F)$ at 120 GeV
could therefore help distinguish between absorption models.  However, BH loss
yields very different expectations at the lower energy. 
At low $x_F$, the BH loss calculations are governed by the decrease 
shown in Fig.~\ref{loss3}.  This effect is enhanced with
singlet absorption since the $\psi$ and $\psi'$ have some probability to
interact with their full cross
sections.  The maximum BH loss is so large that the first component of
Eq.~(\ref{twocom}) is less than the intrinsic charm contribution so that rather
than having a peak in $\alpha$ at low $x_F$, a minimum is seen instead.
Results with the GM model
and the minimum and original BH losses are similar for $x_F > 0.5$ due to
intrinsic charm.

Note that at this energy, the integrated values of $\alpha$, also given
in Tables~\ref{psialfoct}-\ref{psialfcombo}, indicate that the differences
between the $\psi$ and $\psi'$ $A$ dependence are again on the 2\% level
except when pure singlet absorption is considered.  In the pure singlet case, 
the absorption cross section grows more
slowly with $x_F$ at 120 GeV and some portion of the $\psi$ and $\psi'$
resonances are formed inside the target at $x_F >0$.  Even though the
$\psi'$ formation time is larger than the $\psi$ formation time, the
singlet absorption is larger for the $\psi'$ due to its increased cross
section.  High statistics measurements of the $\psi$ and $\psi'$ $A$
dependence at this energy would set bounds on the importance of singlet
production of the resonances since there is a 3\% difference in $\alpha$
between $\psi$ and $\psi'$ at 120 GeV in the singlet case.  
Note also that the $\psi$ and $\psi'$ octet and combination
$A$ dependencies are within 1-2\% of each other at this energy, generally a 
smaller difference than at 800 GeV, because the comover density is lower.

It is typically assumed that the energy dependence of $\alpha$ is small.
This is true for both the GM loss and the minimum and original BH loss
estimates where $\alpha$ changes between 2 and 6\%. 
However, the energy dependence of the maximum BH loss is much
stronger.  There is a 10-25\% decrease of the integrated $\alpha$ with this
estimate for the pure octet and singlet absorption mechanisms, as can be
seen in Tables~\ref{psialfoct} and \ref{psialfsing}.  The energy dependence 
of the combination absorption model is
typically stronger for all BH loss estimates because there is a stronger energy
dependence of BH loss when $\psi$ production is calculated in NRQCD, compare
Figs.~\ref{loss3}(a) and (b).  As seen in Table~\ref{psialfcombo}, the
resulting energy difference in $\alpha$ is as large as 35\% for the maximum BH
estimate but is only 3-7\% for the lower BH loss estimates. 
Precision measurements of $\alpha$ at 800 and 120 GeV would help eliminate 
models.  For example, the maximum BH loss could be ruled out but discerning the
difference between the $A^{1/3}$ dependence of the loss in the original BH
estimate and the $A^{2/3}$ dependence of the minimum BH loss,
Eq.~(\ref{bhminq}), could be difficult.

\begin{figure}[htbp]
\setlength{\epsfxsize=\textwidth}
\setlength{\epsfysize=0.6\textheight}
\centerline{\epsffile{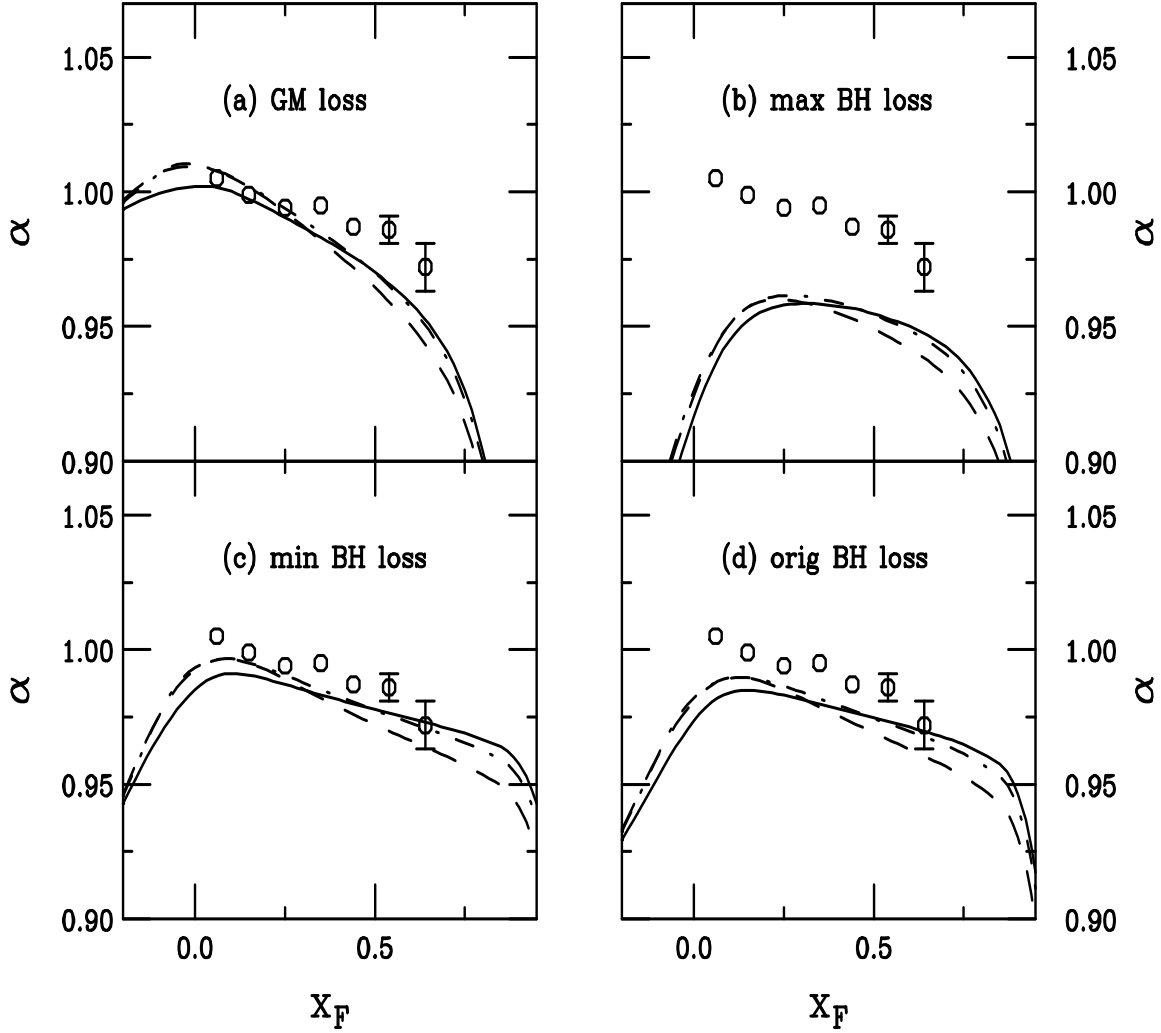}}
\caption[]{Shadowing and energy loss in Drell-Yan production are combined and 
the $A$ dependence calculated for W and D targets, shown at 800 GeV.  
The E772 Drell-Yan $A$ dependence \cite{E772DY} is also shown.
In (a), GM loss is assumed. Energy loss effects associated with the
BH bound are shown in (b), (c), and (d) for the estimated maximum and minimum
loss and the original bound respectively. 
All calculations are with the
MRST LO parton densities.  The curves represent shadowing with the
$S_3$ (solid), $S_2$ (dashed) and $S_1$ (dot-dashed) parameterizations. } 
\label{dyhigh}
\end{figure}

\begin{figure}[htbp]
\setlength{\epsfxsize=\textwidth}
\setlength{\epsfysize=0.6\textheight}
\centerline{\epsffile{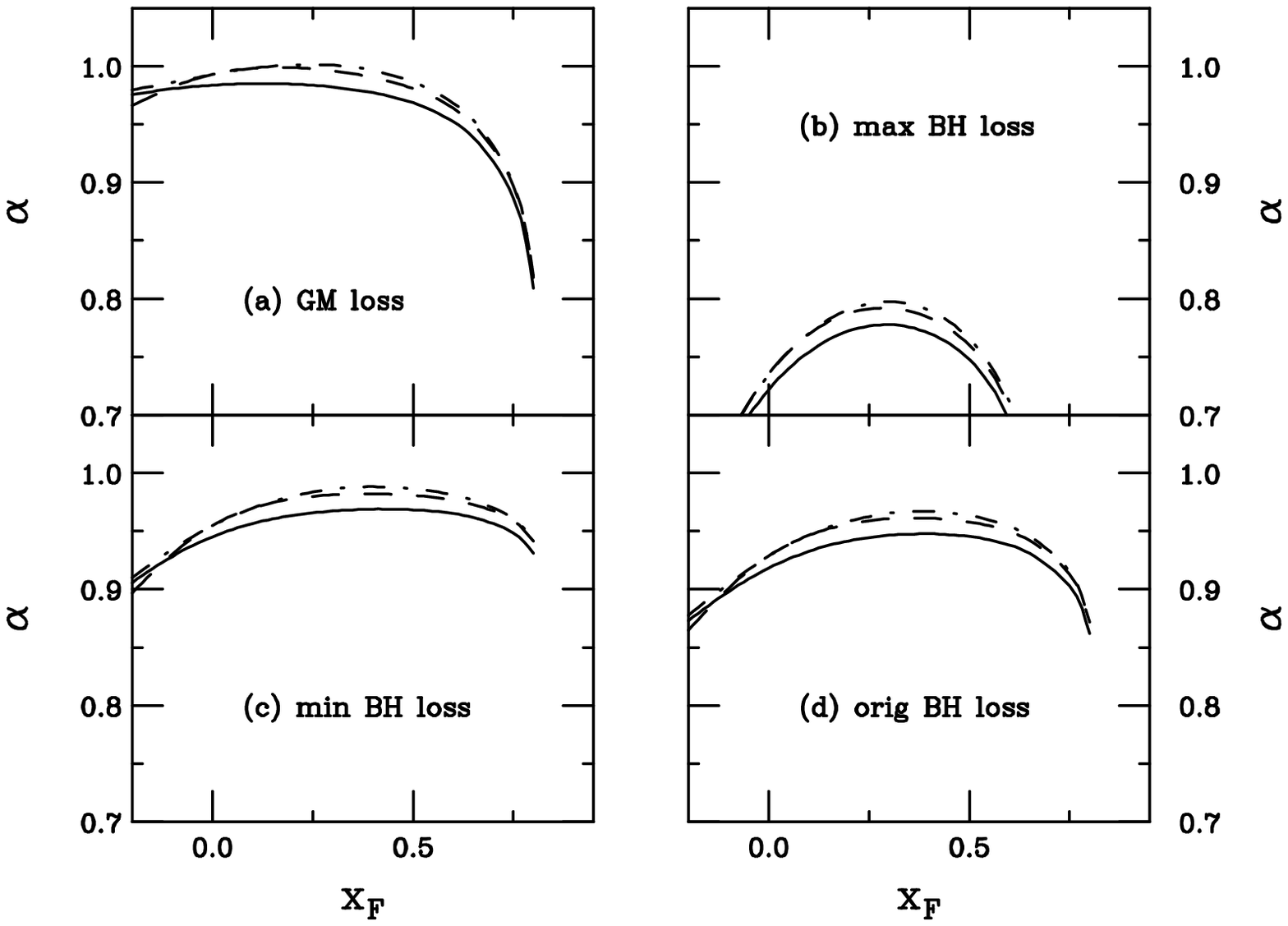}}
\caption[]{Shadowing and energy loss in Drell-Yan production are combined and 
predictions of the $A$ dependence for W and Be targets given at 120 GeV. 
In (a), GM loss is assumed. Energy loss effects associated with the
BH bound are shown in (b), (c), and (d) for the estimated maximum and minimum
loss and the original bound respectively. 
All calculations are with the
MRST LO parton densities.  The curves represent shadowing with the
$S_3$ (solid), $S_2$ (dashed) and $S_1$ (dot-dashed) parameterizations. } 
\label{120dy}
\end{figure}

Finally we compare the difference between the Drell-Yan $A$ dependencies with
shadowing and energy loss at 800
and 120 GeV in Figs.~\ref{dyhigh} and \ref{120dy} respectively.  Recall
that the Drell-Yan mass range is $4<M<9$ GeV.  All three
shadowing parameterizations are shown for each energy loss estimate.
We have included the E772 measurement of the Drell-Yan $A$ dependence
based on D, Ca, Fe and W targets \cite{E772DY} in
Fig.~\ref{dyhigh}.  Since the GM loss parameter, $\epsilon_q$, is tuned to this
data without shadowing, the full model calculation overestimates the $A$
dependence at large $x_F$.  All the calculations with BH loss predict a
significantly less
than linear $A$ dependence for $x_F < 0$ at 800 GeV, as shown in
Fig.~\ref{loss3}.  The maximum BH loss 
estimate is in complete disagreement with the E772 data which shows a rather
mild $A$ dependence.   The deviation of the model from the data 
in Fig.~\ref{dyhigh}(b) suggests that the maximum estimated BH loss is
certainly too large.  Better agreement with the data is found with the original
BH loss in Fig.~\ref{dyhigh}(d) for the $S_1$ and $S_3$ shadowing
parameterizations although the calculations consistently overestimate the $A$
dependence.  When the difference in the color factors is included, the minimum
BH quark loss, Eq.~(\ref{bhminq}), agrees reasonably well with the data.
Indeed, this model is the only one that follows the trend of the E772 data
over all $x_F$. 

The 120 GeV calculation shows a significant energy dependence of $\alpha$ when
BH loss is considered.  The $A$ dependence is now
calculated for W and Be targets, as used in the E866 experiment. 
A precision measurement of $\alpha(x_F)$ at this energy
could reveal if the $A$ dependence is less than linear here.
There is, up to now, no data on the $A$ dependence of Drell-Yan production with
a proton projectile below 200 GeV.  A measurement of $\alpha$ at 200 GeV
is for low masses, $1.7 <M< 2.7$ GeV \cite{na38dy}.  Earlier measurements at
lower energies are only available with pion beams \cite{na10dy,other}.  The
lowest energy pion beam data, at 140 GeV, obtained $\alpha = 0.980 \pm
0.006 \pm 0.013$ \cite{na10dy} which leaves room for a less than linear $A$
dependence at the lower proton beam energy.  Other pion data below 280 GeV
\cite{other} does not have enough statistical significance to determine whether
$\alpha$ deviates from unity at this energy, particularly at large $x_F$.

\begin{figure}[htbp]
\setlength{\epsfxsize=\textwidth}
\setlength{\epsfysize=0.3\textheight}
\centerline{\epsffile{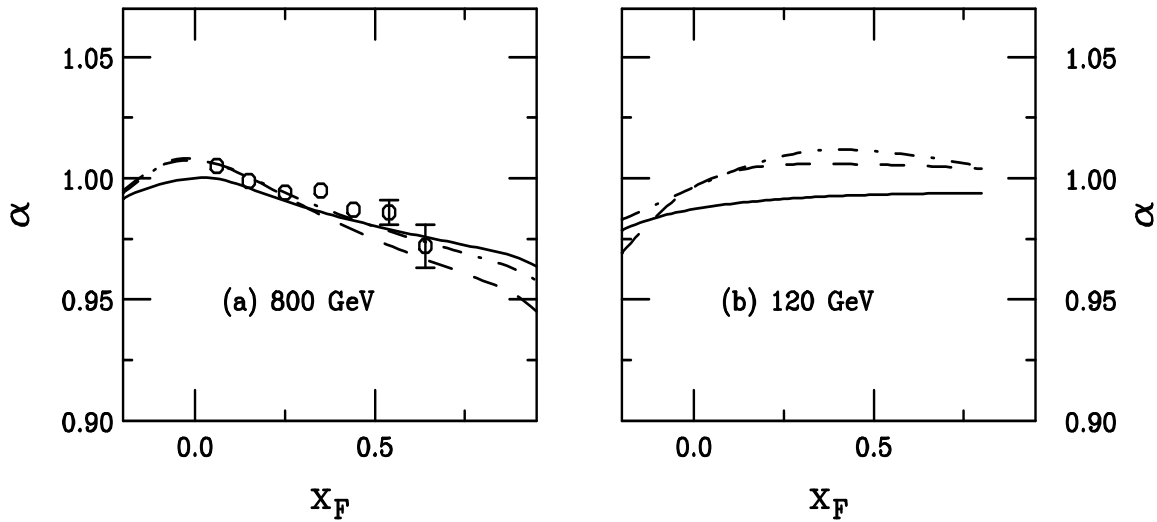}}
\caption[]{Shadowing effects on Drell-Yan production at 800 (a) and 120 (b)
GeV.  All calculations are with the
MRST LO parton densities.  The curves represent shadowing with the
$S_3$ (solid), $S_2$ (dashed) and $S_1$ (dot-dashed) parameterizations. Note
that in (a), $\alpha$ is calculated with W and D targets for comparison with
the E772 \cite{E772DY} data while the calculations at 120 GeV assume W and Be
targets.} 
\label{shaddy}
\end{figure}

Since the E866 Drell-Yan data seems to indicate that only shadowing is
necessary to explain the Drell-Yan $A$ dependence, in Fig.~\ref{shaddy}(a), we
compare the  E772 $A$ dependence with calculations of shadowing effects alone.
All three shadowing parameterizations are in reasonable agreement with
the data.  That should be expected because the $S_2$ and $S_3$
parameterizations included these data in their fits.
A comparison of Figs.~\ref{dyhigh}(c) and \ref{shaddy}(a) shows that the
minimum BH quark loss has a very weak effect on the shape of $\alpha(x_F)$.
The energy loss is only obvious in the curves at $x_F < 0.1$ and $x_F > 0.85$
and has little apparent influence on the agreement with the data.  Thus, at
least for this case, energy loss may be present but the effects nearly
indistinguishable from those due to shadowing alone.
The 120 GeV predictions with shadowing only are shown in Fig.~\ref{shaddy}(b).
Contrary to the results in Fig.~\ref{120dy}, shadowing alone predicts a
negligible influence on the $A$ dependence at the lower energy.  A clear
distinction can be made between models which include energy loss and those
which do not with this measurement.  Even a very small quark loss, such as
that in the minimum BH quark loss estimate, causes the $A$ dependence to be
less than linear at 120 GeV while shadowing alone would suggest that the $A$
dependence is either exactly linear or slightly greater than linear over all
$x_F$.  A high statistics measurement of the Drell-Yan $A$ dependence at 120
GeV could decisively settle the issue. 

The integrated $\alpha$ values 
for all the calculations at 800 GeV and 120 GeV assuming W and Be targets as
in the E866 experiment are given in Table~\ref{dyalf}.  The choice of Be as
the lowest mass target results in lower values of $\alpha$ than for a very
light target like D since nuclear effects are larger in Be than D.  When the GM
loss is considered, all shadowing parameterizations are consistent with a 1\% 
determination of $\alpha = 1$ at 800 GeV.  At the lower energy, $\alpha$ is up
to 2\%
away from unity with the $S_3$ parameterization.  
These results can be expected from the shadowing of the sea
quark distributions in the `antishadowing' range, $0.1<x<0.3$, in the $S_3$
parameterization.  (Compare Figs.~\ref{shad1}(c) and \ref{shad2}(c) with 
Fig.~\ref{shad3}(d).)  If the original BH model is correct, a less than linear
integrated $A$ dependence should already have been observed at 800 GeV.  Since
this is not the case, it seems that all but the minimum estimates 
are too large to explain
the current Drell-Yan data.  We note that even a reduced loss which agreed
with the 800 GeV data would still predict $\alpha<1$ at 120 GeV due
to the energy dependence of the BH model which should be observable with high
precision data.  These results can be compared to the $\alpha$ predicted for
shadowing alone, all compatible with unity at 120 GeV.

\section{Conclusions}

In this paper we have tested a number of different nuclear effects:  nuclear
absorption, comover scattering, nuclear shadowing, energy loss and the $A$
dependence of intrinsic
charm.  We have not attempted to make precise fits but rather check the
resulting shape of $\alpha(x_F)$ for a number of processes that
are expected to contribute to $\psi$ `suppression' in $pA$ collisions.  The
processes with the strongest influence on the $x_F$ dependence are shadowing
and energy loss.

It is clear from the comparisons of the $\psi$ and $\psi'$
calculations with the data that
a single mechanism cannot describe the shape of $\alpha(x_F)$ for all $x_F$.
Combining all effects can explain some portion of the data, depending on which
model of energy loss is assumed.  A constant energy loss, \'{a} la Gavin and
Milana \cite{GM}, can describe the forward data
when combined with the other effects
discussed here but results in values of $\alpha$ too large at low $x_F$ due to
gluon antishadowing.  

An energy dependent model of energy loss, like that proposed by Brodsky
and Hoyer \cite{BH} with refinements by Baier {\it et al.}\
\cite{bdmps1,bdms}, is less influenced by antishadowing 
because the dependence $\Delta x_1 \propto 1/x_1 s$
produces a strong $x_1$ shift at negative $x_F$, even
strong enough to counteract the gluon antishadowing.  At $x_F < 0$, the
application of the model becomes problematic because $\Delta x_1$ grows larger
than $x_1$.  Therefore the results in this region should be treated
cautiously.  This type of model alone fails
to explain the data at larger $x_F$ because the $x_1$ shift becomes 
too small to
cause a large enough change in the parton distributions for $x_F > 0.25$.  
The BH loss estimates are rather crude.  For example,
the minimum BH loss was calculated assuming that the gluon distribution in the
proton remains
relatively constant over the $x_1$ range of the data.  If the nuclear gluon 
density as a function of $x_1$ was considered instead, the results may be
more compatible with the data.  It is unlikely however that such an effect
could be significant at large $x_F$ since the
gluon distribution decreases strongly at large $x_1$, thereby weakening the
effect at large $x_F$. Thus if the BH model of energy loss is 
correct, a combination of BH loss and shadowing alone cannot describe the data,
further strong absorption at large $x_F$ is still needed.  

The apparently 
stronger absorption at large $x_F$ seen in the NA3 data \cite{NA3} was the
motivation for introducing $\psi$ production by hadronization of
intrinsic charm states \cite{VBH1}.  Indeed, without intrinsic
charm, $\alpha(x_F)$ at large $x_F$ would be even further above the data, see
Fig.~\ref{psiictest}.  We
have used an effective intrinsic charm production probability of 1\%,
within the uncertainties of the production probability determined from a fit
to deep-inelastic scattering \cite{hsv}.  

We have primarily used the MRST LO and CTEQ 3L parton densities in our
calculations.  While some deviations in the expectations of $\alpha(x_F)$
appear for other sets of parton densities,
these are not significant enough to change our general conclusions.  The most
important effect in the determination of the $A$ dependence is the nature of
the  energy loss.  Understanding this loss requires correlation of the $\psi$
$A$ dependence with that of other processes.

The preliminary data thus seem to suggest that final-state absorption,
regardless of the mechanism, is not as
strong as previously expected from studies of the integrated $A$ dependence
\cite{klns} when other nuclear effects are included.  
A smaller cross section is needed than determined from
absorption alone.  This would be true even if comover effects are neglected
since they are very small in $pA$ interactions.  If the energy loss of quarks
and gluons is treated on an equal footing, the BH loss mechanism
results in a stronger $A$ dependence than required for the Drell-Yan data.
Indeed, the minimum BH loss shows that treating the energy loss of 
quarks and gluons separately can lead to qualitative agreement with the $\psi$
and Drell-Yan $A$ dependence.

Further data on the $A$ dependence at 120 GeV could clarify 
the relative importance of octet and
singlet states in the production and absorption 
of the charmonium states.  A Drell-Yan measurement at this energy may
decisively determine the importance of energy loss by the initial partons.
In addition, precision
measurements of the $x_F$ dependent absolute cross sections in $pp$ collisions
could show whether
the $x_F$ distribution is closer to that expected from the CEM or NRQCD.

The lower energy data can also provide an additional point of comparison to the
NA50 measurements of $\psi$ suppression in heavy-ion collisions at the CERN SPS
\cite{na50}.  Only nuclear absorption and some comover scattering has been used
to compare to the $x_F$-integrated data \cite{na50,RV,klns}.  Shadowing should
also be included in the analysis \cite{spenpsi}.  Interestingly, the
$\alpha(x_F)$ extracted from the E866 800 GeV data \cite{MJL}
in the NA50 $x_F$ 
region is larger than that obtained by NA50 between 158 GeV and 450 GeV
\cite{na50}.   An independent measurement at a similar energy could be very
valuable.\\[2ex]

{\bf Acknowledgments}
I would like to thank R. Baier, S.J. Brodsky, S. Gavin,
D. Geesaman, D. Kharzeev, M.J. Leitch, A.H. Mueller, H. Satz, and C. Spieles 
for helpful discussions. I thank K.J. Eskola for providing the shadowing
routines and for discussions.  I also
thank the Institute for Nuclear Theory in Seattle and Los Alamos
National Laboratory for hospitality.

\begin{table}[htb]
\begin{center}
\begin{tabular}{|c|c|c|c|c|} \hline
 & octet & singlet & \multicolumn{2}{c|}{combination} \\ 
 & $\sigma_{\psi N}^{\rm o}$ (mb) & $\sigma_{\psi N}^{\rm s}$ (mb) 
& $\sigma_{\psi N}^{\rm
 octet}$  (mb) & $\sigma_{\psi N}^{\rm singlet}$ (mb) \\ \hline
$S_1$ & 2 & 8 & 3 & 2 \\ \hline
$S_2$ & 1 & 5 & 1 & 1 \\ \hline
$S_3$ & 3 & 10 & 3 & 5 \\ \hline
\end{tabular}
\end{center}
\caption[]{The $\psi$ absorption cross sections used with each shadowing
parameterization.  Note that the corresponding $\psi'$ absorption cross
sections are the same as those for the $\psi$ in the octet case and
a factor of 3.7 larger for singlet production.  In all cases the comover cross
sections are $\sigma_{\psi {\rm co}} = 0.67$ mb and $\sigma_{\psi' {\rm co}} =
3.7 \sigma_{\psi {\rm co}}$.}
\label{sigtable}
\end{table}

\begin{table}[htb]
\begin{center}
\begin{tabular}{|c|c|c|c|c|c|} \hline
 & & \multicolumn{2}{c|}{$\psi$} & \multicolumn{2}{c|}{$\psi'$} \\ 
Model & $S$ & 800 GeV & 120 GeV & 800 GeV & 120 GeV \\ \hline
            & $S_1$ & 0.938 & 0.922 & 0.918 & 0.907 \\
GM          & $S_2$ & 0.959 & 0.953 & 0.939 & 0.938 \\
            & $S_3$ & 0.951 & 0.922 & 0.931 & 0.907 \\ \hline
            & $S_1$ & 0.984 & 0.985 & 0.982 & 0.979 \\
KS          & $S_2$ & 0.980 & 0.998 & 0.978 & 0.992 \\
            & $S_3$ & 1.014 & 1.002 & 1.012 & 0.996 \\ \hline
            & $S_1$ & 0.805 & 0.661 & 0.786 & 0.648 \\
max BH      & $S_2$ & 0.823 & 0.685 & 0.803 & 0.672 \\
            & $S_3$ & 0.817 & 0.658 & 0.798 & 0.646 \\ \hline
            & $S_1$ & 0.872 & 0.843 & 0.852 & 0.828 \\
min BH      & $S_2$ & 0.891 & 0.872 & 0.871 & 0.857 \\
            & $S_3$ & 0.884 & 0.842 & 0.865 & 0.827 \\ \hline
            & $S_1$ & 0.890 & 0.850 & 0.870 & 0.836 \\
orig BH     & $S_2$ & 0.908 & 0.875 & 0.888 & 0.864 \\
            & $S_3$ & 0.903 & 0.851 & 0.883 & 0.836 \\ \hline
\end{tabular}
\end{center}
\caption[]{The integrated value of $\alpha$ for $\psi$ and $\psi'$ production
at 800 GeV and 120 GeV assuming pure octet absorption.  Note that 
$\alpha$ is integrated over the range $-0.2 \leq x_F \leq 0.8$ at both
energies.}
\label{psialfoct}
\end{table}

\begin{table}[htb]
\begin{center}
\begin{tabular}{|c|c|c|c|c|c|} \hline
 & & \multicolumn{2}{c|}{$\psi$} & \multicolumn{2}{c|}{$\psi'$} \\ 
Model & $S$ & 800 GeV & 120 GeV & 800 GeV & 120 GeV \\ \hline
            & $S_1$ & 0.958 & 0.916 & 0.933 & 0.882 \\
GM          & $S_2$ & 0.968 & 0.944 & 0.944 & 0.914 \\
            & $S_3$ & 0.983 & 0.924 & 0.957 & 0.886 \\ \hline
            & $S_1$ & 0.829 & 0.658 & 0.805 & 0.631 \\
max BH      & $S_2$ & 0.833 & 0.679 & 0.811 & 0.655 \\
            & $S_3$ & 0.854 & 0.661 & 0.830 & 0.632 \\ \hline
            & $S_1$ & 0.894 & 0.840 & 0.870 & 0.808 \\
min BH      & $S_2$ & 0.901 & 0.865 & 0.878 & 0.836 \\
            & $S_3$ & 0.920 & 0.846 & 0.895 & 0.811 \\ \hline
            & $S_1$ & 0.912 & 0.848 & 0.888 & 0.816 \\
orig BH     & $S_2$ & 0.918 & 0.873 & 0.895 & 0.845 \\
            & $S_3$ & 0.938 & 0.855 & 0.913 & 0.821 \\ \hline
\end{tabular}
\end{center}
\caption[]{The integrated value of $\alpha$ for $\psi$ and $\psi'$ production
at 800 GeV and 120 GeV assuming pure singlet absorption.  Note that 
$\alpha$ is integrated over the range $-0.2 \leq x_F \leq 0.8$ at both
energies.}
\label{psialfsing}
\end{table}

\begin{table}[htb]
\begin{center}
\begin{tabular}{|c|c|c|c|c|c|} \hline
 & & \multicolumn{2}{c|}{$\psi$} & \multicolumn{2}{c|}{$\psi'$} \\ 
Model & $S$ & 800 GeV & 120 GeV & 800 GeV & 120 GeV \\ \hline
            & $S_1$ & 0.948 & 0.939 & 0.926 & 0.917 \\
GM          & $S_2$ & 0.977 & 0.983 & 0.955 & 0.963 \\
            & $S_3$ & 0.978 & 0.956 & 0.954 & 0.929 \\ \hline
            & $S_1$ & 0.804 & 0.597 & 0.788 & 0.583 \\
max BH      & $S_2$ & 0.831 & 0.637 & 0.814 & 0.625 \\
            & $S_3$ & 0.835 & 0.612 & 0.816 & 0.593 \\ \hline
            & $S_1$ & 0.880 & 0.830 & 0.863 & 0.817 \\
min BH      & $S_2$ & 0.907 & 0.872 & 0.890 & 0.860 \\
            & $S_3$ & 0.911 & 0.847 & 0.892 & 0.830 \\ \hline
            & $S_1$ & 0.899 & 0.853 & 0.881 & 0.835 \\
orig BH     & $S_2$ & 0.926 & 0.896 & 0.908 & 0.879 \\
            & $S_3$ & 0.931 & 0.871 & 0.910 & 0.847 \\ \hline
\end{tabular}
\end{center}
\caption[]{The integrated value of $\alpha$ for $\psi$ and $\psi'$ production
at 800 GeV and 120 GeV assuming a combination of octet and 
singlet absorption.  Note that 
$\alpha$ is integrated over the range $-0.2 \leq x_F \leq 0.8$ at both
energies.}
\label{psialfcombo}
\end{table}

\begin{table}[htb]
\begin{center}
\begin{tabular}{|c|c|c|c|} \hline
Model & $S$ & 800 GeV & 120 GeV \\ \hline
          & $S_1$ & 0.998 & 0.994 \\
GM        & $S_2$ & 0.997 & 0.990 \\
          & $S_3$ & 0.993 & 0.980 \\ \hline
          & $S_1$ & 0.916 & 0.757 \\
max BH    & $S_2$ & 0.914 & 0.753 \\
          & $S_3$ & 0.910 & 0.740 \\ \hline
          & $S_1$ & 0.980 & 0.967 \\
min BH    & $S_2$ & 0.979 & 0.964 \\
          & $S_3$ & 0.974 & 0.953 \\ \hline
          & $S_1$ & 0.973 & 0.944 \\
orig BH   & $S_2$ & 0.972 & 0.941 \\
          & $S_3$ & 0.968 & 0.930 \\ \hline
          & $S_1$ & 0.997 & 1.002 \\
no loss   & $S_2$ & 0.997 & 0.999 \\
          & $S_3$ & 0.993 & 0.989 \\ \hline
\end{tabular}
\end{center}
\caption[]{The integrated value of $\alpha$ for Drell-Yan production
at 800 GeV and 120 GeV for different model assumptions.  Note that
$\alpha$ is integrated over the range $-0.2 \leq x_F \leq 0.8$ at both
energies.}
\label{dyalf}
\end{table}

\end{document}